\documentclass[reprint,superscriptaddress,amsmath,amsfonts,amssymb,aps,prb,showkeys]{revtex4-2}
\usepackage{tabularx}
\usepackage{graphicx}
\usepackage{dcolumn}
\usepackage{bm, color}
\usepackage{amsmath}
\usepackage{amsfonts}
\usepackage{amssymb}
\usepackage{amsmath}
\usepackage{here}
\usepackage{multirow}
\usepackage{ulem}
\usepackage{cancel}
\providecommand{\U}[1]{\protect\rule{.1in}{.1in}}

\newcommand{\vv}[1]{\boldsymbol #1}%

\usepackage{hyperref}
\hypersetup{
	colorlinks = true,
}
\usepackage{physics}

\usepackage[whole]{bxcjkjatype}

\begin{document}

\title{Raman response of collective modes in multicomponent superconductors}

\author{Yuki Yamazaki}
\affiliation{Department of Applied Physics, The University of Tokyo, Hongo, Tokyo, 113-8656, Japan}
\author{Takahiro Morimoto}
\affiliation{Department of Applied Physics, The University of Tokyo, Hongo, Tokyo, 113-8656, Japan}
\date{\today}

\date{\today}

\begin{abstract}
We formulate a microscopic theory of the Raman response of superconducting collective modes in multicomponent superconductors. Starting from a general Bogoliubov--de Gennes (BdG) Hamiltonian with a separable pairing interaction, we derive a gauge-invariant expression for the Raman susceptibility, including a long-range Coulomb interaction.
The resulting Raman susceptibility is directly computable for an arbitrary BdG Hamiltonian, which contains single- and multiband systems, spin-singlet and triplet order parameters, and time-reversal-symmetric and time-reversal-symmetry-breaking superconducting states.
Based on the microscopic coupling between a Raman source field and collective modes, we derive a symmetry selection rule for Raman-active collective modes and show a group-theoretical classification for all crystalline point groups.
This classification provides a unified framework based on the ``higher-order Lifshitz-invariant'' to identify Raman-active collective modes such as Leggett mode, Bardasis--Schrieffer (BS) mode, and clapping mode.
As an application, we focus on an effective model of the heavy-fermion superconductor UTe$_2$ with a fully gapped multicomponent odd-parity pairing state. We find sharp in-gap Raman resonances below the quasiparticle continuum, which do not correspond to a conventional Leggett mode but arise from the {\it intraband} relative modes between different pairing components.
\end{abstract} 

\maketitle
      
\makeatletter
\def\ext@table{}
\makeatother
\makeatletter
\def\ext@figure{}
\makeatother

\section{Introduction}\label{sec: Introduction}

Collective excitations of the superconducting order parameter provide a direct window into the internal structure and symmetry of superconductors. In the simplest case of a single-band $s$-wave superconductor, the complex order parameter supports massive amplitude (Higgs) \cite{Anderson1900, Schmid129, Littlewood811, Littlewood4883, Pekker269, Shimano103, Tsuji} and massless phase (Nambu--Goldstone) fluctuations \cite{Nambu345, Goldstone154}, where the phase mode is lifted to the plasma frequency through the Anderson--Higgs mechanism \cite{Anderson439, Higgs508} and the Higgs mode appears at twice the superconducting gap \cite{Littlewood4883}. 
In contrast, superconductors with multicomponent order parameters can host a far richer variety of collective modes. A prominent example in multiband systems is the relative-phase oscillation between condensates on different bands, known as the Leggett mode \cite{Leggett901}. Because it corresponds to a relative phase rather than the overall phase, it is not lifted to the plasma frequency by the Anderson--Higgs mechanism, and it can therefore reside inside the quasiparticle gap and produce a sharp in-gap resonance. 

Multicomponent physics also arises even within a single band when multiple pairing channels exist. In such cases, fluctuations in a subdominant (noncondensed) pairing channel can give rise to the Bardasis--Schrieffer (BS) mode \cite{Bardasis1050}. Another characteristic excitation is the clapping mode \cite{Wolfle1279, Tewordt1007} (also referred to as an orbital-magnetization mode \cite{Balatsky4445}), which appears in chiral superconductors as a fluctuation of the Cooper-pair angular momentum, and this mode can also be regarded as the BS-type mode \cite{Nicholas44}. These collective modes may likewise occur below the quasiparticle continuum and exhibit in-gap resonances. Experimentally, such resonances have indeed been observed in Raman and optical probes: Leggett modes in MgB$_2$ \cite{Blumberg227002}, and BS modes in iron-based superconductors \cite{Kretzschmar187002, Bohm041046, Jost020504, He217002, Matsumoto2025}.

How these collective modes couple to external electromagnetic fields has been extensively discussed. For example, since the Higgs mode does not couple linearly to electromagnetic fields in the presence of the particle-hole symmetry \cite{Varma901, Pekker269, Tsuji064508}, many studies have focused on nonlinear optical responses \cite{Tsuji064508, Kemper224517, Cea180507, Tsuji224519, Jujo024704, Silaevl224511, Schwarz287, Tsuji043029, Seibold014512, Haenel134504, Udina168, Oh2025, Matsunaga057002, Matsunaga1254697, Matsunaga020505, Katsumi117001, Chu1793}. In contrast, Leggett modes can become linearly optically active where inversion symmetry is broken in superconducting states \cite{Kamatani094520, Nagashima013120}. BS and clapping modes also show linear optical response \cite{Lee307, Matsushita2026}. Furthermore, Ref.~\cite{Nagashima013120} developed a group-theoretical classification of possible linear couplings between Leggett modes and electromagnetic fields in terms of so-called Lifshitz invariants \cite{Lifshitz} appearing in the Ginzburg--Landau free energy. In particular, when the normal state preserves inversion symmetry, the two order parameters whose relative phase forms the Leggett mode must belong to irreducible representations of opposite parity in order for such a linear optical coupling to be allowed.

Raman spectroscopy provides another powerful route to detecting collective modes in superconductors \cite{Klein4976, Devereaux175}. For instance, Higgs-mode signatures have been reported in layered two-dimensional superconductors \cite{Sooryakumar660, Measson060503, Grasset094502, Grasset127001, Majumdar084005}. Importantly, Raman scattering is a two-photon process (schematically $\sim A_i A_j$) and is therefore even parity under inversion symmetry. Consequently, even in centrosymmetric crystals, relative modes between multiple order-parameter components of the same parity can, in principle, couple to a Raman source field. Previous studies have addressed Raman responses of collective modes beyond the Higgs mode: in multiband or multicomponent superconductors, Raman scattering couples to the Leggett and BS modes and shapes the $A_{g}$ or $B_{g}$ response \cite{Scalapino140512, Lee445701, Klein014507, Khodas245134, Maiti257001, Cea064512, Maiti014503, Igor104505, Sarkar094515}. 

Despite these theoretical studies, a general and practically computable Raman theory that (i) applies to an arbitrary Bogoliubov--de Gennes (BdG) Hamiltonian and (ii) treats multicomponent order parameters in a basis-independent manner is still lacking, especially for spin-triplet superconductors. Moreover, a systematic symmetry-based selection rule, which is valid for arbitrary crystalline point groups and Raman vertices, has not yet been established for identifying which collective modes can couple to a given Raman vertex.

In this work, we formulate the Raman response of such collective excitations in multicomponent superconductors. Under the assumptions of (i) separable pairing interactions, (ii) Gaussian fluctuations around the mean-field solution, (iii) a nonresonant Raman vertex, and (iv) neglecting couplings to additional bosonic modes (e.g., phonons, magnons, and excitons) as well as impurity-induced damping (clean limit),
we derive a gauge-invariant expression for the Raman susceptibility that is directly computable for an arbitrary BdG Hamiltonian and an arbitrary Raman vertex. In practical terms, once a BdG Hamiltonian is specified as an input, the Raman spectrum follows immediately from the BdG eigenvalues and eigenvectors through explicit kernel expressions. This represents an extension of the previous general formulation for spin-singlet superconductors \cite{Maiti014503, Sarkar094515} to spin-triplet pairings. Within the above assumptions, our framework is applicable irrespective of the choice of basis, the presence or absence of time-reversal symmetry, and the detailed structure of superconducting pairings.

We further provide a group-theoretical classification of the linear coupling between collective modes and Raman source fields. This classification is similar to the symmetry analysis of Lifshitz invariants for linear optical responses \cite{Nagashima013120}. Since the Raman response considered here is proportional to the second order in the vector potential, it may be viewed as a higher-order counterpart of the Lifshitz-invariant (i.e., ``second-order Lifshitz-invariant'') classification \cite{Takasan2025}. In particular, we identify the combinations of order-parameter components that can couple to a given Raman vertex field for all point groups.

Finally, we apply our theory to the recently and extensively studied superconductor UTe$_2$ \cite{Ran684, aoki2022}, which has been discussed as a leading candidate for time-reversal-symmetric spin-triplet superconductivity \cite{Matsumura063701, theuss2024single, Suetsugu3772, Li2419734122, Hayes021029, QGu2025, Wang1555} and topological superconductivity \cite{Ishizuka217001, Tei144517, Yamazaki174507}. Using a fully gapped spin-triplet $A_u$ pairing state with multiple order-parameter components belonging to the same irreducible representation, we find that {\it intraband} relative collective modes between these components can possess resonance frequencies inside the quasiparticle gap. We also show that these collective modes appear as the peak structures in the Raman spectrum. 

This paper is organized as follows. In Sec. \ref{sec: Microscopic model}, we present a microscopic path-integral formalism for the Raman response of superconducting collective modes in multicomponent superconductors.
By introducing a scalar field that represents the long-range Coulomb interaction, we obtain a gauge-invariant expression for the Raman susceptibility. 
In Sec. \ref{sec: Group theoretical classification of Raman response}, we perform a group-theoretical classification of the coupling between Raman source fields and superconducting collective modes, providing systematic criteria and classification tables for all point groups. In Sec. \ref{sec: Application to concrete model}, we apply our formalism to a concrete tight-binding model for the UTe$_2$ superconductor and demonstrate that a fully gapped multicomponent superconducting state can exhibit sharp in-gap Raman resonances associated with relative amplitude/phase oscillations between different order-parameter components.

\section{Microscopic model and Raman response of collective modes}
\label{sec: Microscopic model}

In this section, we formulate a microscopic theory of the Raman response in a general multicomponent superconductor. Our strategy follows the functional-integral formalism developed in Appendix~\ref{app: Microscopic derivation for Raman response}: we start from a BdG Hamiltonian with separable attractive interactions in the pairing channels, couplings among fermions, a Raman source field, and a scalar potential that represents the long-range Coulomb interaction. After integrating out fermionic degrees of freedom, we obtain an effective action for the fluctuations of superconducting order parameters and the scalar field. By expanding this action to quadratic order in the fluctuating bosonic fields and integrating them out, we obtain a gauge-invariant expression for the Raman susceptibility, which is valid for an arbitrary BdG Hamiltonian. 

\subsection{Microscopic action}
\label{subsec: action}

We consider a general multiband system described by fermionic operators $c_{\alpha}(\bm k)$, where $\alpha$ collectively labels spin, orbital, and sublattice degrees of freedom. The non-interacting Euclidean action is
\begin{align}
S_0[c^{*},c]
= \int^{\beta}_0 d\tau \sum_{\bm{k}}\bm{c}^{\dagger}_{\bm{k}}(\tau)
\big[\partial_{\tau}+H_0(\bm{k})\big]\bm{c}_{\bm{k}}(\tau),
\end{align}
where $\bm{c}_{\bm k}=[c_{\alpha,\bm k}]^{\text T}$ and $H_0(\bm{k})$ is the normal-state Hamiltonian. We assume that superconductivity is described by the following BdG Hamiltonian
\begin{align}
\nonumber
H_{\text{BdG}}(\bm{k}) &= \left[
\begin{array}{cc}
H_0(\bm{k}) & \Delta(\bm{k}) \\
\Delta^{\dagger}(\bm{k}) & -H_0^{\text{T}}(-\bm{k})
\end{array}\right]_{2N\times 2N}, \\
\Delta(\bm{k}) &= \sum_{\alpha=1}^{N_{\Delta}} \Delta^{\text{MF}}_{\alpha}\,\hat{\Delta}_\alpha(\bm{k}),
\label{eq-main: BdG-Hamiltonian-multi}
\end{align}
where $N_{\Delta}$ is the number of nonzero order-parameter components, $\hat{\Delta}_{\alpha}(\bm{k})$ are form factors of each order parameter, and the amplitudes $\Delta^{\text{MF}}_{\alpha} \in \mathbb{R}$ are determined by the gap equations (Appendix~\ref{app: gap-equation}). Fermi statistics impose $\hat{\Delta}^{\text{T}}_{\alpha}(\bm k) =-\hat{\Delta}_{\alpha}(-\bm k)$. This BdG Hamiltonian can describe single- and multiband systems, spin-singlet and triplet order parameters, and both time-reversal-symmetric (TRS) and time-reversal-symmetry-breaking (TRSB) superconducting states.

We assume that superconductivity arises from a short-range attractive interaction that is separable in the pairing channel,
\begin{align}
\nonumber
H_{\text{int}}(\tau) &= -\frac{1}{4}\sum_{i=1}^{N'_{\Delta}}\bm{\rho}_i^{\dagger}(\tau)X_i\bm{\rho}_i(\tau), \\ \nonumber
\bm{\rho}_i(\tau) &\equiv (\rho_{i,\bm{q}_1}(\tau),\rho_{i,\bm{q}_2}(\tau),...),\\ \nonumber
\rho_{i,\bm{q}}(\tau)&=U_i\sum_{\bm{k}}\bm{c}^{\text{T}}_{-\bm{k}+\frac{\bm{q}}{2}}(\tau)\hat{\Delta}^{\dagger}_i(\bm{k})\bm{c}_{\bm{k}+\frac{\bm{q}}{2}}(\tau), \\
X_i&=[X_i]_{\bm{q},\bm{q}'}=(1/U_i)\delta_{\bm{q},\bm{q}'},
\label{eq-main: interaction}
\end{align}
with $U_i>0$ and $N'_{\Delta}\ge N_{\Delta}$. The form factors $\hat{\Delta}_i(\bm{k})$ span all pairing channels that are allowed microscopically; only a subset of them acquires a nonzero expectation value $\Delta^{\text{MF}}_\alpha$ in the superconducting phase. We note that even when there is no mean-field solution $\Delta^{\text{MF}}_{i \neq \alpha}$, the fluctuation for the pairing channel $\hat \Delta_{i \neq \alpha}$ can still exist. 

To derive the Raman response, we introduce the nonresonant Raman Hamiltonian \cite{Devereaux175}:
\begin{align}
&H_R(\tau) = \sum_{\bm{k},\bm{q}}
\bm{c}^{\dagger}_{\bm{k}+\frac{\bm{q}}{2}}(\tau)\big[-R(\bm{q},\tau)\gamma(\bm{k})\big]
\bm{c}_{\bm{k}-\frac{\bm{q}}{2}}(\tau), \\
&\gamma(\bm{k}) \equiv 
\sum_{ij}e^{\text{in}}_{i}e^{\text{out}}_{j}\,\partial_{k_i}\partial_{k_j}H_0(\bm{k}), \label{eq: Raman-vertex-main}
\end{align}
where $e^{\text{in}}$ and $e^{\text{out}}$ are the polarization vectors of the incoming and outgoing photons. $R(\bm q,\tau)$ is an external Raman source field. Since Raman scattering is a two-photon scattering process, the Raman vertex $\gamma(\bm{k})$ is inversion symmetric, and the point-group symmetry of $\gamma(\bm{k})$ is determined by the polarization of the incoming and outgoing photons.

Finally, we introduce a scalar field $\phi(\bm q,\tau)$ whose role is twofold. First, it provides a Hubbard--Stratonovich representation of the long-range Coulomb interaction and leads to Coulomb screening of the Raman response \cite{Devereaux12523, Devereaux175, Boyd174521, Sauer014525, Cea064512, Maiti014503, Sarkar094515}. Second, it allows us to implement gauge invariance and the Anderson--Higgs mechanism in a clear way. We couple $\phi$ to the electron density and include the electrostatic energy,
\begin{align}
&H_{\phi}(\tau) = -ie\sum_{\bm{k},\bm{q}}\phi(\bm{q},\tau)
\bm{c}^{\dagger}_{\bm{k}+\frac{\bm{q}}{2}}(\tau)\bm{c}_{\bm{k}-\frac{\bm{q}}{2}}(\tau), 
\label{eq-main: scalar1} \\ \nonumber
&S_{\text{EM}}[\phi] \equiv \frac{1}{8\pi}\int^{\beta}_0 d\tau \int d^3r \, \bm{E}^2(\bm{r},\tau),
\qquad \bm{E}= \bm{\nabla} \phi \\
&\hspace{12mm}
= \frac{1}{8\pi}\sum_{\bm{q}}\int^{\beta}_0 d\tau \,\bm{q}^2
\phi(-\bm q,\tau)\phi(\bm{q},\tau),
\label{eq-main: scalar2}
\end{align}
which reproduces the Coulomb potential $V_C(\bm q)=4\pi e^2/\bm q^2$ upon integrating out $\phi$. We note that the explicit form of $V_C(\bm q)$ affects the collective-mode dispersion at finite $\bm q$ (e.g., plasmons), whereas the $\bm{q} \to \bm{0}$ Raman response is not affected by the Coulomb interaction as $S_{\text{EM}}\to 0$ (corresponding to $V_C^{-1}\to 0$).
In particular, this decoupling (with $S_{\text{EM}}\to 0$) holds for any spatial dimension. We work in Euclidean time, with the relation between Minkowski and Euclidean scalar potentials $\phi_{\mathrm E}=i\phi_{\mathrm M}$ and actions $S_{\mathrm M}=iS_{\mathrm E}$.

Collecting all contributions, the full Euclidean action entering the partition function $\mathcal{Z}=\int D[c^{*},c,\phi]e^{-S[c^{*},c,\phi]}$ is
\begin{align}
\nonumber
S[c^{*},c,\phi] \equiv S_0[c^{*},c]+\int^{\beta}_0d\tau&\big[H_{\text{int}}(\tau)+H_{R}(\tau)\\
&+H_{\phi}(\tau)\big] + S_{\text{EM}}[\phi].
\label{eq-main: action}
\end{align}

\subsection{Gaussian fluctuation action and collective modes}
\label{subsec: fluct_action}

We decouple the attractive interaction~(\ref{eq-main: interaction}) using complex bosonic Hubbard--Stratonovich fields $\Delta_i(\bm q,\tau)$ and integrate out fermions. The detailed steps are given in Appendix~\ref{app: Microscopic derivation for Raman response}; here we summarize the final results.

We introduce the Nambu spinor
\begin{align}
\bm{\Psi}_{\bm{k}}(\tau) \equiv 
\begin{bmatrix}
\bm{c}_{\bm{k}}(\tau)  \\
\bm{c}^{*}_{-\bm{k}}(\tau) 
\end{bmatrix},
\end{align}
in terms of which the fermionic part of the action becomes quadratic. After Hubbard--Stratonovich decoupling and integrating out the fermions, we obtain an effective action for the bosonic fields,
\begin{align}
\nonumber
S[\Delta^{*},\Delta,\phi]=\sum_i\frac{\beta}{U_i}\sum_{\bm q}\sum_{p}
|\Delta_i(Q_p)|^2
\\
-\frac{1}{2}\mathrm{Tr}\,\ln[-\beta G^{-1}]
+ S_{\text{EM}}[\phi],
\label{eq: general-action-main}
\end{align}
where $Q_p \equiv (\bm q,i\Omega_p)$ and $G^{-1}$ is the inverse Nambu Green's function in the presence of the fluctuations of superconducting order parameters, the Raman source, and the scalar field (explicit expressions are given in Appendix~\ref{app: Microscopic derivation for Raman response}).

In the superconducting phase, some of the pairing fields acquire nonzero mean-field values $\Delta^{\text{MF}}_{\alpha}$ ($\alpha=1,\dots, N_\Delta$). We write
\begin{align}
\Delta_{\alpha}(\bm{r},\tau) = \Delta^{\text{MF}}_{\alpha} 
+ \delta \Delta_{\alpha}(\bm{r},\tau),
\end{align}
and fluctuations corresponding to amplitude (Higgs) and phase (Nambu--Goldstone) modes,
\begin{align}
\nonumber
\Delta^{\text{MF}}_{\alpha} + \delta\Delta_{\alpha}(\bm{r},\tau)
&\equiv \big(\Delta^{\text{MF}}_{\alpha} + H_{\alpha}(\bm{r},\tau)\big)
e^{i\theta_{\alpha}(\bm{r},\tau)} \\
&\approx \Delta^{\text{MF}}_{\alpha}+H_{\alpha}(\bm{r},\tau)
+i\Delta^{\text{MF}}_{\alpha}\theta_{\alpha}(\bm{r},\tau),
\label{eq: Higgs-NG-main}
\end{align}
where $H_{\alpha} \in \mathbb{R}$ and $\Theta_{\alpha}\equiv \Delta^{\text{MF}}_{\alpha}\theta_{\alpha} \in \mathbb{R}$ describe the Higgs and Nambu--Goldstone fluctuations, respectively.

Expanding the fermionic determinant $\mathrm{Tr}\ln[-\beta G^{-1}]$ in powers of the fluctuation fields and the external sources,
\begin{align}
\mathrm{Tr}\ln[-\beta G^{-1}]=\mathrm{Tr}\ln[-\beta G_{0}^{-1}]
-\mathrm{Tr}\Big[\sum_{L\ge1}\frac{(G_0\Sigma)^L}{L}\Big],
\end{align}
where $G_0$ is the mean-field Green's function and $\Sigma$ is the self energy accounting for pairing fluctuations and effects of the source fields. We keep terms up to quadratic order in $(\delta\Delta_{\alpha},\delta\Delta_{\alpha}^{*},\phi,R)$, which are sufficient to describe collective modes and their linear coupling to the Raman source field. Linear terms in $\delta\Delta_{\alpha}$ vanish by the mean-field gap equations (Appendix~\ref{app: gap-equation}), and we also drop constant terms proportional to $R(Q_p=0)$ and $\phi(Q_p=0)$.

\begin{figure}[t]
\centering
\includegraphics[width=86mm]{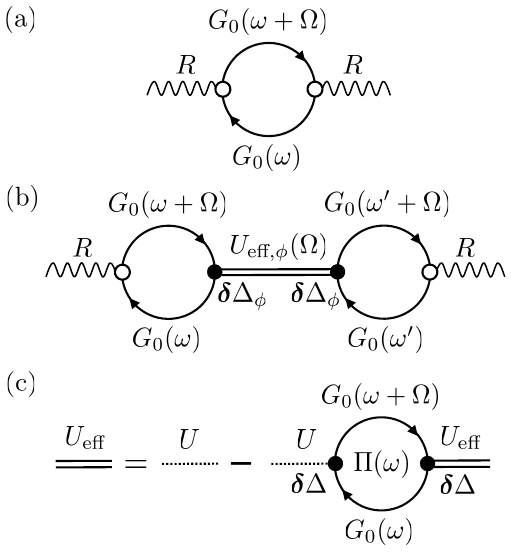}
\caption{Feynman diagram of (a,b) the Raman response in superconductors, and (c) the effective interaction $U_{\text{eff}}(\omega)$ within the random phase approximation. 
The diagram (a) corresponds to the quasiparticle excitation. The diagram (b) corresponds to the collective excitations, including the charge density fluctuation, and we replace $\bm{\delta}\Delta \equiv (\delta\Delta, \delta ^{*}\Delta)$ and $U_{\text{eff}}$ to $\bm{\delta}\Delta_{\phi} \equiv (\bm{\delta}\Delta,\phi)$ and $U_{\text{eff},\phi}$.
$U$ is the bare interaction and $\Pi$ is the polarization bubble.}
\label{fig: diagram}
\end{figure}

After performing the Matsubara-frequency summations, we obtain the quadratic action in the compact form
\begin{align}
S[\delta\Delta^{*},\delta\Delta,\phi]
= S_R + S_{\text{FL}},
\end{align}
where $S_R$ is the purely electronic Raman bubble,
\begin{align}
S_R = \frac{\beta}{4}\sum_{\bm q}\sum_{p} R(Q_p)R(-Q_p)\,\Phi_{RR}(Q_p),
\end{align}
which corresponds to the diagram as shown in Fig.~\ref{fig: diagram} (a) and $S_{\text{FL}}$ contains all fluctuations of the superconducting order parameters and the scalar field,
\begin{align}
\nonumber
S_{\text{FL}} &= 
\frac{\beta}{2}\sum_{\bm q}\sum_{p}
\bm{\delta}\Delta_{\phi}^{\text T}(-Q_p)\,
U^{-1}_{\text{eff},\phi}(-Q_p)\,
\bm{\delta}\Delta_{\phi}(Q_p) \\ \nonumber
&-\frac{\beta}{4}\sum_{\bm q}\sum_{p}
\bm{\delta}\Delta_{\phi}^{\text T}(-Q_p)\,
\bm{Q}_{R,\phi}(Q_p)\,R(Q_p) \\ 
&-\frac{\beta}{4}\sum_{\bm q}\sum_{p}
\big[\bm{Q}_{R,\phi}(-Q_p)R(-Q_p)\big]^{\text T}
\bm{\delta}\Delta_{\phi}(Q_p),
\label{eq: general-action3-main}
\end{align}
where we defined the vector of bosonic fields
\begin{align}
\nonumber
\bm{\delta}\Delta_{\phi}(Q_p)&\equiv 
\big[\delta\Delta_1(Q_p),\dots,\delta\Delta_{N_\Delta}(Q_p), \\
&\quad \quad \delta\Delta_1^{*}(-Q_p),\dots,\delta\Delta_{N_\Delta}^{*}(-Q_p),
\phi(Q_p)\big]^{\text T},
\end{align}
and the corresponding source vector
\begin{align}
\bm{Q}_{R,\phi}(Q_p) \equiv
\begin{bmatrix}
\bm{Q}_{R,\Delta}(Q_p) \\
\bm{Q}_{R,\Delta^{\dagger}}(Q_p) \\
-ie\Pi_{R,\phi}(Q_p)
\end{bmatrix},
\end{align}
where the corresponding diagram is shown in Fig.~\ref{fig: diagram} (b).
The matrix $U^{-1}_{\text{eff},\phi}(Q_p)$ is the inverse propagator (kernel) for the coupled system of order-parameter fluctuations and the scalar field,
\begin{align}
\nonumber
&U^{-1}_{\text{eff},\phi}(Q_p) \\
&\equiv
\left[
\begin{array}{c|c}
U^{-1}_{\text{eff}}(Q_p) & \begin{matrix}  \frac{-ie}{2}\bm{Q}_{\phi,\Delta}(-Q_p) \\[2pt] \frac{-ie}{2}\bm{Q}_{\phi,\Delta^{\dagger}}(-Q_p) \end{matrix} \\
\hline
\begin{matrix} \frac{-ie}{2}\bm{Q}^{\text{T}}_{\phi,\Delta}(Q_p) & \frac{-ie}{2}\bm{Q}^{\text{T}}_{\phi,\Delta^{\dagger}}(Q_p) \end{matrix} &  \frac{1}{4\pi}(\bm{q}^2-2\pi e^2\kappa(Q_p))
\end{array}
\right], \label{eq: general-Anderson-Higgs-main}
\end{align}
where $U^{-1}_{\text{eff}}(Q_p)$ encodes the dynamics of amplitude and phase modes and $\kappa(Q_p)$ is the density--density polarization bubble. Explicit expressions for the kernels $\Phi_{RR}$, $\kappa$, $U^{-1}_{\text{eff}}$, $\bm{Q}_{R,Y},\bm{Q}_{\phi,Y} \ (Y=\Delta,\Delta^{\dagger})$, and $\Pi_{R,\phi}$ in terms of the BdG eigenvalues $E_m(\bm k)$ and eigenvectors $|u_m(\bm k)\rangle$ are listed in Eq.~(\ref{eq: total}) of Appendix~\ref{app: Microscopic derivation for Raman response}. In particular, the effective interaction $U_{\text{eff}}$ is given within the random-phase approximation (RPA),
\begin{align}
U_{\mathrm{eff}} = U -U\Pi U + \cdots = (1+U\Pi)^{-1}U,
\label{U_eff-main}
\end{align}
as shown diagrammatically in Fig.~\ref{fig: diagram}(c).

The poles of the matrix propagator $U_{\text{eff},\phi}(Q_p)$, or equivalently the zeros of $\det \left[U^{-1}_{\text{eff},\phi}(Q_p)\right]$, determine the dispersion of the collective modes, including the Anderson--Higgs mass associated with the overall phase mode (whereas this Anderson--Higgs mass vanishes at $\bm{q}=\bm{0}$ as shown in Appendix \ref{app: Relationship between gauge invariance and Anderson-Higgs mass}).

\subsection{General expression for the Raman susceptibility}
\label{subsec: Raman susceptibility}

The Raman susceptibility is obtained from the effective action by integrating out the bosonic fields $\bm{\delta}\Delta_{\phi}$ and differentiating with respect to the Raman source field. Performing the Gaussian integration over $\bm{\delta}\Delta_{\phi}$ in Eq.~(\ref{eq: general-action3-main}) yields
\begin{align}
\nonumber
&\int D[\bm{\delta} \Delta_{\phi}]\,e^{-S_{\text{FL}}} \\
&= \exp\left[
\frac{\beta}{8}\sum_{\bm q}\sum_p
\bm{T}^{\text T}(-Q_p)\,U_{\text{eff},\phi}(-Q_p)\,\bm{T}(Q_p)
\right],
\end{align}
with $\bm{T}(Q_p)\equiv \bm{Q}_{R,\phi}(Q_p)R(Q_p)$. The resulting effective action for the Raman source field is
\begin{align}
\nonumber
S_{\text{eff}}[R] = &\frac{\beta}{4}\sum_{\bm q}\sum_{p} R(Q_p)R(-Q_p)\,\Phi_{RR}(Q_p) \\
&-\frac{\beta}{8}\sum_{\bm q}\sum_{p}
\bm{T}^{\text T}(-Q_p)\,U_{\text{eff},\phi}(-Q_p)\,\bm{T}(Q_p).
\end{align}
The corresponding Raman susceptibility is defined by
\begin{align}
\chi_{RR}(Q_p) \equiv \frac{1}{\beta}
\frac{\delta^2 S_{\text{eff}}}{\delta R(-Q_p)\,\delta R(Q_p)},
\end{align}
which gives
\begin{align}
\nonumber
&\chi_{RR}(Q_p) \\
&=\frac{1}{4}\Phi_{RR}(Q_p) 
-\frac{1}{8}\bm{Q}^{\text{T}}_{R,\phi}(Q_p)\,U_{\text{eff},\phi}(Q_p)\,
\bm{Q}_{R,\phi}(-Q_p).
\label{eq: raman-main}
\end{align}
After analytic continuation $i\Omega_p\to\omega+i0^{+}$, we obtain the retarded Raman susceptibility $\chi_{RR}(\bm q,\omega)$, and the Raman intensity is $I(\omega)\propto\mathrm{Im}\,\chi_{RR}(\bm q\to\bm 0,\omega)$.

Eq.~(\ref{eq: raman-main}) is our central microscopic result. For any given BdG Hamiltonian $H_{\text{BdG}}(\bm k)$, one can evaluate the kernels in Eq.~(\ref{eq: total}) and obtain the gauge-invariant Raman response including all superconducting collective modes and the coupling to the scalar field. This formula applies to any pairing symmetry including both spin-singlet and triplet pairings. 
In this regard, the present result extends the formula for multiband spin-singlet superconductors in Ref.~\cite{Maiti014503, Sarkar094515} (although here we incorporate only a long-range Coulomb interaction while neglecting other particle-hole interactions).
Furthermore, our formalism admits systematic extensions. Additional particle--hole channels and other bosons (e.g., phonons or magnons) are incorporated by enlarging the Hubbard--Stratonovich fluctuation sector, i.e., by extending $\bm{\delta}\Delta_{\phi}$ and the corresponding kernel matrix $U^{-1}_{\mathrm{eff},\phi}$, which captures hybridization between superconducting and these collective modes.
Subdominant pairing fluctuations, such as Bardasis--Schrieffer and clapping modes, are included analogously by adding the corresponding pairing fields. In contrast to condensed channels, where the gap equation allows one to eliminate $U_{\alpha}$ in favor of $\Delta^{\mathrm{MF}}_{\alpha}$ as shown in Eq.~(\ref{eq: total}) of Appendix \ref{app: Microscopic derivation for Raman response}, subdominant channels have no gap equation, and the coupling $U_{\beta}$ must be kept as an additional input in the enlarged kernel.

\section{Group-theoretical classification of Raman-active collective modes}
\label{sec: Group theoretical classification of Raman response}

In this section, we use group theory to classify which superconducting collective modes can be active in a given Raman scattering geometry. The microscopic expression for the Raman susceptibility, Eq.~(\ref{eq: raman-main}), shows that collective modes contribute to $\chi_{RR}$ via the vertices $\bm{Q}_{R,\Delta^{\dagger}}$ which describe the coupling between the Raman source field and fluctuations of the superconducting order parameters. These vertices are nonzero only if they are allowed by the point-group symmetry of the system. We therefore formulate a symmetry-based selection rule for Raman-active collective modes, and use it to classify the possible couplings for all point group symmetries. The final results are summarized in Tables~\ref{tab:classification1} and~\ref{tab:classification2}. The details of the calculations are shown in Appendix \ref{app: Symmetry operators in superconducting states and group theoretical classification}.

\subsection{Symmetry of the BdG Hamiltonian and Raman vertex}

We begin with the BdG Hamiltonian in Eq.~(\ref{eq-main: BdG-Hamiltonian-multi}):
\begin{align}
H_{\text{BdG}} &= \frac{1}{2}\sum_{\bm{k}} \bm{c}^{\dagger}_{\bm{k}} H_{\text{BdG}}(\bm k) \bm{c}_{\bm{k}}, \notag \\
\nonumber
H_{\text{BdG}}(\bm{k}) &= 
\begin{bmatrix}
H_0(\bm{k}) & \Delta(\bm{k}) \\
\Delta^{\dagger}(\bm{k}) & -H_0^{\text{T}}(-\bm{k})
\end{bmatrix}, \\
\Delta(\bm{k}) &= \sum_{\alpha=1}^{N_{\Delta}} \Delta^{\text{MF}}_{\alpha} \hat{\Delta}_\alpha(\bm{k}),
\end{align}
where $\bm{c}_{\bm{k}} = [c_{\bm{k}},c^{\dagger}_{-\bm{k}}]^{\text{T}}$ is the Nambu spinor and the indices for spin, orbital, and sublattice degrees of freedom are implicit. 

We assume that the normal state has point group symmetry $G$. For each $g\in G$, the action on $H_0(\bm k)$ is
\begin{align}
D(g) H_0(\bm k) D^{\dagger}(g) = H_0(g\bm k),
\label{eq:normalstateg-main}
\end{align}
where $D(g)$ is a (generally spinful) unitary representation of $g$. We further assume that each gap function $\hat{\Delta}_{\alpha}(\bm k)$ transforms according to a one-dimensional (1D) irreducible representation (IR) of $G$ \footnote{Higher-dimensional IRs are regarded as 1D IRs of the subgroup of $G$},
\begin{align}
D(g) \hat{\Delta}_{\alpha}(\bm k) D^{\rm T}(g) = \eta^{\alpha}_g \hat{\Delta}_{\alpha}(g \bm k),
\qquad \eta^{\alpha}_g\in U(1).
\label{eq:gapfunctiong-main}
\end{align}
The factor $\eta^{\alpha}_g$ is the character of 1D IRs that the gap function $\hat{\Delta}_{\alpha}$ belongs to. When all order-parameter components $\hat{\Delta}_{\alpha}$ belong to the same IR, the BdG Hamiltonian has the point-group symmetry $G$,
\begin{align}
\tilde{D}(g) &H_{\text{BdG}}(\bm k) \tilde{D}^{\dagger}(g) = H_{\text{BdG}}(g\bm k), \\
&\tilde{D}(g) \equiv  
\begin{bmatrix}
D(g) & 0 \\ 0 & \eta_g D^{\ast}(g)
\end{bmatrix}.
\label{eq:nambug-main} 
\end{align}

Similarly, the Raman vertex $\gamma(\bm{k})$ belongs to a one-dimensional IR associated with the chosen scattering geometry,
\begin{align}
D(g)\gamma(\bm{k})D^\dagger(g) = \eta^R_g\gamma(g\bm{k}),
\label{eq:RamannR-main} 
\end{align}
with $\eta^R_g\in U(1)$. For example, in a tetragonal $D_{4h}$ crystal, $\gamma(\bm{k})\propto \partial_{k_x}^2+\partial_{k_y}^2$ transforms as $A_{1g}$, $\gamma(\bm{k})\propto \partial_{k_x}^2-\partial_{k_y}^2$ transforms as $B_{1g}$, and $\gamma(\bm{k})\propto \partial_{k_x}\partial_{k_y}$ transforms as $B_{2g}$.

\subsection{Selection rule for Raman coupling to collective modes}

\begin{table*}[t]
\caption{
Group-theoretical classification of Raman-active collective modes in noncentrosymmetric superconductors.
For each point group $G$ (first column), we list the irreducible representations (IRs) that can appear in the Raman vertex $\gamma(\bm{k})$ (second column), and all inequivalent pairs of order-parameter IRs $(\Delta_{\alpha},\Delta_{\beta})$ (third column) that satisfy the selection rule
$\Gamma_{\Delta_{\alpha}^{\dagger}} \otimes \Gamma_{\gamma} \otimes \Gamma_{\Delta_{\beta}} \supset A$.
Pairs of the form $(\Gamma_i,\Gamma_i)$ are not listed since they belong to the trivial representation and always can couple to the trivial representation of the Raman vertex.
}
\begin{tabular}{ccc}
\hline \hline
Point group & IR of $\gamma(\bm{k})$ & Possible pairs of $(\Delta_{\alpha},\Delta_{\beta})$.
\\
\hline    
$C_{1}$ & $A$ &  \\
$C_{2}$ & $A\oplus B$ & $(A,B)$  \\
$C_{s}$ & $A'\oplus A''$ & $(A',A'')$  \\
$C_{3}$ & $A\oplus E$ & $(A,E)$  \\
$C_{4}$ & $A\oplus B\oplus E$ & $(A,E)$, $(B,E)$  \\
$S_{4}$ & $A\oplus B\oplus E$ & $(A,E)$, $(A,B)$, $(B,E)$  \\
$D_{2}$ & $A\oplus B_{1}\oplus B_{2} \oplus B_{3}$ & $(A,B_{1})$, $(A,B_{2})$, $(A,B_{3})$, $(B_{1}, B_{2})$, $(B_{1}, B_{3})$, $(B_{2}, B_{3})$  \\
$C_{2v}$ & $A_{1}\oplus A_{2}\oplus B_{1} \oplus B_{2}$ & $(A_{1},A_{2})$, $(A_{1},B_{1})$, $(A_{1},B_{2})$, $(A_{2}, B_{1})$, $(A_{2}, B_{2})$, $(B_{1}, B_{2})$\\
$C_{6}$ &  $A\oplus E_{1}\oplus E_{2}$ & $(A, E_{1})$, $(A, E_{2})$, $(B, E_{1})$, $(B, E_{2})$, $(E_{1}, E_{2})$ \\
$C_{3h}$ &  $A'\oplus E'\oplus E''$ & $(A',E')$, $(A',E'')$, $(A'', E')$, $(A'', E'')$, $(E', E'')$ \\
$D_{3}$ &  $A_{1}\oplus E$ & $(A_{1}, E)$, $(A_{2},E)$ \\
$C_{3v}$ &  $A_{1}\oplus E$ & $(A_{1}, E)$, $(A_{2},E)$ \\
$D_{4}$ & $A_{1}\oplus B_{1}\oplus B_{2}\oplus E$ & $(A_{1}, B_1)$, $(A_{1}, B_2)$, $(A_{1}, E)$, $(A_{2}, B_1)$, $(A_{2}, B_2)$, $(A_{2}, E)$, $(B_{1}, E)$, $(B_{2}, E)$ \\
$C_{4v}$ & $A_{1}\oplus B_{1}\oplus B_{2}\oplus E$ & $(A_{1}, B_1)$, $(A_{1}, B_2)$, $(A_{1}, E)$, $(A_{2}, B_1)$, $(A_{2}, B_2)$, $(A_{2}, E)$, $(B_{1}, E)$, $(B_{2}, E)$ \\
$D_{2d}$ & $A_{1}\oplus B_{1}\oplus B_{2}\oplus E$ & $(A_{1}, B_1)$, $(A_{1}, B_2)$, $(A_{1}, E)$, $(A_{2}, B_1)$, $(A_{2}, B_2)$, $(A_{2}, E)$, $(B_{1}, E)$, $(B_{2}, E)$ \\
$T$ & $A\oplus E \oplus T$ & $(A,T)$, $(A,E)$, $(E,T)$ \\
$D_{6}$ & $A_{1}\oplus E_{1}\oplus E_{2}$ & $(A_{1}, E_{1})$, $(A_{1}, E_{2})$, $(A_{2}, E_{1})$, $(A_{2}, E_{2})$, $(B_{1}, E_{1})$, $(B_{1}, E_{2})$, $(B_{2}, E_{1})$, $(B_{2}, E_{2})$, $(E_{1}, E_{2})$ \\
$C_{6v}$ & $A_{1}\oplus E_{1}\oplus E_{2}$ & $(A_{1}, E_{1})$, $(A_{1}, E_{2})$, $(A_{2}, E_{1})$, $(A_{2}, E_{2})$, $(B_{1}, E_{1})$, $(B_{1}, E_{2})$, $(B_{2}, E_{1})$, $(B_{2}, E_{2})$, $(E_{1}, E_{2})$ \\
$D_{3h}$ & $A'_{1}\oplus E'\oplus E''$ & $(A'_{1}, E')$, $(A'_{1}, E'')$, $(A'_{2}, E')$, $(A'_{2}, E'')$, $(A''_{1}, E')$, $(A''_{1}, E'')$, $(A''_{2}, E')$, $(A''_{2}, E'')$, $(E', E'')$ \\
$O$ & $A_1\oplus E \oplus T_{2}$ & $(A_{1}, E)$, $(A_{1}, T_{2})$, $(A_{2}, E)$, $(A_{2}, T_{1})$, $(E, T_{1})$, $(E, T_{2})$, $(T_{1}, T_{2})$
\\
$T_{d}$ & $A_1\oplus E \oplus T_{2}$ & $(A_{1}, E)$, $(A_{1}, T_{2})$, $(A_{2}, E)$, $(A_{2}, T_{1})$, $(E, T_{1})$, $(E, T_{2})$, $(T_1, T_{2})$
\\
\hline \hline
\end{tabular}	
\label{tab:classification1}
\end{table*}

\begin{table*}[t]
\caption{
Same classification as in Table~\ref{tab:classification1} but for centrosymmetric superconductors.
}
\begin{tabular}{ccc}
\hline \hline
Point group & IR of $\gamma(\bm{k})$ & Possible pairs of $(\Delta_{\alpha},\Delta_{\beta})$.
\\
\hline    
$C_{i}$ & $A_{g}$ & \\
$C_{2h}$ & $A_{g}\oplus B_{g}$ & $(A_{g},B_{g})$, $(A_{u},B_{u})$  \\
$C_{3i}=S_{6}$ & $A_{g}\oplus E_{g}$ & $(A_g, E_{g})$, $(A_u, E_{u})$ \\
$D_{2h}$ &  $A_g \oplus B_{1g}\oplus B_{2g}\oplus B_{3g}$ & $(A_{g},B_{1g})$ , $(A_{g}, B_{2g})$, $(A_{g},B_{3g})$, $(B_{1g}, B_{2g})$, $(B_{1g}, B_{3g})$, $(B_{2g}, B_{3g})$, \\
& & $(A_{u}, B_{1u})$, $(A_{u}, B_{2u})$, $(A_{u},B_{3u})$, $(B_{1u}, B_{2u})$, $(B_{1u}, B_{3u})$, $(B_{2u}, B_{3u})$ \\
$C_{4h}$ & $A_{g}\oplus B_{g}\oplus E_{g}$ &  $(A_{g}, B_{g})$, $(A_{g}, E_{g})$, $(B_{g}, E_{g})$, $(A_{u}, B_{u})$, $(A_{u}, E_{u})$, $(B_{u}, E_{u})$ \\
$C_{6h}$ & $A_{g}\oplus E_{1g}\oplus E_{2g}$ & $(A_{g}, E_{1g})$, $(A_{g}, E_{2g})$, $(B_{g}, E_{1g})$, $(B_{g}, E_{2g})$, $(E_{1g}, E_{2g})$, \\
& & $(A_{u}, E_{1u})$, $(A_{u}, E_{2u})$, $(B_{u}, E_{1u})$, $(B_{u}, E_{2u})$, $(E_{1u}, E_{2u})$\\
$D_{3d}$ & $A_{1g}\oplus E_{g}$ & $(A_{1g}, E_{g})$, $(A_{2g}, E_{g})$, $(A_{1u}, E_{u})$, $(A_{2u}, E_{u})$ \\
$D_{4h}$ & $A_{1g}\oplus B_{1g}\oplus B_{2g}\oplus E_{g}$ & $(A_{1g}, B_{1g})$, $(A_{1g}, B_{2g})$, $(A_{1g}, E_{g})$, $(A_{2g}, B_{1g})$, $(A_{2g}, B_{2g})$, $(A_{2g}, E_{g})$, $(B_{1g}, E_{g})$, $(B_{2g}, E_g)$, \\
 & & $(A_{1u}, B_{1u})$, $(A_{1u}, B_{2u})$, $(A_{1u}, E_{u})$, $(A_{2u}, B_{1u})$, $(A_{2u}, B_{2u})$, $(A_{2u}, E_{u})$, $(B_{1u}, E_{u})$, $(B_{2u}, E_u)$ \\
$D_{6h}$ & $A_{1g}\oplus E_{1g}\oplus E_{2g}$ & $(A_{1g}, E_{1g})$, $(A_{1g}, E_{2g})$, $(A_{2g}, E_{1g})$, $(A_{2g}, E_{2g})$, $(B_{1g}, E_{1g})$, $(B_{1g}, E_{2g})$,  $(B_{2g}, E_{1g})$, $(B_{2g}, E_{2g})$, 
\\ & &$(A_{1u}, E_{1u})$, $(A_{1u}, E_{2u})$, $(A_{2u}, E_{1u})$, $(A_{2u}, E_{2u})$, $(B_{1u}, E_{1u})$, $(B_{1u}, E_{2u})$,  $(B_{2u}, E_{1u})$, $(B_{2u}, E_{2u})$, 
\\ & & $(E_{1g}, E_{2g})$, $(E_{1u}, E_{2u})$ \\
$T_{h}$ & $A_{g}\oplus E_{g}\oplus T_{g}$ & $(A_{g}, E_{g})$, $(A_{g}, T_{g})$, $(E_{g}, T_{g})$, $(A_{u}, E_{u})$, $(A_{u}, T_{u})$, $(E_{u}, T_{u})$ \\
$O_{h}$ & $A_{1g}\oplus E_{g}\oplus T_{2g}$ & $(A_{1g}, E_{g})$, $(A_{1g}, T_{2g})$, $(A_{2g}, E_{g})$, $(A_{2g}, T_{1g})$, $(E_{g}, T_{1g})$, $(E_{g}, T_{2g})$, $(T_{1g}, T_{2g})$ \\
 & & $(A_{1u}, E_{u})$, $(A_{1u}, T_{2u})$, $(A_{2u}, E_{u})$, $(A_{2u}, T_{1u})$, $(E_{u}, T_{1u})$, $(E_{u}, T_{2u})$, $(T_{1u}, T_{2u})$, $(E_{u}, T_{2u})$
 \\
\hline \hline
\end{tabular}	
\label{tab:classification2}
\end{table*}

Microscopically, the coupling between the Raman source field and superconducting fluctuations is encoded in $\bm{Q}_{R,\Delta^{\dagger}}$ [see Eq.~(\ref{eq: total})]. The relevant structure can be read off from the Feynman diagrams in Fig.~\ref{fig: diagram}(b) and from Eq.~(\ref{eq: diagram}). At $\bm q=0$, a representative contribution to the coupling between the Raman source field and fluctuations of component $\alpha$ is 
\begin{align}
\nonumber
&[\bm{Q}_{R,\Delta^{\dagger}}(i\Omega_n)]_{\alpha} \\
&=
\sum_{\bm{k}}\sum_m
\mathrm{tr}\left[U_{21}\hat{\Delta}^{\dagger}_{\alpha}(\bm{k})G_0(\bm{k},i\omega_m+i\Omega_n)\Gamma(\bm{k})G_0(\bm{k},i\omega_m)\right],
\label{eq: condition-coupling-main}
\end{align}
where $\Gamma(\bm k) \equiv \text{diag}[\gamma(\bm{k}), -\gamma^{\text{T}}(-\bm{k})]$ is the Nambu Raman vertex. Using the Nambu structure of $G_0$ one can show (Appendices \ref{app: Microscopic derivation for Raman response} and \ref{app: Symmetry operators in superconducting states and group theoretical classification}) that this expression is equivalent to
\begin{align}
\nonumber
&[\bm{Q}_{R,\Delta^{\dagger}}(i\Omega_n)]_{\alpha}  \\
&=
2\sum_{\bm{k}}\sum_m
\mathrm{tr}\Big[\hat{\Delta}^{\dagger}_{\alpha}(\bm{k})G^{(11)}_0(\bm{k},i\omega_m+i\Omega_n)\gamma(\bm{k})G^{(12)}_0(\bm{k},i\omega_m)\Big],
\label{eq: condition-coupling2-main}
\end{align}
where $G^{(11)}_0$ and $G^{(12)}_0$ are the normal and anomalous blocks of $G_0$ in Nambu space. The quantity in Eq.~(\ref{eq: condition-coupling2-main}) vanishes unless it is invariant under all elements of $G$.

In what follows, we focus on the Ginzburg--Landau regime close to the superconducting transition, where the condition $\beta \Delta^{\text{MF}}_{\alpha} \ll 1$ is satisfied. Then, we can approximate $G^{(11)}_0$ by the normal-state Green's function and expand $G^{(12)}_0$ to linear order in $\Delta^{\text{MF}}_{\beta}\hat{\Delta}_\beta(\bm k)$,
\begin{align}
\nonumber
G^{(12)}_0(\bm k,i\omega_m)&\approx \sum_{\beta}[G^{(12)}_0(\bm k,i\omega_m)]_{\beta}, \\
[G^{(12)}_0(g\bm{k},i\omega_m)]_{\beta} &\propto \Delta^{\text{MF}}_{\beta}\hat{\Delta}_\beta(\bm k),
\end{align}
where the details are shown in Appendix \ref{app: Symmetry operators in superconducting states and group theoretical classification}. As a result, under a symmetry operation $g\in G$ we then have
\begin{align}
\nonumber
&D(g) G^{(11)}_0(\bm{k},i\omega_m)D^{\dagger}(g) = G^{(11)}_0(g\bm{k},i\omega_m), \\
&D(g)[G^{(12)}_0(\bm{k},i\omega_m)]_{\beta}D^{\text T}(g) 
= \eta^{\beta}_g [G^{(12)}_0(g\bm{k},i\omega_m)]_{\beta}.
\label{eq: trasnform-Green's function}
\end{align}
Using Eqs.~(\ref{eq:gapfunctiong-main}), (\ref{eq:RamannR-main}), and  Eq.~(\ref{eq: trasnform-Green's function}), Eq.~(\ref{eq: condition-coupling2-main}) becomes
\begin{align}
\nonumber
&[\bm{Q}_{R,\Delta^{\dagger}}(i\Omega_n)]_{\alpha} 
\\ \nonumber
&=\frac{2}{|G|}\sum_{g\in G}\sum_{\beta}
(\eta^{\alpha}_g)^{*}\,\eta^R_g\,\eta^{\beta}_g \sum_{\bm{k}}\sum_m\\
&\quad \quad \mathrm{tr}\Big[\hat{\Delta}^{\dagger}_{\alpha}(\bm{k})G^{(11)}_0(\bm{k},i\omega_m+i\Omega_n)\gamma(\bm{k})[G^{(12)}_0(\bm{k},i\omega_m)]_{\beta}\Big],
\end{align}
where $|G|$ is the order of the point group. The sum over $g$ vanishes unless the product of characters $(\eta^{\alpha}_g)^{*}\eta^R_g\eta^{\beta}_g$ is trivial for all $g$, i.e., unless the product representation
\begin{align}
\Gamma_{\hat{\Delta}^{\dagger}_\alpha}\otimes\Gamma_{\gamma}\otimes\Gamma_{\hat{\Delta}_\beta}
\label{eq: condition}
\end{align}
contains the totally symmetric representation of $G$, i.e., the Raman vertex must transform as an IR contained in a product of IRs of the two gap functions. Although the group-theoretical condition is symmetric for two pairing channels, a nonzero Raman coupling requires that at least one of the two pairing channels corresponds to a condensed order parameter. In Eq.~(\ref{eq: condition}), $\beta$ labels a condensed component $\Delta^{\text{MF}}_\beta \neq 0$, while $\alpha$ labels a fluctuating channel. Otherwise, the coupling is absent since it is proportional to the mean-field order parameter.

This symmetry criterion holds near the SC transition ($T \sim T_c$) where the Raman source field linearly couples with a collective mode associated with two different fluctuations at $\bm{q}=\bm{0}$. In this case, the corresponding collective mode (for example, the Leggett, the Bardasis--Schrieffer/clapping modes \footnote{While our explicit calculations focus on fluctuations of the condensed components included in the mean-field BdG Hamiltonian, the selection rule and the classification tables are more general. In particular, by taking one of $(\Delta_\alpha,\Delta_\beta)$ to be condensed and the other to represent a fluctuation in a subdominant (noncondensed) pairing channel, the same classification also contains Raman-active noncondensed pairing modes (with the corresponding subdominant coupling kept as an additional input in the kernel).}
, or relative phase/amplitude, mixtures of those modes) can produce a sharp resonance in $\chi_{RR}(\bm 0,\omega)$ via the second term in Eq.~(\ref{eq: raman-main}), when they appear below the quasiparticle continuum. Conversely, if the product representation does not contain the identity, the Raman source field cannot excite these superconducting collective modes.

\subsection{Classification for crystalline point groups}

The above selection rule depends only on the IRs of the gap functions and the Raman vertex, and is independent of microscopic details such as the band structure or interaction strengths. It allows us to classify possible Raman-active collective modes for each point group symmetry by examining the tensor products of IRs in the corresponding character tables.

For noncentrosymmetric point groups, the IRs are labelled by $\Gamma\in\{A,B,E,\dots\}$, and the Raman vertex transforms as
\begin{align}
\Gamma_{\gamma}=\bigoplus_i \Gamma_i,
\end{align}
where the sum runs over the symmetry channels that are accessible by choosing appropriate light polarizations (e.g. $A_{1g}$, $B_{1g}$, $B_{2g}$ in a tetragonal crystal). For each $\Gamma_\gamma$, we list all pairs of gap-function IRs $(\Gamma_{\hat{\Delta}^{\dagger}_\alpha},\Gamma_{\hat{\Delta}_\beta})$ such that
\begin{align}
\Gamma_{\hat{\Delta}^{\dagger}_\alpha}\otimes\Gamma_{\gamma}\otimes\Gamma_{\hat{\Delta}_\beta}
\ni A_{\text{tot}},
\end{align}
where $A_{\text{tot}}$ denotes the totally symmetric representation. The results for all 21 noncentrosymmetric point groups are summarized in Table~\ref{tab:classification1}. Likewise, the classification for all 11 centrosymmetric point groups, where we distinguish gerade ($g$) and ungerade ($u$) IRs, is given in Table~\ref{tab:classification2}. As concrete examples, consider a tetragonal crystal with point group $D_{4h}$ and Raman vertex in the $B_{1g}$ geometry. From Table~\ref{tab:classification2} we see that a $B_{1g}$ Raman vertex can couple to collective modes built from pairs $(A_{1g},B_{1g})$, $(A_{2g},B_{2g})$, $(A_{1u},B_{1u})$, and $(A_{2u},B_{2u})$, among others.

The group-theoretical classification presented here is in correspondence with the microscopic expressions in Eq.~(\ref{eq: total}): whenever the product of IRs does not contain the identity, the corresponding microscopic vertex $[Q_{R,\Delta^{\dagger}}]_{\alpha}$ vanishes identically. Conversely, if the symmetry condition is satisfied, the strength and detailed spectral shape of the Raman resonance are determined by the microscopic band structure and interaction parameters, but the presence or absence of the coupling is fixed solely by point group symmetry. In this sense, our classification is similar to the classification of Lifshitz invariants in Ginzburg--Landau theories for the linear optical response of multiband superconductors \cite{Nagashima013120}.
In particular, since the Raman source field considered here is proportional to second order in the vector potential, the resulting classification of Raman responses can be viewed as a “second-order Lifshitz-invariant” classification \cite{Takasan2025}.
Our result provides a symmetry-based framework for identifying Raman-active collective modes in candidate multicomponent superconductors.

\section{Application to concrete model: UT$\text{e}$$_2$ superconductor}\label{sec: Application to concrete model}

In the previous sections, we have derived a microscopic, gauge-invariant expression for the Raman susceptibility of a general multicomponent superconductor [Eq.~(\ref{eq: raman-main})], and formulated symmetry-based selection rules for Raman-active collective modes. In this section, we illustrate how our formalism applies to a concrete model of a multiband and multicomponent superconductor.
Specifically, we consider a minimal tight-binding model for the heavy-fermion superconductor UTe$_2$ \cite{Ran684, aoki2022}, whose point group symmetry is $D_{2h}$. We find that relative collective modes between multiple components have in-gap resonance frequencies. We also show that the magnitude of these in-gap resonances depends strongly on the form of the Raman vertex. In the following, we consider the fully gapped case, which corresponds to the $A_u$ pairing supported by experiments on NMR Knight shift \cite{Matsumura063701} and thermal conductivity \cite{Suetsugu3772}. The nodal case, which corresponds to the $B_{3u}$ pairing supported by experiments on quasiparticle interference measurements \cite{Wang1555}, is shown in Appendix \ref{app: B3u}.

\begin{figure}[t]
\centering
    \includegraphics[scale=0.06]{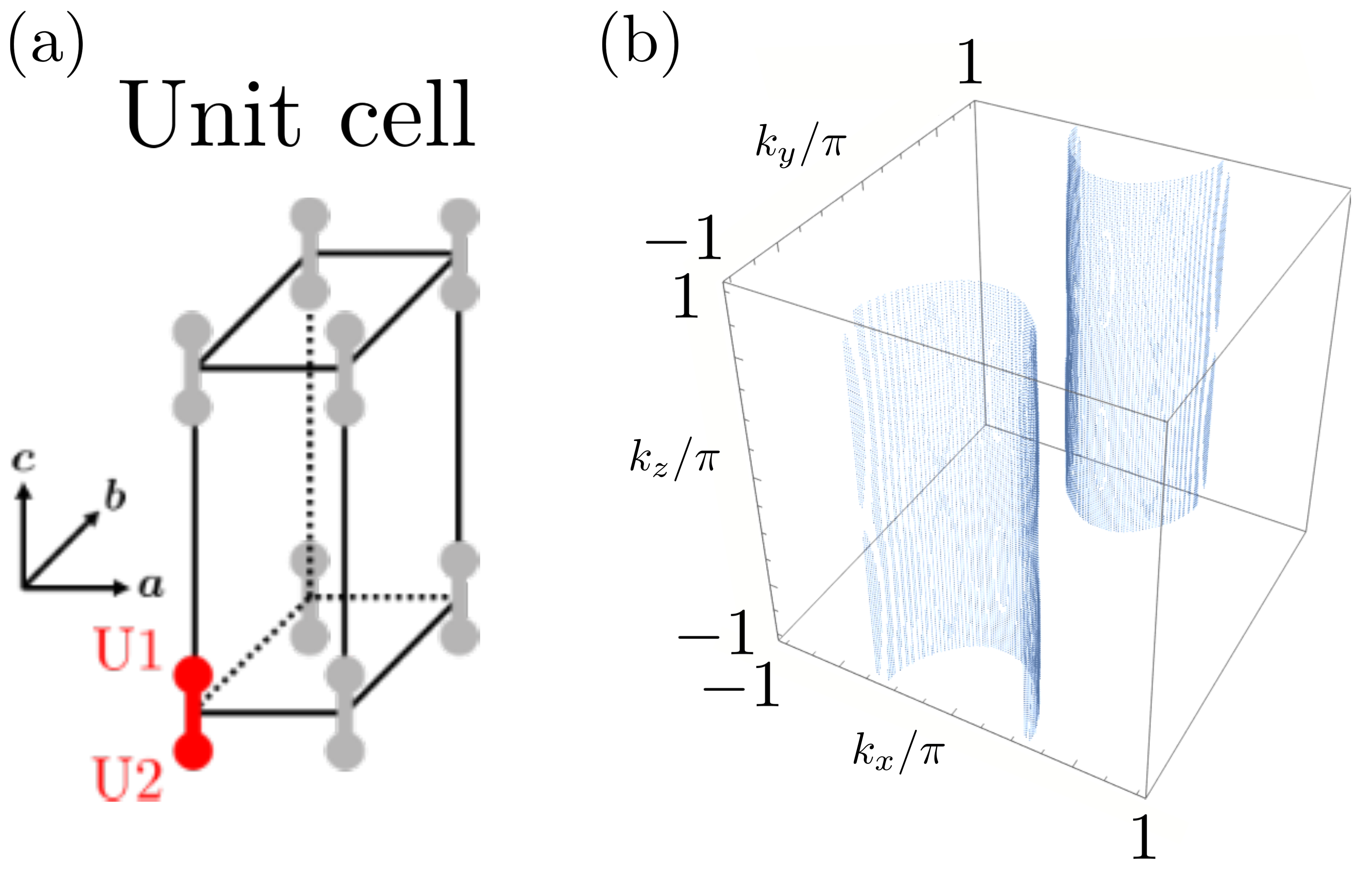}
    \caption{(a) Crystal structure (point group $D_{2h}$) of UTe$_2$, where a unit cell (red) includes two U atoms: U1 and U2. (b) Cylindrical Fermi surface of UTe$_2$.}
    \label{Fig:UTe2-symmetry-operation}
\end{figure}

\subsection{Model}\label{sec:level5-2}

We consider a tight-binding model of UTe$_2$ proposed in Ref. \cite{Shishidou104504}, which takes into account the crystal symmetry of UTe$_2$ (point group $D_{2h}$) and sublattice structures of U atoms.
The tight-binding Hamiltonian is given by
\begin{align} 
H_0(\bm k) &= c(\bm k) + t_3\sigma_x + f_y(\bm k)\sigma_y + \bm{g(\bm{k})} \cdot \bm{s}\sigma_z, \label{normalpart1} 
\end{align}
with
\begin{subequations}
\label{normalpart2}
\begin{align} 
c(\bm k) &= t_1 \cos k_x + t_2 \cos k_y - \mu, \\ 
f_y(\bm k) &= t_4\sin(k_z), \\
g_x(\bm{k}) &= R_x \sin(k_y), \\
g_y(\bm{k}) &= R_y \sin(k_x), \\
g_z(\bm{k}) &= R_z \sin(k_x)\sin(k_y)\sin(k_z),
\end{align}
\end{subequations}
where $\bm{s}$ and $\bm{\sigma}$ denote the spin and sublattice degrees of freedom (U1 and U2) as shown in Fig.~\ref{Fig:UTe2-symmetry-operation} (a), respectively. The $t_{1-4}$ terms are the intra and inter sublattice hopping terms, $\mu$ is the chemical potential, and the $R_i$ terms $(i=x,y,z)$ are the spin-orbit coupling terms. 
We set the parameters as $t_1=-2.0, \ t_2=1.5, \ t_3=-1.4, \ t_4=0.3, \ \mu=-3.6, \ R_x=0.6,\ R_y=0.4, \ R_z=0.2$ to reproduce the cylindrical Fermi surface as shown in Fig.~\ref{Fig:UTe2-symmetry-operation} (b) \cite{Aoki083704, Weinberger2024, Aoki123702}. 

\begin{figure}[t]
\centering
    \includegraphics[scale=0.12]{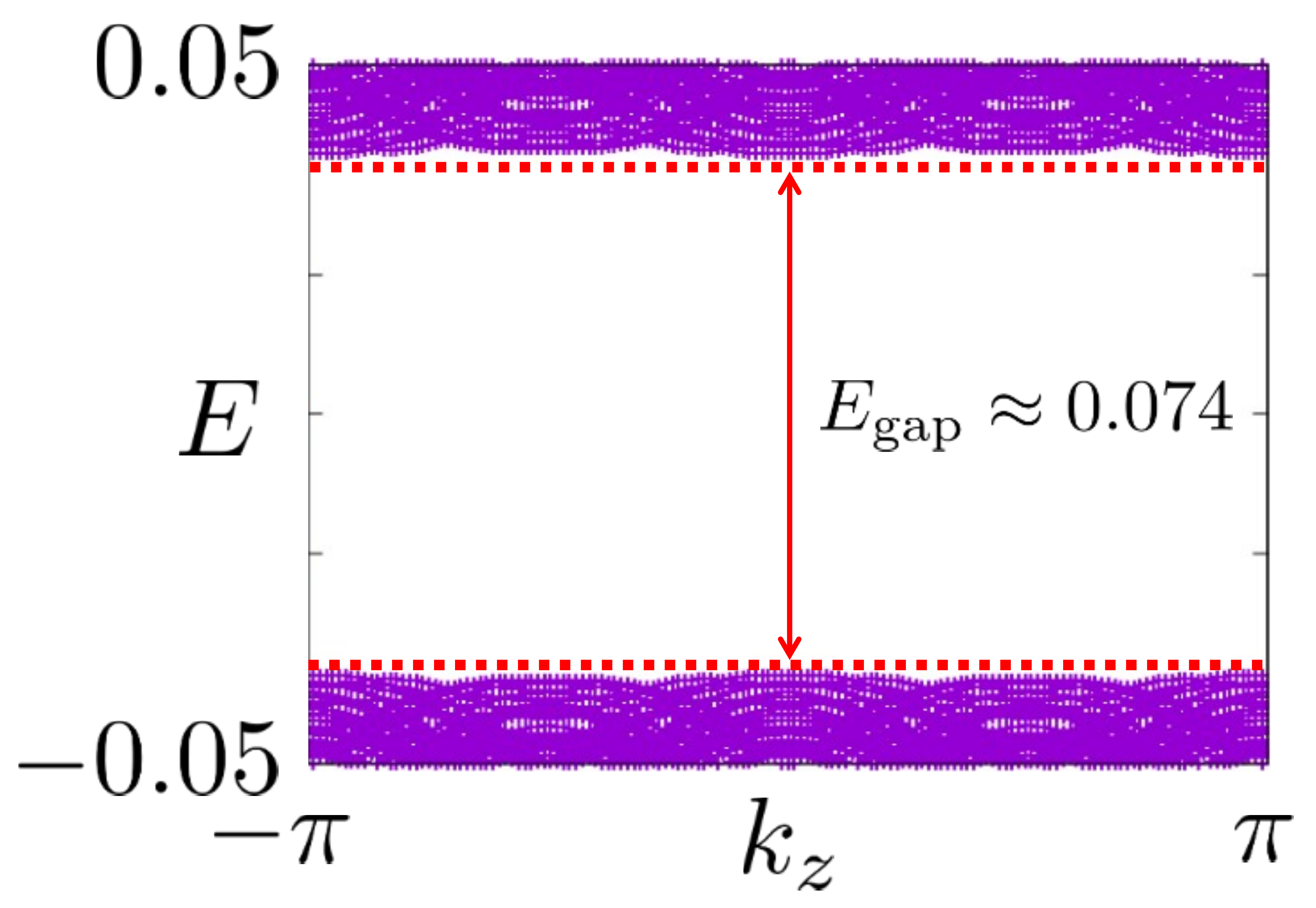}
    \caption{Energy spectrum of the BdG Hamiltonian in Eq.~(\ref{eq-main: BdG-Hamiltonian-multi}) for the UTe$_2$, plotted as a function of $k_z$. For each $k_z$, all eigenvalues $E_n(\bm{k})$ obtained by scanning $(k_x,k_y)$ over the Brillouin zone are shown, so that the vertical spread reflects the residual $(k_x,k_y)$ dispersion. The energy gap is given by $E_{\text{gap}} \approx 0.074$.}
    \label{fig: BdG}
\end{figure}

\begin{figure*}[t]
\centering
\includegraphics[scale=0.125]{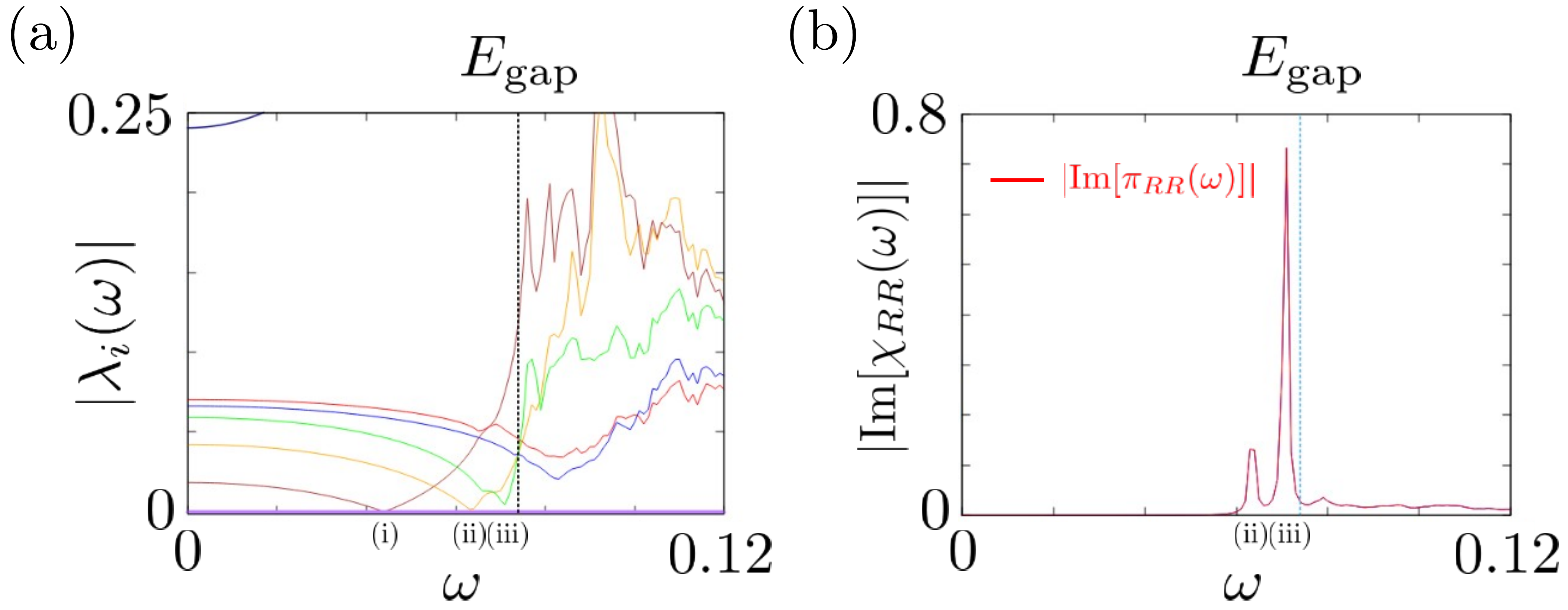}
\caption{(a) Frequency dependence of the eigenvalues of the inverse fluctuation matrix  $U^{-1}_{\mathrm{eff},\phi}(\bm{q}=\bm{0},\omega)$ in Eq.~(\ref{eq: general-Anderson-Higgs-main}) with the tight-binding model of UTe$_2$. The vertical dotted line indicates $\omega\simeq E_{\text{gap}}$.
(b) Raman susceptibility for $\bm{e}^{\text{in}}_x$ and $\bm{e}^{\text{out}}_x$ at $\bm{q}=0$. The blue line shows the full $\chi_{RR}(\bm{0},\omega)$ of Eq.~(\ref{eq: raman-numerical}), which includes the scalar field and Coulomb interaction, while the red line shows the bare response $\pi_{RR}(\bm{0},\omega)$ [Eq.~(\ref{eq: Raman_without_scalar-app})] without the scalar field. The two curves are indistinguishable, numerically confirming that long-range Coulomb screening does not affect the Raman response: $\chi_{RR}(\bm{0},\omega)=\pi_{RR}(\bm{0},\omega)$. At $\omega \approx 0.063 \ (\textrm{ii}), 0.071 \ (\textrm{iii})$, the peak structures appear corresponding to the relative collective modes in (a).}
\label{fig: Results}
\end{figure*}

For the superconducting state, we consider odd-parity three-component order parameters which belong to the $A_u$ IR of $D_{2h}$. The pair potential is written as
\begin{align}
\Delta(\vv k) &= \sum_{\alpha=x,y,z}\Delta^{\text{MF}}_{\alpha}\hat{\Delta}_{\alpha}(\bm{k}),  
\label{eq: UTe2_gap}
\end{align}
with basis functions,
\begin{align}
\hat\Delta_\alpha(\bm k) = \bigl[\sin(k_\alpha)s_{\alpha}\bigr]i s_y.
\label{eq: UTe2_basis}
\end{align}
For simplicity, we take all mean-field amplitudes to be real and choose
\begin{align}
\Delta^{\text{MF}}_x=\Delta^{\text{MF}}_y = 0.04,\ \Delta^{\text{MF}}_z = 0.01,
\end{align}
which realizes a multicomponent superconducting state. The resulting BdG spectrum in Eq.~(\ref{eq-main: BdG-Hamiltonian-multi}) is shown in Fig.~\ref{fig: BdG}. The quasiparticle continuum in the Raman response occurs at $\omega\simeq E_{\text{gap}}$, indicated by the vertical black dotted line in Fig.~\ref{fig: Results}.

The Raman vertex in Eq.~(\ref{eq: Raman-vertex-main}) with incoming and outgoing light polarizations $e^{\text{in}}_i$ and $e^{\text{out}}_j$ are transformed in IRs of the $D_{2h}$ point group symmetry. In the numerical results below, we focus on a Raman vertex $\gamma(\bm{k})$ in $A_g$ IR for which our group-theoretical analysis in Table \ref{tab:classification2} predicts that it can couple to a pair of different components of the $A_u$ order parameters. Hereafter, we consider $e^{\text{in}}_x$ and $e^{\text{out}}_x$ and thus $\gamma(\bm{k}) = \partial^2_{k_x}H_0(\bm{k})$.

\subsection{Collective modes and Raman response}
\label{subsec:UTe2_results}

We now use the general formalism of Sec. \ref{sec: Microscopic model} to compute the collective-mode spectrum and the Raman susceptibility for the $A_u$ state of UTe$_2$. Throughout this section, we consider the $\bm q=\bm 0$ limit relevant for Raman scattering and perform the analytic continuation $i\Omega_p\to\omega+i0^+$. We set $e=1$.

\subsubsection{Spectrum of collective modes}

The coupled fluctuations of the superconducting order parameters and the scalar field are described by the effective interaction in Eq.~(\ref{eq: general-Anderson-Higgs-main}). At fixed real frequency $\omega$ and $\bm q=\bm 0$, we evaluate the inverse fluctuation kernel $U^{-1}_{\text{eff},\phi}(\bm 0,\omega)$ and diagonalize it as
\begin{align}
U^{-1}_{\text{eff},\phi}(\bm 0,\omega)\,\bm{v}_i(\omega) = \lambda_i(\omega)\bm{v}_i(\omega).
\label{eq: fluctuation-kernel-equation}
\end{align}
Here $i$ labels the eigenmodes built from the amplitude and phase fluctuations of the three gap components and the scalar potential: $(\delta\Delta_x,\delta\Delta_y,\delta\Delta_z,\delta\Delta^{*}_x,\delta\Delta^{*}_y,\delta\Delta^{*}_z,\phi)$.
Fig.~\ref{fig: Results} (a) shows the absolute values $|\lambda_i(\omega)|$ for all modes.
The vertical dotted line indicates the quasiparticle threshold $E_{\text{gap}}$ extracted from the BdG spectrum in Fig.~\ref{fig: BdG}.

One noticeable feature of Fig.~\ref{fig: Results} (a) is the presence of one eigenvalue that vanishes for all frequencies (purple line). The corresponding eigenvector is numerically found to coincide, up to an overall normalization, with the vector $\bm\delta(-\omega)$ given in Eq.~(\ref{eq: Ward-Takahashi-vector}) after the analytic continuation $i\Omega_p\to\omega$: its components are proportional to $
\bm\delta(-\omega) \propto
\left(\Delta^{\text{MF}}_x,\Delta^{\text{MF}}_y,\Delta^{\text{MF}}_z,-\Delta^{\text{MF}}_x,-\Delta^{\text{MF}}_y,-\Delta^{\text{MF}}_z,-\frac{i\omega}{2e}\right)^{\text{T}}$.
As shown in Appendix \ref{app: Relationship between gauge invariance and Anderson-Higgs mass}, this vector generates  the gauge transformation of both overall phase and scalar field, and the Ward–Takahashi identity guarantees the corresponding eigenvalue to be zero.
Our numerical results with Eq.~(\ref{eq: general-Anderson-Higgs-main}) thus confirms this identity.

The remaining eigenvalues in Fig.~\ref{fig: Results} (a) correspond to physical collective modes. At $\omega =0$, there exists an eigenvalue which takes $~0.25$, and the corresponding eigenmode is the plasma mode: $\bm{v} \propto (0,0,0,0,0,0,1)^{\text{T}}$.
Also, at $\omega \approx 0.083 \approx E_{\text{gap}}$, there is a cusp structure corresponding to the center-of-mass Higgs mode: $
\bm{v} \propto
\left(\Delta^{\text{MF}}_x,\Delta^{\text{MF}}_y,\Delta^{\text{MF}}_z,\Delta^{\text{MF}}_x,\Delta^{\text{MF}}_y,\Delta^{\text{MF}}_z,0\right)^{\text{T}}$.
On the other hand, three of the eigenmodes exhibit a minimum at a frequency:
\begin{align}
\omega \approx 0.044 \ (\textrm{i}), \ 0.063 \ (\textrm{ii}), \ 0.071 \ (\textrm{iii})
\end{align}
below the quasiparticle continuum. These eigenmodes are dominated by oscillations of the relative amplitudes and phases between different components of the $A_u$ order parameter. We emphasize that these are {\it intraband} relative modes, not interband relative modes such as the Leggett mode. In particular, when $\lambda(\omega \approx 0.044) = 0$, the corresponding eigenmode is the relative Higgs mode which is proportional to $\bm{v} \propto (0,0,\Delta^{\text{MF}}_z,0,0,\Delta^{\text{MF}}_z,0)^{\text{T}}$.
While the details of the mode composition depend on the microscopic parameters, the existence of these relative modes is a general feature of multicomponent superconductors.

Let us remark on the masses of phase modes arising from three spin-triplet pairings $\hat{\Delta}_{\alpha}(\bm{k}) \ (\alpha=x,y,z)$. 
If each pairing channel is decoupled and $U(1)$ symmetry exists for each pairing component as $U(1)_x \times U(1)_y \times U(1)_z$, three gapless phase modes appear for each $U(1)$ symmetry. 
In reality, symmetry-allowed couplings between different components are present. Thus the relative phases of three pairing components are locked, resulting in one gapless (NG) mode for the overall phase, and two relative-phase modes that are massive.
When the breaking of $U(1)_x \times U(1)_y \times U(1)_z$ is weak, the masses for the two relative-phase modes are small, indicating that the relative-phase modes typically appear below the quasiparticle continuum.
Such locking of the relative phase can occur at zero spin-orbit coupling (SOC) through fourth-order terms in the Ginzburg--Landau functional, for example $(\Delta_i^{*})^{2}(\Delta_j)^{2}+\text{c.c.}$, which reduce any accidental $U(1)^3$ symmetry to the $U(1)$ symmetry. 
Also, introducing SOC enables further inter-component mixing at quadratic order, e.g., Josephson-type terms $\Delta_i^{*}\Delta_j+\mathrm{c.c.}$. Thus, SOC directly contributes to $U(1)^3$-symmetry-breaking terms, and the masses of the relative-phase modes also depend on SOC (although their detailed dependence is model-specific).

We further note that while 
the symmetry of the normal state [Eq.~(\ref{normalpart1})] seems to be $SO(3)^{\text{spin}} \times D_{2h}^{\text{orbital}} \times U(1)$ in the absence of SOC, the lattice form of the triplet gap $\Delta(\bm{k})=\sum_{\alpha=x,y,z}\Delta_\alpha \sin k_\alpha\, s_\alpha$ invalidates the spin rotation symmetry as the spin rotation alone does not generally map the set $\{\Delta_\alpha\}$ into itself because of the form factors $\sin k_\alpha$. Therefore, additional Nambu--Goldstone (NG) modes beyond the overall phase mode are not guaranteed solely by the normal-state $SO(3)^{\text{spin}}$ symmetry in our setting.

\subsubsection{Raman susceptibility}

Using Eq.~(\ref{eq: raman-main}), we compute the full Raman susceptibility 
\begin{align}
\nonumber
&\chi_{RR}(\bm 0,\omega) \\
&=\frac{1}{4}\Phi_{RR}(\bm 0,\omega)
-\frac{1}{8}\bm{Q}^{\text{T}}_{R,\phi}(\bm 0,\omega)\,U_{\text{eff},\phi}(\bm 0,\omega)\,
\bm{Q}_{R,\phi}(\bm 0,-\omega).
\label{eq: raman-numerical}
\end{align}
where the vertices $\bm Q_{R,\phi}$ encode the coupling of the Raman source field to the superconducting fluctuations and to the scalar potential. For $\bm{e}^{\text{in}}_x$ and $\bm{e}^{\text{out}}_x$, Fig.~\ref{fig: Results} (b) shows the imaginary parts of the full Raman susceptibility $\chi_{RR}(\bm 0,\omega)$ and the Raman susceptibility without a scalar field $\pi_{RR}(\bm 0,\omega)$ (defined with $U_{\text{eff}}$ instead of $U_{\text{eff},\phi}$ as in Eq.~(\ref{eq: Raman_without_scalar-app}) in Appendix \ref{app: Alternative representation and Coulomb screening}).
The blue solid curve shows $\text{Im}[\chi_{RR}(\bm 0,\omega)]$ obtained from Eq.~(\ref{eq: raman-numerical}), and the red solid curve shows $\text{Im}[\pi_{RR}(\bm 0,\omega)]$. The two curves are indistinguishable, indicating that $\text{Im}[\chi_{RR}(\bm 0,\omega)] = \text{Im}[\pi_{RR}(\bm 0,\omega)]$ [Eq.~(\ref{eq: raman2})]. This result is consistent with vanishing of the Coulomb screening at $\bm{q}=\bm{0}$ due to the Ward-Takahashi identity (the details are shown in Appendix \ref{app: Relationship between gauge invariance and Anderson-Higgs mass}).

The spectrum in Fig.~\ref{fig: Results} (b) also shows that sharp collective-mode peaks below the quasiparticle continuum, whose positions track two of the zero eigenvalues of $U^{-1}_{\text{eff},\phi}$ in Fig.~\ref{fig: Results} (a) at $\omega \approx 0.063$(ii), $0.071$(iii). The quasiparticle continuum above $E_{\text{gap}}$ gives rise to a broad background. 
In contrast, the relative Higgs mode at $\omega \approx 0.044$(i) does not appear in the Raman spectrum in Fig.~\ref{fig: Results} (b), although it is symmetry-allowed. This is because the overlap between the Raman coupling vertex $\bm{Q}_{R,\phi}$ and the corresponding eigenvector $\bm{v}$ is small in the present case. 
We find that a larger spin-orbit coupling component $R_z$ in the Raman vertex $\gamma(\bm{k})$ can enhance the intensity of the peak structure (i) in Fig.~\ref{fig: Results} (b), while $R_y$ does not. This feature reflects that the gap functions have spin structures and indicates that the detailed spin structure of the Raman vertex, including possible resonant Raman vertices, is also an important factor, which may affect the experimental visibility of a resonance in Raman spectra in addition to a quantitative change.

\section{Conclusions}\label{sec: Conclusions}

In this paper, we have established a practically computable framework for Raman spectroscopy of superconducting collective modes in multicomponent superconductors. Starting from a general BdG Hamiltonian with separable pairing interactions, we derived a gauge-invariant Raman susceptibility by integrating out fermions and treating order-parameter fluctuations together with the Coulomb scalar field. The resulting expression provides a clear separation between quasiparticle and collective-mode contributions: the poles of the effective interaction matrix determine the peak structure in the Raman spectrum.

Beyond the microscopic formalism, we developed symmetry selection rules and a group-theoretical classification of Raman-active collective modes based on ``second-order Lifshitz-invariant''. The criterion for the presence/absence of linear Raman excitation is obtained solely by point group symmetry, independent of microscopic coupling details: a Raman source field couples to a collective mode built from two superconducting fluctuations only when the product representation contains the Raman-vertex symmetry.
This provides a symmetry-based framework to identify which relative phase/amplitude modes (including Leggett- and Bardasis--Schrieffer-type fluctuations as well as more general multicomponent relative modes) can be Raman active for a given Raman vertex. We summarized the symmetry-allowed combinations for all crystalline point groups.

Applying our formalism to UTe$_2$ with a fully gapped multicomponent odd-parity pairing state, we found in-gap Raman resonances associated with relative collective modes between different order-parameter components below the quasiparticle continuum, where each order-parameter component belongs to the same IR $A_{u}$ of the point group $D_{2h}$.
These results demonstrate that fully gapped multicomponent superconductivity can host sharp in-gap collective excitations that couple to a symmetry-allowed Raman vertex and hence yield distinct polarization-dependent fingerprints in Raman spectra. Our calculation highlights the complementary role of microscopic ingredients; while symmetry dictates whether a given mode can be Raman active, its spectral weight depends on the detailed structure of the normal Hamiltonian, Raman vertex, and the superconducting gap structure.
In this sense, our framework allows direct computation of Raman signatures in multiband and multicomponent systems that are experimentally accessible.

Finally, our formulation also allows systematic extensions. Additional particle--hole interaction channels and other bosonic degrees of freedom (e.g., phonons or magnons) can be incorporated by enlarging the Hubbard--Stratonovich fluctuation sector and kernel matrix, enabling hybridization effects between superconducting and these additional collective modes. Noncondensed pairing fluctuations, including Bardasis--Schrieffer and clapping modes in subdominant channels, are likewise included by adding the corresponding pairing fields, where those extensions are left for future work.

\section{ACKNOWLEDGMENTS}
We thank fruitful discussions with
Sota Kitamura, Hikaru Watanabe, Ken Shiozaki, Taiki Matsushita, Jushin Tei, and Youichi Yanase. 
This work was supported by JSPS KAKENHI, Grants No. 24H02231 (YY and TM), 25K23353 (YY) and 23K25816, 23K17665, 24K00568 (TM).

\clearpage

\onecolumngrid
\appendix
\section{Microscopic derivation for Raman response}
\label{app: Microscopic derivation for Raman response}

In this appendix, we derive a microscopic formalism for the Raman response in superconductors with multiple bands and multiple order-parameter components. To this end, we proceed as follows. First, we write down an effective Euclidean action that couples fermions to an external Raman source field and to a scalar potential which encodes the long-range Coulomb interaction. Next, we decouple the attractive interaction in the pairing channel via a Hubbard--Stratonovich transformation, and integrate out the fermions to obtain an effective action for the superconducting order-parameter fields and the scalar potential. The mean-field gap equation is obtained as the saddle-point condition of this action. By expanding the action to quadratic order in the fluctuations of the order parameter and the scalar potential, and then
integrating out these bosonic fields, we finally obtain a compact expression for the Raman susceptibility given in Eq.~(\ref{eq: raman}) which is valid for a general Bogoliubov--de Gennes (BdG) Hamiltonian.

\subsection{General action}

We start from the Euclidean action related to the partition function $\mathcal{Z}=\int D[c^{*},c,\phi]e^{-S[c^{*},c,\phi]}$:
\begin{align}
&S[c^{*},c,\phi] \equiv S_0[c^{*},c]+\int^{\beta}_0d\tau[H_{\text{int}}(\tau)+H_{R}(\tau)+H_{\phi}(\tau)]+ S_{\text{EM}}[\phi]
\label{eq: action}
\end{align}
where $S_0[c^{*},c]$ is the non-interacting part of the action given by
\begin{align}
&S_0[c^{*},c] = \int^{\beta}_0 d\tau \sum_{\bm{k}}\bm{c}^{\dagger}_{\bm{k}}(\tau)[\partial_{\tau}+H_0(\bm{k})]\bm{c}_{\bm{k}}(\tau)
\end{align}
in the basis of $\bm{c}=[c_{\alpha}, c_{\beta},...]^{\text{T}}$, where $\alpha,\beta,...$ denotes the spin, orbital, and sublattice degrees of freedom. $H_0(\bm{k})$ is the normal Hamiltonian.
In this paper, we consider the following Bogoliubov-de Gennes (BdG) Hamiltonian:
\begin{align}
\nonumber
H_{\text{BdG}}(\bm{k}) &= \left[ 
\begin{array}{cc}
H_0(\bm{k}) & \Delta(\bm{k}) \\
\Delta^{\dagger}(\bm{k}) & -H_0^{\text{T}}(-\bm{k})
\end{array}
\right]_{2N\times2N}, \\
\Delta(\bm{k}) &= \sum_{\alpha=1,2,...,N_{\Delta}} \Delta^{\text{MF}}_{\alpha} \hat{\Delta}_\alpha(\bm{k}),
\label{eq: BdG-Hamiltonian-multi}
\end{align}
where $N_{\Delta}$ is the number of nonzero order parameters and $\Delta^{\text{MF}}_{\alpha}  \in \mathbb{R}$ describes the mean-field solution obtained by the gap equation (see Appendix \ref{app: gap-equation}). $\hat{\Delta}_\alpha(\bm{k})$ is the gap function for each order parameter, which satisfies $\hat{\Delta}^{\text{T}}_{\alpha}(\vv k) =-\hat{\Delta}_{\alpha}(-\vv k)$ due to the Fermi-Dirac statistics. For this BdG Hamiltonian, we assume the following simple attractive interaction: 
\begin{align}
\nonumber
H_{\text{int}}(\tau) &=-\frac{1}{4}\sum^{N'_{\Delta}}_{i=1}\bm{\rho}_i^{\dagger}(\tau)X_i\bm{\rho}_i(\tau), \\ \nonumber
\bm{\rho}_i(\tau) &\equiv (\rho_{i,\bm{q}_1}(\tau),\rho_{i,\bm{q}_2}(\tau),...),\\ \nonumber
\rho_{i,\bm{q}}(\tau)&=U_i\sum_{\bm{k}}\bm{c}^{\text{T}}_{-\bm{k}+\frac{\bm{q}}{2}}(\tau)\hat{\Delta}^{\dagger}_i(\bm{k})\bm{c}_{\bm{k}+\frac{\bm{q}}{2}}(\tau), \\
X_i&=[X_i]_{\bm{q},\bm{q}'}=(1/U_i)\delta_{\bm{q},\bm{q}'},
\label{eq: interaction}
\end{align}
where $U_i>0$ for all $i$ and $N'_{\Delta} \ge N_{\Delta}$. Note that Eqs.~(\ref{eq: BdG-Hamiltonian-multi}) and (\ref{eq: interaction}) are general in that they can treat both single- and multi-band systems with single- or multi-component order parameters, and both time-reversal-symmetric (TRS) and time-reversal-broken (TRSB) systems. To calculate the Raman response function, we introduce the nonresonant Raman Hamiltonian \cite{Devereaux175}:
\begin{align}
&H_R(\tau) = \sum_{\bm{k},\bm{q}}\bm{c}^{\dagger}_{\bm{k}+\frac{\bm{q}}{2}}(\tau)[-R(\bm{q},\tau)
\gamma(\bm{k})]\bm{c}_{\bm{k}-\frac{\bm{q}}{2}}(\tau), \\
&\gamma(\bm{k}) \equiv \sum_{ij}e^{\text{in}}_{i}e^{\text{out}}_{j}\,\partial_{k_i}\partial_{k_j}H_0(\bm{k}),
\end{align}
where $R(\bm{q},\tau)$ is the source term coupled to the Raman density operator $\Phi(\bm{q},\tau) \equiv \sum_{\bm{k}}\bm{c}^{\dagger}_{\bm{k}-\frac{\bm{q}}{2}}(\tau)
\gamma(\bm{k})\bm{c}_{\bm{k}+\frac{\bm{q}}{2}}(\tau)$ and $\gamma(\bm{k})$ is the Raman vertex. $e^{\text{in}}_{i}/e^{\text{out}}_{j}$ indicates the direction of the incident/outgoing light. 

Finally, we consider a scalar field that plays an important role in the Raman response of superconducting collective modes. First, it couples to the totally symmetric Raman vertex, leading to Coulomb screening of the Raman response \cite{Devereaux12523, Devereaux175, Boyd174521, Sauer014525, Cea064512, Maiti014503, Sarkar094515}. Second, it couples to the collective modes of the superconductor; in particular, coupling to the overall phase (Nambu--Goldstone) mode implements the Anderson--Higgs mechanism. This is also important for the gauge invariance of electromagnetic responses in superconductors.
Accordingly, we add the following terms to the action:
\begin{align}
&H_{\phi}(\tau) =-ie\sum_{\bm{k},\bm{q}}\phi(\bm{q},\tau)\bm{c}^{\dagger}_{\bm{k}+\frac{\bm{q}}{2}}(\tau)\bm{c}_{\bm{k}-\frac{\bm{q}}{2}}(\tau), \label{eq: scalar1} \\ \nonumber
&S_{\text{EM}}[\phi] \equiv \frac{1}{8\pi}\int^{\beta}_0 d\tau \int d^3r \, [\bm{E}^2(\bm{r},\tau)], \quad \bm{E}= \bm{\nabla} \phi, \\
&\hspace{11.3mm} =  \frac{1}{8\pi}\sum_{\bm{q}}\int^{\beta}_0 d\tau \, \bm{q}^2\phi(-\bm{q,\tau})\phi(\bm{q},\tau).
\label{eq: scalar2}
\end{align}
Note that Eqs.~(\ref{eq: scalar1}) and (\ref{eq: scalar2}) are equivalent to the long-range Coulomb interaction through the Hubbard–Stratonovich transformation for the scalar field. 
We summarize our notation as follows. We define the scalar potentials in real (Minkowski) time and imaginary (Euclidean) time as
$\phi_{\mathrm{M}}$ and $\phi_{\mathrm{E}}$, respectively, and relate them by $\phi_{\text{E}} \equiv i\,\phi_{\text{M}}$. In real time, the electric field is $\bm{E}_{\text{M}} = -\partial_t \bm{A}_{\text{M}} - \nabla \phi_{\text{M}}$. Hence, if we set \(\bm{E}_{\text{E}} \equiv -i\,\bm{E}_{\text{M}}\) and \(\bm{A}_{\text{E}} \equiv \bm{A}_{\text{M}}\), we obtain $\bm{E}_{\text{E}} = -\partial_{\tau} \bm{A}_{\text{E}} + \nabla \phi_{\text{E}}$ under the Wick rotation $t \to -i\tau$. Unless otherwise noted, we use the Euclidean electromagnetic fields $(\bm{A}_{\text{E}}, \phi_{\text{E}})$ and drop the subscripts. Also, the relationship between the Euclidean and Minkowski action is given by $S_{\text{M}} = iS_{\text{E}}$.

By using the Hubbard–Stratonovich transformation for the attractive interaction term $H_{\text{int}}(\tau)$, we get
\begin{align}
\nonumber
\mathcal{Z} &= \int D[c^{*},c,\phi] \ e^{-S[c^{*},c,\phi]} \\ \nonumber
&=\int dc^{*} dc \int D\phi \int \Pi_{i=1}D\Delta^{*}_iD\Delta_i\exp\Bigg(-S_0[c^{*},c]-\int^{\beta}_0d\tau \ [H_{R}(\tau)+H_{\phi}(\tau)] -S_{\text{EM}}[\phi]\\
&\hspace{30mm}-\int^{\beta}_0 d\tau \ \sum_i\left[\bm{\Delta}^{\dagger}_i(\tau)X_i\bm{\Delta}_i(\tau)+\frac{1}{2}\bm{\Delta}^{\dagger}_i(\tau)X_i\bm{\rho}_i(\tau)+\frac{1}{2}\bm{\rho}^{\dagger}_i(\tau)X_i\bm{\Delta}_i(\tau)\right]\Bigg),
\end{align}
where $\bm{\Delta}_i(\tau) \equiv (\Delta_{i,\bm{q}_1},\Delta_{i,\bm{q}_2},...)^{\text{T}}$ is the auxiliary boson field. Here, we introduce the Nambu spinor:
\begin{align}
\bm{\Psi}_{\bm{k}}(\tau) \equiv \left[
\begin{array}{c}
\bm{c}_{\bm{k}}(\tau)  \\
\bm{c}^{*}_{-\bm{k}}(\tau) 
\end{array}
\right].
\end{align}
Then, we obtain
\begin{align}
\nonumber
\mathcal{Z} 
&=\int dc^{*} dc \int D\phi \int \Pi_{i=1}D\Delta^{*}_iD\Delta_i\\
&\hspace{20mm} \times \exp\Bigg(
 -\int^{\beta}_0 d\tau\sum_i\bm{\Delta}^{\dagger}_i(\tau)X_i\bm{\Delta}_i(\tau) + \frac{1}{2}\int^{\beta}_0 d\tau \ \sum_{\bm{k}_1,\bm{k}_2} \bm{\Psi}^{\dagger}_{\bm{k}_1}(\tau)G^{-1}(\bm{k}_1,\bm{k}_2,\tau)\bm{\Psi}_{\bm{k}_2}(\tau)-S_{\text{EM}}[\phi]\Bigg),
\end{align}
where 
\begin{align}
\nonumber
G^{-1}(\bm{k}_1,\bm{k}_2,\tau) &\equiv -\left[
\begin{array}{cc}
(\partial_{\tau}+H_0(\bm{k}_1))\delta_{\bm{k}_1,\bm{k_2}} & \sum_i\sum_{\bm{q}}\Delta_{i}(\bm{q},\tau)\hat{\Delta}_i\left(\frac{\bm{k}_1+\bm{k_2}}{2}\right)\delta_{\bm{q},\bm{k}_1-\bm{k}_2} \\
\sum_i\sum_{\bm{q}}\Delta^{*}_{i}(\bm{q},\tau)\hat{\Delta}^{\dagger}_i\left(\frac{\bm{k}_1+\bm{k_2}}{2}\right)\delta_{\bm{q},-\bm{k}_1+\bm{k}_2} & (\partial_{\tau}-H^{\text{T}}_0(-\bm{k}_2))\delta_{\bm{k}_1,\bm{k_2}}
\end{array}
\right] \\
&\hspace{5mm} + \sum_{\bm{q}}\left[R(\bm{q},\tau)\Gamma\left(\frac{\bm{k}_1+\bm{k_2}}{2}\right)+ie \phi(\bm{q},\tau)\tau_z\right]\delta_{\bm{q},\bm{k}_1-\bm{k}_2}, \quad \Gamma(\bm{k}) \equiv \left[
\begin{array}{cc}
\gamma\left(\bm{k}\right) & 0 \\
0 & -\gamma^{\text{T}}\left(-\bm{k}\right)
\end{array}
\right],
\end{align}
where $\tau$'s denote the Pauli matrices of the Nambu space.
In this paper, we use the following Fourier transformations:
\begin{align}
\nonumber
\bm{\Psi}_{\bm{k}_a}(i\omega_n) &= \int^{\beta}_0d\tau \ \bm{\Psi}_{\bm{k}_a}(\tau)e^{i\omega_n\tau}, \ \bm{\Psi}_{\bm{k}_a}(\tau) = \frac{1}{\beta}\sum_{n}\bm{\Psi}_{\bm{k}_a}(i\omega_n)e^{-i\omega_n \tau},   \ \omega_n=\frac{(2n+1)\pi}{\beta}, \\ \nonumber
\Delta_{i}(\bm{q},i\Omega_n) &= \frac{1}{\beta}\int^{\beta}_0 d\tau\ \Delta_{i}(\bm{q},\tau)e^{i\Omega_n\tau}, \  \Delta_{i}(\bm{q},\tau) = \sum_{n}\Delta_{i}(\bm{q},i\Omega_n)e^{-i\Omega_n \tau}, \ \Omega_n=\frac{2n\pi}{\beta}, \\
F(\bm{q},i\Omega_n) &= \frac{1}{\beta}\int^{\beta}_0 d\tau\ F(\bm{q},\tau)e^{i\Omega_n\tau}, \  F(\bm{q},\tau) = \sum_{n}F(\bm{q},i\Omega_n)e^{-i\Omega_n \tau}, \ \Omega_n=\frac{2n\pi}{\beta}, \label{Fourier2}
\end{align}
where $a=1,2$ and $F=R,\phi$. As a result, we find
\begin{align}
\nonumber
Z&=\int dc^{*} dc \int D\phi \int \Pi_{i=1}D\Delta^{*}_iD\Delta_i\exp\Bigg(
 -\sum_i\sum_{n}\sum_{\bm{q}}\frac{\beta}{U_i}|\Delta_{i}(\bm{q},i\Omega_n)|^2 \\
&\hspace{30mm}+ \frac{1}{2\beta^2}\sum_{m,n}\sum_{\bm{k}_1,\bm{k}_2}  \bm{\Psi}^{\dagger}_{\bm{k}_1}(i\omega_m)\beta G^{-1}(\bm{k}_1,\bm{k}_2,i\omega_m,i\omega_n)\bm{\Psi}_{\bm{k}_2}(i\omega_n)-S_{\text{EM}}[\phi]\Bigg),
\end{align}
where 
\begin{align}
\nonumber
&G^{-1}(\bm{k}_1,\bm{k}_2,i\omega_m,i\omega_n) \equiv \\ \nonumber
&\hspace{10mm}-\left[
\begin{array}{cc}
(-i\omega_m+H_0(\bm{k}_1))\delta_{\bm{k}_1,\bm{k_2}} & \sum_i\sum_{\bm{q}}\Delta_{i}(\bm{q},i\omega_m-i\omega_n)\hat{\Delta}_i\left(\frac{\bm{k}_1+\bm{k_2}}{2}\right)\delta_{\bm{q},\bm{k}_1-\bm{k}_2} \\
\sum_i\sum_{\bm{q}}\Delta^{*}_{i}(\bm{q},-i\omega_m+i\omega_n)\hat{\Delta}^{\dagger}_i\left(\frac{\bm{k}_1+\bm{k_2}}{2}\right)\delta_{\bm{q},-\bm{k}_1+\bm{k}_2} & (-i\omega_m-H^{\text{T}}_0(-\bm{k}_2))\delta_{\bm{k}_1,\bm{k_2}}
\end{array}
\right] \\
&\hspace{10mm} + \sum_{\bm{q}}\left[R(\bm{q},i\omega_m-i\omega_n)\Gamma\left(\frac{\bm{k}_1+\bm{k_2}}{2}\right)+ie \phi(\bm{q},i\omega_m-i\omega_n)\tau_z\right]\delta_{\bm{q},\bm{k}_1-\bm{k}_2}.
\end{align}
Performing the fermionic path integral gives the action $S[\Delta^{*},\Delta,\phi]$ ($\mathcal{Z}=\int D\phi \int \Pi_{i=1}D\Delta^{*}_iD\Delta_i \ e^{-S[\Delta^{*},\Delta,\phi]}$):
\begin{align}
S[\Delta^{*},\Delta,\phi] =\sum_{i}\frac{\beta}{U_{i}}\sum_{n}\sum_{\bm{q}}|\Delta_{i}(\bm{q},i\Omega_n)|^2 - \frac{1}{2}\text{Tr }\text{ln }[-\beta G^{-1}(\bm{k}_1,\bm{k}_2,i\omega_m,i\omega_n)]+S_{\text{EM}}[\phi],
\label{eq: general-action}
\end{align}
where Tr denotes a trace and summation over the momenta and frequencies $(\bm{k}_1,\bm{k}_2,\omega_m,\omega_n)$.

\subsection{Gap equation}\label{app: gap-equation}

In this section, we derive the mean-field gap equation starting from the general action~(\ref{eq: general-action}). We consider a static and uniform order parameter, $\Delta_{i}(Q)=\Delta_{i}\delta_{\Omega,0}\delta_{\bm{q},0}$ with $Q\equiv(\bm{q},i\Omega)$, and switch off the external fields, $R(Q)=\phi(Q)=0$. Using the Fourier convention in Eq.~(\ref{Fourier2}), the quadratic bosonic term in Eq.~(\ref{eq: general-action}) reduces to $\sum_i (\beta/U_i)|\Delta_i|^2$, and the action becomes
\begin{align}
\nonumber
S[\Delta^{*},\Delta] &=\sum_{i}\frac{\beta}{U_{i}}|\Delta_{i}|^2 - \frac{1}{2}\text{Tr }\text{ln }[-\beta G^{-1}(\bm{k}_1,i\omega_m)], \\
G^{-1}(\bm{k},i\omega_m) &= -\left[
\begin{array}{cc}
(-i\omega_m+H_0(\bm{k})) & \sum_i\Delta_i\hat{\Delta}_i(\bm{k}) \\
\sum_i\Delta^{*}_i\hat{\Delta}^{\dagger}_i(\bm{k}) & (-i\omega_m-H^{\text{T}}_0(-\bm{k}))
\end{array}
\right].
\end{align}
The mean-field gap equation is obtained as the saddle-point condition of the effective action, i.e., by taking the functional derivative with respect to $\Delta^{*}_i$:
\begin{align}
\nonumber
\frac{\delta S[\Delta^{*},\Delta]}{\delta \Delta^{*}_i} &= \frac{\beta}{U_i}\Delta_i-\frac{1}{2}\sum_{m}\sum_{\bm{k}}\text{tr}\left[G(\bm{k},i\omega_m)\frac{\delta}{\delta \Delta^{*}_i}G^{-1}(\bm{k},i\omega_m)\right] \\ \nonumber
&=\frac{\beta}{U_i}\Delta_i+\frac{1}{2}\sum_{m}\sum_{\bm{k}}\text{tr}\left[G(\bm{k},i\omega_m)\left[
\begin{array}{cc}
0 & 0 \\
\hat{\Delta}^{\dagger}_i(\bm{k}) & 0
\end{array}
\right]\right] \\
&=\frac{\beta}{U_i}\Delta_i+\frac{1}{2}\sum_{m}\sum_{\bm{k}}\text{tr}\left[G(\bm{k},i\omega_m)U_{21}\hat{\Delta}^{\dagger}_i(\bm{k})\right]=\frac{\beta}{U_i}\Delta_i+\frac{1}{2}\sum_{m}\sum_{\bm{k}}\text{tr}\left[G^{(12)}(\bm{k},i\omega_m)\hat{\Delta}^{\dagger}_i(\bm{k})\right],
\end{align}
where $(12)$ implies the block matrix element for the Nambu space:
\begin{align}
\nonumber
&G(\bm{k},i\omega_m) \equiv \left[
\begin{array}{cc}
G^{(11)}(\bm{k},i\omega_m) & G^{(12)}(\bm{k},i\omega_m) \\
G^{(21)}(\bm{k},i\omega_m) & G^{(22)}(\bm{k},i\omega_m)
\end{array}
\right], \\ &U_{11} = \left[
\begin{array}{cc}
1_{N\times N} & 0 \\
0 & 0
\end{array}
\right], \ U_{12} = \left[
\begin{array}{cc}
0 & 1_{N\times N} \\
0 & 0
\end{array}
\right], \ U_{21} = \left[
\begin{array}{cc}
0 & 0 \\
1_{N\times N} & 0
\end{array}
\right], \ U_{22} = \left[
\begin{array}{cc}
0 & 0 \\
0 & 1_{N\times N}
\end{array}
\right], 
\label{eq: Green-function-conventional}
\end{align}
As a result, we find the gap equation:
\begin{align}
\frac{\delta S[\Delta^{*},\Delta]}{\delta \Delta^{*}_i} = 0 \ \ \leftrightarrow \ \ \Delta_i=-\frac{U_i}{2\beta }\sum_{m}\sum_{\bm{k}}\text{tr}\left[G^{(12)}(\bm{k},i\omega_m)\hat{\Delta}^{\dagger}_i(\bm{k})\right].
\label{eq: gap-equation-multi}
\end{align}
If there exist multiple order parameters, we have to solve a system of gap equations.
Near the transition temperature $T_{\mathrm{c}}$, the order parameter is small, and we can construct a linearized gap equation. More precisely, we assume $\beta |\Delta_i| \ll 1$ and expand the Green's function to first order in the order parameters:
\begin{align}
\nonumber
&G = \left(G^{-1}_0 + \Sigma \right)^{-1} 
\approx G_0 - G_0 \Sigma G_0 + \cdots, \\
&G^{-1}_0 = -\left[
\begin{array}{cc}
-i\omega_m+H_0(\bm{k}) & 0 \\
0 & -i\omega_m-H^{\text{T}}_0(-\bm{k})
\end{array}
\right], \quad 
\Sigma =-\left[
\begin{array}{cc}
0 & \sum_i\Delta_i\hat{\Delta}_i(\bm{k}) \\
\sum_i\Delta^{*}_i\hat{\Delta}^{\dagger}_i(\bm{k}) & 0
\end{array}
\right].
\end{align}
Then, the gap equation becomes
\begin{align}
\Delta_i &\approx -\frac{U_i}{2\beta}\sum_{n}\sum_{\bm{k}}\text{tr}\left[g^{\text{e}}(\bm{k},i\omega_n) 
\left( \sum_j \Delta_j \hat{\Delta}_j(\bm{k}) \right) 
g^{\text{h}}(\bm{k},i\omega_n)\hat{\Delta}^{\dagger}_i(\bm{k})\right] = \sum_{j}M_{ij}(T)\Delta_j
\end{align}
where we define the normal part of Green's functions:
\begin{align}
g^{\text{e}}(\bm{k},i\omega_n) \equiv \frac{1}{i\omega_n-H_0(\bm{k})}, \quad g^{\text{h}}(\bm{k},i\omega_n) \equiv \frac{1}{i\omega_n+H^{\text{T}}_0(-\bm{k})}.
\label{eq: normal-green-function}
\end{align}
Here, $M_{ij}(T)$ is given by
\begin{align}
\nonumber
M_{ij}(T) &\equiv -\frac{U_i}{2\beta}\sum_{n}\sum_{\bm{k}}\text{tr}\left[g^{\text{e}}(\bm{k},i\omega_n) 
\hat{\Delta}_j(\bm{k}) 
g^{\text{h}}(\bm{k},i\omega_n)\hat{\Delta}^{\dagger}_i(\bm{k})\right] \\ 
&=-\frac{U_i}{2}\sum_{n,m}\sum_{\bm{k}}\left(\frac{f(\epsilon_n(\bm{k}))}{\epsilon_n(\bm{k})+\epsilon_m(-\bm{k})}-\frac{f(-\epsilon_m(-\bm{k}))}{\epsilon_m(-\bm{k})+\epsilon_n(\bm{k})}\right) \langle \epsilon_n(\bm{k})|\hat{\Delta}_j(\bm{k})|\epsilon^{\text{T}}_m(-\bm{k})\rangle\langle \epsilon^{\text{T}}_m(-\bm{k})|\hat{\Delta}^{\dagger}_i(\bm{k})|\epsilon_n(\bm{k})\rangle,
\end{align}
where $H_0(\bm{k})|\epsilon_n(\bm{k})\rangle=\epsilon_n(\bm{k})|\epsilon_n(\bm{k})\rangle, \ H^{\text{T}}_0(\bm{k})|\epsilon^{\text{T}}_n(\bm{k})\rangle=\epsilon_n(\bm{k})|\epsilon^{\text{T}}_n(\bm{k})\rangle$ and we use the residue theorem:
\begin{align}
\frac{1}{\beta}\sum_{n}g(i\omega_n) &= -\int_C \frac{dz}{2\pi i}F(z)g(z) = \sum_{i}\text{Res}_{z=a_i}[F(z)g(z)]
\label{eq: residue theorem}
\end{align}
with $F(z)=1/(e^{\beta z}+1)$ and $z=a_i$ denoting the pole of $g(z)$. We take the integral path $C$ counter-clockwise. As a result, we obtain the linearized gap equation: 
\begin{align}
M(T)\bm{\Delta}=\bm{\Delta}.
\end{align}
Here, we consider the case where the above equation has a general eigenvalue $\lambda(T)$.
\begin{align}
M(T)\bm{\Delta}=\lambda(T)\bm{\Delta}.
\end{align}
In this case, $\bm{\Delta} \neq \bm 0$ occurs when $\lambda(T) = 1$. If $\lambda(T) < 1$, from the above gap equation, $\lambda(T)\bm{\Delta} = \bm{\Delta} \leftrightarrow (1-\lambda(T))\bm{\Delta} = \bm 0$, and only the trivial solution $\bm{\Delta} = \bm 0$ appears. Furthermore, below the transition temperature, $\lambda(T) > 1$, the linear approximation breaks down, where we have to solve the original gap equation (\ref{eq: gap-equation-multi}) self-consistently.

\subsection{Raman response of collective modes}

Having set up the formalism, we now consider the Raman response of collective modes in the case where multiple components of the order parameter are simultaneously nonzero. Among the order-parameter components introduced above, we label those that are ordered by $\alpha=1,2,...,N_{\Delta}$. We define the fluctuations of the order parameter as
\begin{align}
\Delta_{\alpha}(\bm{r},\tau) \equiv \Delta^{\text{MF}}_{\alpha} + \delta \Delta_{\alpha}(\bm{r},\tau),
\end{align}
where $\Delta^{\text{MF}}_{\alpha}$ is obtained from the gap equation. As shown below, within the first-order fluctuation regime, the real and imaginary parts of the order-parameter fluctuation $\delta\Delta_{\alpha}(\bm r,\tau)$ correspond to the amplitude (Higgs) and phase (Nambu--Goldstone) modes, respectively:
\begin{align}
\nonumber
\Delta^{\text{MF}}_{\alpha} \to (\Delta^{\text{MF}}_{\alpha} + H_{\alpha}(\bm{r},\tau))e^{i\theta_{\alpha}(\bm{r},\tau)} &\approx \Delta^{\text{MF}}_{\alpha}+H_{\alpha}(\bm{r},\tau)+i\Delta^{\text{MF}}_{\alpha}\theta_{\alpha}(\bm{r},\tau),\\ \nonumber
H_{\alpha}(\bm{r},\tau) &\equiv \text{Re}[\delta\Delta_{\alpha}(\bm{r},\tau)], \\
\Theta_{\alpha}(\bm{r},\tau) &\equiv \text{Im}[\delta\Delta_{\alpha}(\bm{r},\tau)]=\Delta^{\text{MF}}_{\alpha}\theta_{\alpha}(\bm{r},\tau), \label{eq: Higgs-GS}
\end{align}
where $\Delta^{\text{MF}}_{\alpha},H_{\alpha}(\bm{r},\tau),\Theta_{\alpha}(\bm{r},\tau)\in \mathbb{R}$. Revisiting the action given in Eq.~(\ref{eq: general-action}), we separate the Green function $G^{-1}$ to mean-field part $G^{-1}_0$ and fluctation and external fields $\Sigma$:
\begin{align}
&S[\Delta^{*},\Delta,\phi] =\sum_{\alpha}\frac{\beta}{U_{\alpha}}\sum_{n}\sum_{\bm{q}}|\Delta_{\alpha}(Q_n)|^2 - \frac{1}{2}\text{Tr }\text{ln }[-\beta G^{-1}(\bm{k}_1,\bm{k}_2,i\omega_m,i\omega_n)]+S_{\text{EM}}[\phi], \label{eq: general-action2} \\
\nonumber
&G^{-1}(\bm{k}_1,\bm{k}_2,i\omega_m,i\omega_n) = G^{-1}_{0}(\bm{k}_1,i\omega_m) - \Sigma(\bm{k}_1,\bm{k}_2,i\omega_m,i\omega_n), \\ \nonumber
&G^{-1}_{0}(\bm{k}_1,i\omega_m) = -\left[
\begin{array}{cc}
(-i\omega_m+H_0(\bm{k}_1)) & \sum_{\alpha}\Delta^{\text{MF}}_{\alpha}\hat{\Delta}_{\alpha}\left(\bm{k}_1\right) \\
\sum_{\alpha}\Delta^{\text{MF}}_{\alpha}\hat{\Delta}^{\dagger}_{\alpha}\left(\bm{k}_1\right) & (-i\omega_m-H^{\text{T}}_0(-\bm{k}_1))
\end{array}
\right] = \left[\frac{1}{i\omega_m1_{2N\times 2N}-H_{\text{BdG}}(\bm{k}_1)}\right]^{-1},\\ 
\nonumber
&\Sigma(\bm{k}_1,\bm{k}_2,i\omega_m,i\omega_n) = \Sigma_{\Delta}(\bm{k}_1,\bm{k}_2,i\omega_m,i\omega_n) + \Sigma_{R,\phi}(\bm{k}_1,\bm{k}_2,i\omega_m,i\omega_n), \\ \nonumber
&\Sigma_{\Delta}\equiv
\left[
\begin{array}{cc}
0 & \sum_{\alpha}\sum_{\bm{q}}\delta\Delta_{\alpha}(\bm{q},i\omega_m-i\omega_n)\hat{\Delta}_{\alpha}\left(\frac{\bm{k}_1+\bm{k_2}}{2}\right)\delta_{\bm{q},\bm{k}_1-\bm{k}_2} \\
\sum_{\alpha}\sum_{\bm{q}}\delta\Delta^{*}_{\alpha}(\bm{q},-i\omega_m+i\omega_n)\hat{\Delta}^{\dagger}_{\alpha}\left(\frac{\bm{k}_1+\bm{k_2}}{2}\right)\delta_{\bm{q},-\bm{k}_1+\bm{k}_2} & 0
\end{array}
\right], \\ \nonumber
&\Sigma_{R,\phi}\equiv \sum_{\bm{q}}\Bigg[-R(\bm{q},i\omega_m-i\omega_n)\Gamma\left(\frac{\bm{k}_1+\bm{k}_2}{2}\right)  -ie\phi(\bm{q},i\omega_m-i\omega_n)\tau_z \Bigg]\delta_{\bm{q},\bm{k}_1-\bm{k}_2}, \\ 
&S_{\text{EM}}[\phi] = \frac{\beta}{8\pi}\sum_{\bm{q}}\sum_{n}\bm{q}^2\phi(-Q_n)\phi(Q_n),
\end{align}
where $Q_n \equiv (\bm{q},i\Omega_n)$. We next expand the fermionic determinant around the mean-field solution. Using $G^{-1}=G_0^{-1}-\Sigma$ and the identity $\ln(1-X)=-\sum_{L\ge 1}X^L/L$, we obtain
\begin{align}
\nonumber
\text{Tr }\text{ln }[-\beta G^{-1}] &= \text{Tr }\text{ln }[-\beta G^{-1}_{0}(1 - G_{0}\Sigma)] = \text{Tr }\text{ln }[-\beta G^{-1}_{0}]+\text{Tr }\text{ln }[1 - G_{0}\Sigma] \\
&=\text{Tr }\text{ln }[-\beta G^{-1}_{0}]-\text{Tr }\left[\sum_{L\ge1}\frac{(G_0\Sigma)^L}{L}\right].
\end{align}
This series expansion can be applied in the regime $\beta|\delta\Delta_\alpha|\ll 1$ and will be truncated at quadratic order in the fluctuation fields. For $L=1$, the linear terms of the fluctations $\delta\Delta^{*},\delta\Delta$ appear, but they are cancelled by the linear terms of those in the first term of Eq.~(\ref{eq: general-action2}). Also, we neglect the constant terms which are proportional to $R(Q=0)$ and $\phi(Q=0)$. For $L=2$, we obatin
\begin{align}
\nonumber
&\frac{1}{2}\text{Tr }\left[G_0\Sigma G_0\Sigma\right] \\ \nonumber
&=\sum_{\alpha,\beta}\sum_{\bm{k},\bm{q}}\sum_{m,n}\delta\Delta_{\alpha}(Q_n)\delta\Delta^{*}_{\beta}(Q_n)\text{tr }\Big(U_{21}\hat{\Delta}^{\dagger}_{\beta}(\bm{k}+\bm{q}/2)G_0(\bm{k}+\bm{q},i\omega_m+i\Omega_n)U_{12}\hat{\Delta}_{\alpha}(\bm{k}+\bm{q}/2) G_0(\bm{k},i\omega_m)\Big) \\ \nonumber
&+\frac{1}{2}\sum_{\alpha,\beta}\sum_{\bm{k},\bm{q}}\sum_{m,n}\delta\Delta^{*}_{\alpha}(-Q_n)\delta\Delta^{*}_{\beta}(Q_n)\text{tr }\Big(U_{21}\hat{\Delta}^{\dagger}_{\beta}(\bm{k}+\bm{q}/2)G_0(\bm{k}+\bm{q},i\omega_m+i\Omega_n)U_{21}\hat{\Delta}^{\dagger}_{\alpha}(\bm{k}+\bm{q}/2) G_0(\bm{k},i\omega_m)\Big) \\ \nonumber
&+\frac{1}{2}\sum_{\alpha,\beta}\sum_{\bm{k},\bm{q}}\sum_{m,n}\delta\Delta_{\alpha}(Q_n)\delta\Delta_{\beta}(-Q_n)\text{tr }\Big(U_{12}\hat{\Delta}_{\beta}(\bm{k}+\bm{q}/2)G_0(\bm{k}+\bm{q},i\omega_m+i\Omega_n)U_{12}\hat{\Delta}_{\alpha}(\bm{k}+\bm{q}/2) G_0(\bm{k},i\omega_m)\Big) \\ \nonumber
&+ \frac{1}{2}\sum_{\bm{k},\bm{q}}\sum_{m,n}R(Q_n)R(-Q_n)\text{tr }\Big(\Gamma(\bm{k}+\bm{q}/2)G_0(\bm{k}+\bm{q},i\omega_m+i\Omega_n)\Gamma(\bm{k}+\bm{q}/2)G_0(\bm{k},i\omega_m)\Big) 
\\ \nonumber
&+ie\sum_{\bm{k},\bm{q}}\sum_{m,n}R(Q_n)\phi(-Q_n)\text{tr }\Big(\tau_zG_0(\bm{k}+\bm{q},i\omega_m+i\Omega_n)\Gamma(\bm{k}+\bm{q}/2)G_0(\bm{k},i\omega_m)\Big) 
\\ \nonumber
&-\sum_{\alpha}\sum_{\bm{k},\bm{q}}\sum_{m,n}R(Q_n)\delta \Delta^{*}_{\alpha}(Q_n)\text{tr }\Big(U_{21}\hat{\Delta}^{\dagger}_{\alpha}(\bm{k}+\bm{q}/2)G_0(\bm{k}+\bm{q},i\omega_m+i\Omega_n)\Gamma(\bm{k}+\bm{q}/2)G_0(\bm{k},i\omega_m)\Big) 
\\ \nonumber
&-\sum_{\alpha}\sum_{\bm{k},\bm{q}}\sum_{m,n}R(Q_n)\delta \Delta_{\alpha}(-Q_n) \text{tr }\Big(U_{12}\hat{\Delta}_{\alpha}(\bm{k}+\bm{q}/2)G_0(\bm{k}+\bm{q},i\omega_m+i\Omega_n)\Gamma(\bm{k}+\bm{q}/2)G_0(\bm{k},i\omega_m)\Big) 
\\ \nonumber
&- \frac{e^2}{2}\sum_{\bm{k},\bm{q}}\sum_{m,n}\phi(Q_n)\phi(-Q_n)\text{tr }\Big(\tau_z G_0(\bm{k}+\bm{q},i\omega_m+i\Omega_n)\tau_z G_{0}(\bm{k},i\omega_m)\Big)
\\ \nonumber
&-ie\sum_{\alpha}\sum_{\bm{k},\bm{q}}\sum_{m,n}\phi(Q_n)\delta \Delta^{*}_{\alpha}(Q_n)\text{tr }\Big(U_{21}\hat{\Delta}^{\dagger}_{\alpha}(\bm{k}+\bm{q}/2) G_0(\bm{k}+\bm{q},i\omega_m+i\Omega_n)\tau_zG_0(\bm{k},i\omega_m)\Big) 
\\
&-ie\sum_{\alpha}\sum_{\bm{k},\bm{q}}\sum_{m,n}\phi(Q_n)\delta \Delta_{\alpha}(-Q_n) \text{tr }\Big(U_{12}\hat{\Delta}_{\alpha}(\bm{k}+\bm{q}/2)G_0(\bm{k}+\bm{q},i\omega_m+i\Omega_n)\tau_z G_0(\bm{k},i\omega_m)\Big),
\label{eq: diagram}
\end{align}
where the corresponding diagrams are shown in Fig.~\ref{fig: diagram} (b) and (c). 
By using Eq.~(\ref{eq: residue theorem}), we find
\begin{align}
\nonumber
&\sum_{r}\text{tr }\Big(AG_0(\bm{k}+\bm{q},i\omega_r+i\Omega_p)BG_0(\bm{k},i\omega_r)\Big) =\beta\sum_{i}\text{Res}_{z=a_i}\left[F(z)\text{tr }\left(AG_0(\bm{k}+\bm{q},z+i\Omega_p)BG_0(\bm{k},z)\right)\right] \\ 
=\beta&\sum_{m,n}\frac{f(E_m(\bm{k}))-f(E_n(\bm{k}+\bm{q}))}{i\Omega_p+(E_m(\bm{k})-E_n(\bm{k}+\bm{q}))} \langle u_m(\bm{k})|A|u_{n}(\bm{k}+\bm{q})\rangle \langle u_n(\bm{k}+\bm{q})|B|u_{m}(\bm{k})\rangle,
\end{align}
where $A,B$ are arbitrary operators and $H_{\text{BdG}}(\bm{k})|u_m(\bm{k})\rangle = E_m(\bm{k})|u_m(\bm{k})\rangle$. 
Collecting the quadratic terms in the Raman source field $R(Q_p)$ and in the bosonic fluctuation fields, we obtain
\begin{align}
&S[\delta\Delta^{*},\delta\Delta,\phi] = S_R + S_{\text{FL}}, \\ \nonumber
&S_R \equiv 
 \frac{\beta}{4}\sum_{\bm{q}}\sum_p R(Q_p)R(-Q_p)\Phi_{RR}(Q_p),
\\ \nonumber
&S_{\text{FL}} \equiv \frac{\beta}{2}\sum_{\bm{q}}\sum_p \ \bm{\delta}\Delta^{\text{T}}_{\phi}(-Q_p) U^{-1}_{\text{eff},\phi}(-Q_p) \bm{\delta}\Delta_{\phi}(Q_p)  \\  &\quad \quad \quad - \frac{\beta}{4}\sum_{\bm{q}}\sum_p\bm{\delta}\Delta^{\text{T}}_{\phi}(-Q_p)\cdot \bm{Q}_{R,\phi}(Q_p)R(Q_p) - \frac{\beta}{4}\sum_{\bm{q}}\sum_p [\bm{Q}_{R,\phi}(-Q_p)R(-Q_p)]^{\text{T}} \cdot\bm{\delta} \Delta_{\phi}(Q_p), \label{eq: general-action3} \\
&U^{-1}_{\text{eff},\phi}(Q_p) \equiv
\left[
\begin{array}{c|c}
U^{-1}_{\text{eff}}(Q_p) & \begin{matrix}  \frac{-ie}{2}\bm{Q}_{\phi,\Delta}(-Q_p) \\[2pt] \frac{-ie}{2}\bm{Q}_{\phi,\Delta^{\dagger}}(-Q_p) \end{matrix} \\
\hline
\begin{matrix} \frac{-ie}{2}\bm{Q}^{\text{T}}_{\phi,\Delta}(Q_p) & \frac{-ie}{2}\bm{Q}^{\text{T}}_{\phi,\Delta^{\dagger}}(Q_p) \end{matrix} &  \frac{1}{4\pi}(\bm{q}^2-2\pi e^2\kappa(Q_p))
\end{array}
\right], \quad \bm{Q}_{R,\phi}(Q_p) \equiv \left[
\begin{array}{c}
\bm{Q}_{R,\Delta}(Q_p) \\
\bm{Q}_{R,\Delta^{\dagger}}(Q_p) \\
-ie\Pi_{R,\phi}(Q_p)
\end{array}
\right],\label{eq: general-Anderson-Higgs} 
\end{align}
where $\bm{\delta}\Delta_{\phi}(Q_p)\equiv \big[\delta\Delta_1(Q_p),\dots,\delta\Delta_{N_\Delta}(Q_p),\delta\Delta_1^{*}(-Q_p),\dots,\delta\Delta_{N_\Delta}^{*}(-Q_p),\phi(Q_p)\big]^{\text T}$. $U^{-1}_{\text{eff},\phi}(Q_p)$ plays the role of an inverse propagator for the coupled system of amplitude/phase fluctuations of the order parameters and the scalar potential.  The poles of $U_{\text{eff},\phi}(Q_p)$, or equivalently the zeros of $\det [U^{-1}_{\text{eff},\phi}(Q_p)]$, determine the mass (the excitation energy) of each collective mode including the Anderson--Higgs mass.
Also, we can explicitly write down
\begin{align}
\nonumber
&\Phi_{RR}(Q_p) \equiv \sum_{m,n}\sum_{\bm{k}}\frac{f(E_m(\bm{k}))-f(E_n(\bm{k}+\bm{q}))}{i\Omega_p+(E_m(\bm{k})-E_n(\bm{k}+\bm{q}))} \langle u_m(\bm{k})|\Gamma(\bm{k}_{+})|u_{n}(\bm{k}+\bm{q})\rangle \langle u_n(\bm{k}+\bm{q})|\Gamma(\bm{k}_{+})|u_{m}(\bm{k})\rangle, \\ \nonumber
&\kappa(Q_p) \equiv \sum_{m,n}\sum_{\bm{k}}\frac{f(E_m(\bm{k}))-f(E_n(\bm{k}+\bm{q}))}{i\Omega_p+(E_m(\bm{k})-E_n(\bm{k}+\bm{q}))} \langle u_m(\bm{k})|\tau_z|u_{n}(\bm{k}+\bm{q})\rangle \langle u_n(\bm{k}+\bm{q})|\tau_z|u_{m}(\bm{k})\rangle, \\ 
\nonumber
U^{-1}_{\text{eff}}(Q_p) &\equiv \left[
\begin{array}{cc}
\frac{1}{2}\Pi^{(2)}_{\text{O}}(Q_p) &  \frac{1}{U}+\frac{1}{2}\Pi_{\text{D}}(Q_p) \\
\frac{1}{U}+\frac{1}{2}\Pi^{\text{T}}_{\text{D}}(-Q_p) & \frac{1}{2}\Pi^{(1)}_{\text{O}}(Q_p)
\end{array}
\right], \quad \left[1/U\right]_{\alpha\beta}=G^{\alpha}_{\Delta}\delta_{\alpha\beta},
 \\ \nonumber
G^{\alpha}_{\Delta} &= -\frac{1}{2\Delta^{\text{MF}}_{\alpha}} \sum_{\bm{k}}\left[\sum_{m}f(E_m(\bm{k}))\langle u_m(\bm{k})|U_{21}\hat{\Delta}^{\dagger}_{\alpha}(\bm{k})|u_{m}(\bm{k})\rangle\right], \\ \nonumber
\left[\Pi_{\text{D}}(Q_p)\right]_{\alpha \beta} &= \sum_{m,n}\sum_{\bm{k}}\frac{f(E_m(\bm{k}))-f(E_n(\bm{k}+\bm{q}))}{i\Omega_p+(E_m(\bm{k})-E_n(\bm{k}+\bm{q}))}\langle u_m(\bm{k})|U_{21}\hat{\Delta}^{\dagger}_{\beta}(\bm{k}_+) |u_{n}(\bm{k}+\bm{q})\rangle \langle u_n(\bm{k}+\bm{q})|U_{12}\hat{\Delta}_{\alpha}(\bm{k}_+)|u_{m}(\bm{k})\rangle, \\ \nonumber
\left[\Pi^{(1)}_{\text{O}}(Q_p)\right]_{\alpha \beta}  &= \sum_{m,n}\sum_{\bm{k}}\frac{f(E_m(\bm{k}))-f(E_n(\bm{k}+\bm{q}))}{i\Omega_p+(E_m(\bm{k})-E_n(\bm{k}+\bm{q}))}\langle u_m(\bm{k})|U_{21}\hat{\Delta}^{\dagger}_{\beta}(\bm{k}_+)|u_{n}(\bm{k}+\bm{q})\rangle\langle u_n(\bm{k}+\bm{q})|U_{21}\hat{\Delta}^{\dagger}_{\alpha}(\bm{k}_+)|u_{m}(\bm{k})\rangle, \\ \nonumber
\left[\Pi^{(2)}_{\text{O}}(Q_p)\right]_{\alpha \beta}  &=\sum_{m,n}\sum_{\bm{k}}\frac{f(E_m(\bm{k}))-f(E_n(\bm{k}+\bm{q}))}{i\Omega_p+(E_m(\bm{k})-E_n(\bm{k}+\bm{q}))}\langle u_m(\bm{k})|U_{12}\hat{\Delta}_{\beta}(\bm{k}_+)|u_{n}(\bm{k}+\bm{q})\rangle\langle u_n(\bm{k}+\bm{q})|U_{12}\hat{\Delta}_{\alpha}(\bm{k}_+)|u_{m}(\bm{k})\rangle, \\ \nonumber
\bm{Q}_{R,Y}(Q_p) &\equiv \left[[Q_{R,Y}(Q_p)]_{1},[Q_{R,Y}(Q_p)]_{2},...,[Q_{R,Y}(Q_p)]_{N_{\alpha}}\right]^\text{T}, \quad Y=\Delta,\Delta^{\dagger}, \\ \nonumber
[Q_{R,\Delta}(Q_p)]_{\alpha} &=\sum_{m,n}\sum_{\bm{k}}\frac{f(E_m(\bm{k}))-f(E_n(\bm{k}+\bm{q}))}{i\Omega_p+(E_m(\bm{k})-E_n(\bm{k}+\bm{q}))}\langle u_m(\bm{k})|U_{12}\hat{\Delta}_{\alpha}(\bm{k}_+)|u_{n}(\bm{k}+\bm{q})\rangle \langle u_n(\bm{k}+\bm{q})|\Gamma(\bm{k}_{+})|u_{m}(\bm{k})\rangle, \\ \nonumber
[Q_{R,\Delta^{\dagger}}(Q_p)]_{\alpha} &=\sum_{m,n}\sum_{\bm{k}}\frac{f(E_m(\bm{k}))-f(E_n(\bm{k}+\bm{q}))}{i\Omega_p+(E_m(\bm{k})-E_n(\bm{k}+\bm{q}))}\langle u_m(\bm{k})|U_{21}\hat{\Delta}^{\dagger}_{\alpha}(\bm{k}_+)|u_{n}(\bm{k}+\bm{q})\rangle \langle u_n(\bm{k}+\bm{q})|\Gamma(\bm{k}_{+})|u_{m}(\bm{k})\rangle, \\ \nonumber
\bm{Q}_{\phi,Y}(Q_p) &\equiv \left[[Q_{\phi,Y}(Q_p)]_{1},[Q_{\phi,Y}(Q_p)]_{2},...,[Q_{\phi,Y}(Q_p)]_{N_{\alpha}}\right]^\text{T}, \quad Y=\Delta,\Delta^{\dagger}, \\ \nonumber
[Q_{\phi,\Delta}(Q_p)]_{\alpha} &=\sum_{m,n}\sum_{\bm{k}}\frac{f(E_m(\bm{k}))-f(E_n(\bm{k}+\bm{q}))}{i\Omega_p+(E_m(\bm{k})-E_n(\bm{k}+\bm{q}))}\langle u_m(\bm{k})|U_{12}\hat{\Delta}_{\alpha}(\bm{k}_+)|u_{n}(\bm{k}+\bm{q})\rangle \langle u_n(\bm{k}+\bm{q})|\tau_z|u_{m}(\bm{k})\rangle, \\ \nonumber
[Q_{\phi,\Delta^{\dagger}}(Q_p)]_{\alpha} &=\sum_{m,n}\sum_{\bm{k}}\frac{f(E_m(\bm{k}))-f(E_n(\bm{k}+\bm{q}))}{i\Omega_p+(E_m(\bm{k})-E_n(\bm{k}+\bm{q}))}\langle u_m(\bm{k})|U_{21}\hat{\Delta}^{\dagger}_{\alpha}(\bm{k}_+)|u_{n}(\bm{k}+\bm{q})\rangle \langle u_n(\bm{k}+\bm{q})|\tau_z|u_{m}(\bm{k})\rangle, \\
\Pi_{R,\phi}(Q_p) &=\sum_{m,n}\sum_{\bm{k}}\frac{f(E_m(\bm{k}))-f(E_n(\bm{k}+\bm{q}))}{i\Omega_p+(E_m(\bm{k})-E_n(\bm{k}+\bm{q}))}\langle u_m(\bm{k})|\tau_z|u_{n}(\bm{k}+\bm{q})\rangle \langle u_n(\bm{k}+\bm{q})|\Gamma(\bm{k}_+)|u_{m}(\bm{k})\rangle
\label{eq: total}
\end{align}
where $\bm{k}_{\pm} \equiv \bm{k} \pm \bm{q}/2$. Here $U_{\mathrm{eff}}$ represents the effective interaction in the pairing channel obtained within the random-phase approximation (RPA), i.e., as the geometric series of ladder diagrams shown in Fig.~\ref{fig: diagram}(a).
\begin{align}
U_{\mathrm{eff}} = U -U\Pi U + \cdots = (1+U\Pi)^{-1}U, 
\label{U_eff}
\end{align}
Also, by using the gap equation given in Eq.~(\ref{eq: gap-equation-multi}), we can replace the interaction $1/U_{\alpha}$ to $G^{\alpha}_{\Delta}$, which is given by the mean-field solution $\Delta^{\text{MF}}_{\alpha}$. We consider the Gaussian integral of bosonic fields $\bm{\delta} \Delta_{\phi}$:
\begin{align}
\nonumber
&\int D[\bm{\delta} \Delta_{\phi}] \text{exp}\Bigg\{-\Bigg[\frac{\beta}{2}\sum_{\bm{q}}\sum_p \ \bm{\delta}\Delta^{\text{T}}_{\phi}(-Q_p) U^{-1}_{\text{eff},\phi}(-Q_p) \bm{\delta}\Delta_{\phi}(Q_p)  \\ \nonumber  &\quad- \frac{\beta}{4}\sum_{\bm{q}}\sum_p\bm{\delta}\Delta^{\text{T}}_{\phi}(-Q_p)\cdot \bm{Q}_{R,\phi}(Q_p)R(Q_p) - \frac{\beta}{4}\sum_{\bm{q}}\sum_p [\bm{Q}_{R,\phi}(-Q_p)R(-Q_p)]^{\text{T}} \cdot\bm{\delta} \Delta_{\phi}(Q_p)\Bigg]\Bigg\} \\
&=\text{exp}\left(\frac{\beta}{8}\sum_{\bm{q}}\sum_{p}\bm{T}^{\text{T}}(-Q_p)U_{\text{eff},\phi}(-Q_p)\bm{T}(Q_p)\right), \quad \bm{T}(Q_p) \equiv \bm{Q}_{R,\phi}(Q_p)R(Q_p).
\end{align}
As a result, we obtain
\begin{align}
S_{\text{eff}}[R] = \frac{\beta}{4}\sum_{\bm{q}}\sum_p R(Q_p)R(-Q_p)\Phi_{RR}(Q_p) -\frac{\beta}{8}\sum_{\bm{q}}\sum_{p}\bm{T}^{\text{T}}(-Q_p)U_{\text{eff},\phi}(-Q_p)\bm{T}(Q_p)
\end{align}
and hence the Raman susceptibility is given by
\begin{align}
\nonumber
\chi_{RR}(Q_p) &\equiv \frac{1}{\beta}\frac{\delta S_{\text{eff}}}{\delta R(-Q_p)\delta R(Q_p)} \\ \nonumber
    &=\frac{1}{4}\Phi_{RR}(Q_p) -\frac{1}{8}\bm{Q}^{\text{T}}_{R,\phi}(-Q_p)U_{\text{eff},\phi}(-Q_p)\bm{Q}_{R,\phi}(Q_p) \\
    &=\frac{1}{4}\Phi_{RR}(Q_p) -\frac{1}{8}\bm{Q}^{\text{T}}_{R,\phi}(Q_p)U_{\text{eff},\phi}(Q_p)\bm{Q}_{R,\phi}(-Q_p),
\end{align}
where we use $U^{\text{T}}_{\text{eff},\phi}(Q_p)=U_{\text{eff},\phi}(-Q_p)$.
After analytical continuation $i\Omega_p \to \omega+i0^{+}$, the retarded Raman susceptibility is given by
\begin{align}
\chi_{RR}(\bm{q},\omega) &=\frac{1}{4}\Phi_{RR}(\bm{q},\omega) -\frac{1}{8}\bm{Q}^{\text{T}}_{R,\phi}(\bm{q},\omega)U_{\text{eff},\phi}(\bm{q},\omega)\bm{Q}_{R,\phi}(-\bm{q},-\omega),
\label{eq: raman}
\end{align}
We note that for any given BdG Hamiltonian $H_{\text{BdG}}(\bm{k})$, we can immediately calculate the above Raman susceptibility through Eq.~(\ref{eq: total}) by $i\Omega_p \to \omega+i0^{+}$.

\section{Alternative representation of the Raman susceptibility and vanishing of the Coulomb screening at  \texorpdfstring{$\bm q=0$}{q=0}}\label{app: Alternative representation and Coulomb screening}

In this section, we rewrite the Raman susceptibility in Eq.~(\ref{eq: raman}) in a form that separates the contribution of the scalar field from the purely superconducting part. Our goal is to make explicit which pieces of $\chi_{RR}$ survive when the scalar field is switched off and which arise solely from its coupling to the superconducting collective modes. To this end, we go back to the Gaussian action in Eq.~(\ref{eq: general-action3}) and regard the fluctuation vector $\bm{\delta}\Delta_{\phi}$ as a multi-component bosonic field with quadratic kernel $U^{-1}_{\text{eff},\phi}(Q)$. This kernel contains (i) a part acting purely in the superconducting sector $(\delta\Delta,\delta\Delta^{*})$, (ii) a part describing the dynamics of the scalar field $\phi$, and (iii) off-diagonal terms which hybridize these two sectors. It is therefore convenient to decompose this effective interaction into a $2\times 2$ block structure,
\begin{align}
&U^{-1}_{\text{eff},\phi}(\bm{q},\omega) \equiv \left[\begin{array}{cc} 
B(\bm{q},\omega) & C(\bm{q},\omega) \\
C^{\text{T}}(-\bm{q},-\omega) & D(\bm{q},\omega) 
\end{array}
\right] \quad \to \quad  U_{\text{eff},\phi}(\bm{q},\omega) \equiv \left[\begin{array}{cc} 
B^{-1}+B^{-1}CS^{-1}C^{\text{T}}B^{-1} & -B^{-1}CS^{-1} \\
-S^{-1}C^{\text{T}}B^{-1} & S^{-1}
\end{array}
\right], \\
&S(\bm{q},\omega) \equiv D(\bm{q},\omega)-C^{\text{T}}(-\bm{q},-\omega)B^{-1}(\bm{q},\omega)C(\bm{q},\omega),
\end{align}
where \begin{align}
&B(\bm{q},\omega) \equiv U^{-1}_{\text{eff}}(\bm{q},\omega), \quad B^{\text{T}}(\bm{q},\omega) = B(-\bm{q},-\omega), \\
&C(\bm{q},\omega) \equiv \left[
\begin{array}{c}
\frac{-ie}{2}\bm{Q}_{\phi,\Delta}(-\bm{q},-\omega) \\
\frac{-ie}{2}\bm{Q}^{*}_{\phi,\Delta}(\bm{q},\omega)
\end{array}
\right], \quad \bm{Q}_{\phi,\Delta^{\dagger}}(\bm{q},\omega) =\bm{Q}^{*}_{\phi,\Delta}(-\bm{q},-\omega), \\
&D(\bm{q},\omega) \equiv \frac{1}{4\pi}(\bm{q}^2-2\pi e^2\kappa(\bm{q},\omega)).
\end{align}
We also rewrite the source vector as
\begin{align}
\bm{Q}_{R,\phi}(\bm{q},\omega) \equiv \left[
\begin{array}{c}
\alpha(\bm{q},\omega) \\
\beta(\bm{q},\omega)
\end{array}
\right], \quad \alpha(\bm{q},\omega) \equiv \left[
\begin{array}{c}
\bm{Q}_{R,\Delta}(\bm{q},\omega) \\
\bm{Q}^{*}_{R,\Delta}(-\bm{q},-\omega)
\end{array}
\right],  \quad \beta(\bm{q},\omega) \equiv -ie\Pi_{R,\phi}(\bm{q},\omega).
\label{eq: definitions-vertices}
\end{align}
From Eq.~(\ref{eq: raman}), we find 
\begin{align}
\nonumber
\chi_{RR}(\bm{q},\omega) &=\frac{1}{4}\Phi_{RR}(\bm{q},\omega) -\frac{1}{8}\bm{Q}^{\text{T}}_{R,\phi}(\bm{q},\omega)U_{\text{eff},\phi}(\bm{q},\omega)\bm{Q}_{R,\phi}(-\bm{q},-\omega) \label{eq: raman-another-app} \\
&=\pi_{RR}(\bm{q},\omega) -\frac{1}{8}X^{\text{T}}(\bm{q},\omega)S^{-1}(\bm{q},\omega)X(-\bm{q},-\omega), \\
\pi_{RR}(\bm{q},\omega) &\equiv \frac{1}{4}\Phi_{RR}(\bm{q},\omega) - \frac{1}{8}\alpha^{\text{T}}(\bm{q},\omega)B^{-1}(\bm{q},\omega)\alpha(-\bm{q},-\omega), \label{eq: Raman_without_scalar-app} \\
X(\bm{q},\omega) &\equiv \beta(\bm{q},\omega) - C^{\text{T}}(\bm{q},\omega)B^{-1}(-\bm{q},-\omega)\alpha(\bm{q},\omega).
\end{align}
$\pi_{RR}$ is the Raman susceptibility obtained after integrating out the superconducting fluctuations but without the scalar field, whereas the second term of Eq.~(\ref{eq: raman-another-app}) contains all scalar-field contributions.

From the above definitions, we obtain the effective coupling between the Raman source field and the scalar field,
\begin{align}
\nonumber
X(\bm{q},\omega) &= -e\Pi_{\text{RC}}(\bm{q},\omega), \\
\Pi_{\text{RC}}(\bm{q},\omega) &\equiv i\left(\Pi_{R,\phi}(\bm{q},\omega)-\frac{1}{2}[\bm{Q}^{\text{T}}_{\phi,\Delta}(-\bm{q},-\omega), \bm{Q}^{\dagger}_{\phi,\Delta}(\bm{q},\omega)]U_{\text{eff}}(-\bm{q},-\omega)\left[
\begin{array}{c}
\bm{Q}_{R,\Delta}(\bm{q},\omega) \\
\bm{Q}^{*}_{R,\Delta}(-\bm{q},-\omega)
\end{array}
\right]\right), 
\end{align}
and the renormalized effective Coulomb interaction
\begin{align}
\nonumber
S^{-1}(\bm{q},\omega) &= \frac{1}{e^2}\frac{V_C(\bm{q})}{1-V_{C}(\bm{q})\pi_{CC}(\bm{q},\omega)}, \quad V_C(\bm{q}) \equiv \frac{4\pi e^2}{\bm{q}^2}, \\ \nonumber
\pi_{CC}(\bm{q},\omega) &\equiv \frac{1}{2}\left(\kappa(\bm{q},\omega)-\frac{1}{2}\tilde{\bm{Q}}^{\text{T}}_{\phi,\Delta}(-\bm{q},-\omega)U_{\text{eff}}(\bm{q},\omega)\tilde{\bm{Q}}_{\phi,\Delta}(\bm{q},\omega)\right), \\
\tilde{\bm{Q}}_{\phi,\Delta}(\bm{q},\omega) &\equiv \left[
\bm{Q}_{\phi,\Delta}(-\bm{q},-\omega), \bm{Q}^{*}_{\phi,\Delta}(\bm{q},\omega)
\right]^{\text{T}},
\end{align}
where $\pi_{CC}(\bm{q},\omega)$ is the density–density susceptibility including superconducting fluctuations. As a result, we obtain the representation
\begin{align}
\chi_{RR}(\bm{q},\omega) &=\pi_{RR}(\bm{q},\omega) -\frac{1}{8}\Pi^{\text{T}}_{\text{RC}}(\bm{q},\omega)\frac{V_C(\bm{q})}{1-V_{C}(\bm{q})\pi_{CC}(\bm{q},\omega)}\Pi_{\text{RC}}(-\bm{q},-\omega).
\label{eq: alternative-suspensibility}
\end{align}

Finally, we consider the limit $\bm{q}\to\bm{0}$. The renormalized density susceptibility must satisfy
\begin{align}
\pi_{CC}(\bm{0},\omega)=0, \quad (\omega\neq 0),
\label{eq: pi_CC=0}
\end{align}
as a consequence of charge conservation, i.e., the Ward--Takahashi identity. The same identity also constrains the mixed Raman-charge vertex such that
\begin{align}
\Pi_{\text{RC}}(\bm{0},\omega)=0,
\label{eq: Pi_RC=0_q0}
\end{align}
(the details are shown in Appendix \ref{app: Relationship between gauge invariance and Anderson-Higgs mass}). In this limit, the screened Coulomb propagator behaves as
\begin{align}
S^{-1}(\bm{q},\omega)
=\frac{1}{e^2}\frac{V_C(\bm{q})}{1-V_C(\bm{q})\pi_{CC}(\bm{q},\omega)}
\sim \frac{4\pi}{\bm{q}^2}, \quad (\bm{q}\to\bm{0}).
\end{align}
Moreover, when inversion symmetry is present, $\Pi_{\text{RC}}(\bm{q},\omega)$ is an even analytic function of $\bm{q}$, and together with Eq.~(\ref{eq: Pi_RC=0_q0}), this implies
\begin{align}
\Pi_{\text{RC}}(\bm{q},\omega)=\mathcal{O}(\bm{q}^2)\quad (\bm{q}\to\bm{0}).
\label{eq: Pi_RC=0}
\end{align}
Therefore, the Coulomb-screening contribution in Eq.~(\ref{eq: alternative-suspensibility}) vanishes as $\bm{q}\to\bm{0}$ since $V_C(\bm{q})\Pi_{\text{RC}}(\bm{q},\omega)\Pi_{\text{RC}}(-\bm{q},-\omega)\sim \bm{q}^2$,
and we obtain
\begin{align}
\chi_{RR}(\bm{0},\omega)=\pi_{RR}(\bm{0},\omega).
\label{eq: raman2}
\end{align}
This result is consistent with Ref. \cite{Maiti014503}.

\section{Relationship between gauge invariance, Anderson--Higgs mass, and Coulomb screening}\label{app: Relationship between gauge invariance and Anderson-Higgs mass}

In this section, we clarify the relationship between gauge invariance, Anderson--Higgs mass, and Coulomb screening in our formalism. In particular, we show that gauge invariance enforces the Anderson--Higgs mass to vanish at $\bm{q}=0$ so that the Nambu--Goldstone mode is a purely gauge mode for all frequencies. We revisit the fluctuation action $S_{\text{FL}}$ in Eq.~(\ref{eq: general-action3}):
\begin{align}
\nonumber
&S_{\text{FL}}[\delta \Delta^{*}, \delta \Delta,\phi] \equiv \frac{\beta}{2}\sum_p \ \bm{\delta}\Delta^{\text{T}}_{\phi}(-\Omega_p) U^{-1}_{\text{eff},\phi}(-\Omega_p) \bm{\delta}\Delta_{\phi}(\Omega_p)  \\  &\quad \quad \quad - \frac{\beta}{4}\sum_p\bm{\delta}\Delta^{\text{T}}_{\phi}(-\Omega_p)\cdot \bm{Q}_{R,\phi}(\Omega_p)R(\Omega_p) - \frac{\beta}{4}\sum_p [\bm{Q}_{R,\phi}(-\Omega_p)R(-\Omega_p)]^{\text{T}} \cdot\bm{\delta} \Delta_{\phi}(\Omega_p).
\end{align}
The gauge transformation in Euqulid space at $\bm{q}=0$ is given by
\begin{align}
&\phi(i\Omega_p) \to \phi(i\Omega_p) - i\Omega_p\chi(i\Omega_p).
\end{align}
Since the phase of the electron field is also transformed as $c \to e^{ie\chi}c$, the overall phase of the order parameter is also transformed
\begin{align}
\theta(i\Omega_p) \to \theta(i\Omega_p) + 2e\chi(i\Omega_p).
\end{align}
Since each order parameter is given by Eq.~(\ref{eq: Higgs-GS}), we find
\begin{align}
&\delta \Delta_{\alpha}(i\Omega_p) \to \delta \Delta_{\alpha}(i\Omega_p) + 2ie\Delta^{\text{MF}}_{\alpha}\chi(i\Omega_p), \\
&\delta \Delta^{*}_{\alpha}(i\Omega_p) \to \delta \Delta^{*}_{\alpha}(i\Omega_p) -2ie\Delta^{\text{MF}}_{\alpha}\chi(-i\Omega_p),
\end{align}
where $\delta \Delta^{*}_{\alpha}(i\Omega_p)=H(-i\Omega_p)-i\Theta(-i\Omega_p)$ from Eqs.~(\ref{Fourier2}) and (\ref{eq: Higgs-GS}). As a result, the action is changed to
\begin{align}
\nonumber
&S_{\text{FL}}[\delta \Delta^{*}-\epsilon (2ie\Delta^{\text{MF}}_{\alpha}\chi), \delta \Delta+\epsilon (2ie\Delta^{\text{MF}}_{\alpha}\chi),\phi-\epsilon (i\Omega_p \chi)] \\ \nonumber
=&S_{\text{FL}}[\delta \Delta^{*}, \delta \Delta,\phi] + 2ie\epsilon\beta \sum_p \chi(i\Omega_p)\Bigg[ \bm{\delta}\Delta^{\text{T}}_{\phi}(-i\Omega_p) U^{-1}_{\text{eff},\phi}(-i\Omega_p) \bm{\delta}(i\Omega_p) \\ &\quad - \frac{1}{2}\sum_p [\bm{Q}_{R,\phi}(-i\Omega_p)R(-i\Omega_p)]^{\text{T}} \cdot\bm{\delta}(i\Omega_p)\Bigg] - 4e^2\epsilon^2\frac{\beta}{2}\sum_p \chi(i\Omega_p)\chi(-i\Omega_p) \bm{\delta}^{\text{T}}(-i\Omega_p) U^{-1}_{\text{eff},\phi}(-i\Omega_p) \bm{\delta}(i\Omega_p)
\end{align}
where we use $[U^{-1}_{\text{eff},\phi}(i\Omega_p)]^{\text{T}}=U^{-1}_{\text{eff},\phi}(-i\Omega_p)$ and
\begin{align}
\bm{\delta}(i\Omega_p) \equiv \left(\bm\delta_{\text{NG}},-\frac{\Omega_p}{2e}\right)^{\text{T}}, \quad \bm\delta_{\text{NG}} \equiv \left(\Delta^{\text{MF}}_{1}, \Delta^{\text{MF}}_{2},...,\Delta^{\text{MF}}_{N_\Delta},-\Delta^{\text{MF}}_{1},-\Delta^{\text{MF}}_{2},...,-\Delta^{\text{MF}}_{N_\Delta}\right)^{\text{T}}.
\label{eq: Ward-Takahashi-vector}
\end{align}
Because $\chi(i\Omega_p)$ is an arbitrary gauge parameter, gauge invariance requires
\begin{align}
\left.\frac{d}{d\epsilon}S_{\rm FL}[\delta \Delta^{*}-\epsilon (2ie\Delta^{\rm MF}\chi),
\delta \Delta+\epsilon (2ie\Delta^{\rm MF}\chi),
\phi-\epsilon (i\Omega_p \chi)]\right|_{\epsilon=0}=0, \quad (\forall\,\chi).
\end{align}
In order that the above equation holds for arbitrary $\bm{\delta}\Delta_\phi$ and the Raman source $R$,
the coefficient of $\chi$ must vanish separately for $\bm{\delta}\Delta_\phi$ and $R$, which yields
\begin{align}
U^{-1}_{\text{eff},\phi}(i\Omega_p) \bm{\delta}(-i\Omega_p) = \bm{0}, \  \bm{Q}^{\text{T}}_{R,\phi}(i\Omega_p) \cdot\bm{\delta}(-i\Omega_p)=0 \quad
\leftrightarrow  \quad  U^{-1}_{\text{eff},\phi}(\omega) \bm{\delta}(-\omega) = \bm{0}, \  \bm{Q}^{\text{T}}_{R,\phi}(\omega) \cdot\bm{\delta}(-\omega)=0,
\label{eq: Ward-Takahashi}
\end{align}
where we replace $i\Omega_p \to \omega +i0^{+}$. This set of relations is precisely the two-point-function part of the Ward--Takahashi identity at $\bm{q}=\bm{0}$. The first equation of the right side shows that for each frequency $\omega$, the vector $\bm{\delta}(-\omega)$ is a zero mode of $U^{-1}_{\text{eff},\phi}(\omega)$, corresponding to the overall phase mode of the superconducting order parameter dressed by the scalar potential. The second equation guarantees that this gauge mode does not lead to a spurious divergence in the Raman susceptibility in Eq.~(\ref{eq: raman}).

Next, we consider the relationship between these gauge invariance conditions and alternative representation of the Raman susceptibility in Appendix \ref{app: Alternative representation and Coulomb screening}. We first show Eqs.~(\ref{eq: pi_CC=0}) and (\ref{eq: Pi_RC=0}) by using the right side of Eq.~(\ref{eq: Ward-Takahashi}). Writing
$\bm\delta(-\omega)=\left(\bm\delta_{\mathrm{NG}},-\frac{i\omega}{2e}\right)^{\text{T}}$, the equations $U^{-1}_{\text{eff},\phi}(\omega) \bm{\delta}(-\omega) = \bm{0}$ and $\bm{Q}^{\text{T}}_{R,\phi}(\omega) \cdot\bm{\delta}(-\omega)=0$ yield
\begin{align}
U^{-1}_{\text{eff}}(\omega)\bm\delta_{\mathrm{NG}} + \left[
\begin{array}{c}
\frac{-ie}{2}\bm{Q}_{\phi,\Delta}(-\omega) \\
\frac{-ie}{2}\bm{Q}^{*}_{\phi,\Delta}(\omega)
\end{array}
\right]\left(-\frac{i\omega}{2e}\right) = 0, \quad \leftrightarrow \quad \bm\delta_{\mathrm{NG}} =-U_{\text{eff}}(\omega)&\left[
\begin{array}{c}
\frac{-ie}{2}\bm{Q}_{\phi,\Delta}(-\omega) \\
\frac{-ie}{2}\bm{Q}^{*}_{\phi,\Delta}(\omega)
\end{array}
\right]\left(-\frac{i\omega}{2e}\right),
\label{eq: first-eq-Ward-Takahashi1} \\ 
\left[
\begin{array}{c}
\frac{-ie}{2}\bm{Q}_{\phi,\Delta}(\omega) \\
\frac{-ie}{2}\bm{Q}^{*}_{\phi,\Delta}(-\omega)
\end{array}
\right]^{\text{T}}\cdot \bm\delta_{\mathrm{NG}} + \left(-\frac{e^2}{2}\kappa(\omega)\right)\left(-\frac{i\omega}{2e}\right) &= 0, \label{eq: first-eq-Ward-Takahashi2} \\
\left[
\begin{array}{c}
\bm{Q}_{R,\Delta}(\omega) \\
\bm{Q}^{*}_{R,\Delta}(-\omega)
\end{array}
\right]^{\text{T}}\cdot \bm\delta_{\mathrm{NG}} + \left(-ie\Pi_{R,\phi}(\omega)\right)\left(-\frac{i\omega}{2e}\right) &= 0 \label{eq: first-eq-Ward-Takahashi3}.
\end{align}
As a result, substituting $\bm\delta_{\text{NG}}$ of Eq.~(\ref{eq: first-eq-Ward-Takahashi1}) into Eqs.~(\ref{eq: first-eq-Ward-Takahashi2}) and (\ref{eq: first-eq-Ward-Takahashi3}) yields $\pi_{CC}(\bm{q}=0,\omega)=0$ and $\Pi_{RC}(\bm{q}=0,\omega)=0$.
Thus, we obtain Eqs.~(\ref{eq: pi_CC=0}) and (\ref{eq: Pi_RC=0}) with $\omega \neq 0$.

We also note that in Eq.~(\ref{eq: raman2}), there exists the pole of the Nambu--Goldstone mode at $\omega=0$ and hence the susceptibility is divergent at first glance. However, from Eqs.~(\ref{eq: definitions-vertices}) and (\ref{eq: Ward-Takahashi}), we find 
\begin{align}
&\bm{Q}^{\text{T}}_{R,\phi}(0) \cdot\bm{\delta}(0)=0 \quad  \longleftrightarrow \quad  \bm{\alpha}^{\text{T}}(0) \cdot\bm{\delta}_{\text{NG}}=0.
\end{align}
This indicates that the contribution of the Nambu--Goldstone mode to $\alpha^\textrm{T}(0) U_\textrm{eff} \alpha(0)$ vanishes due to the orthogonality condition $\bm{\alpha}^{\text{T}}(0) \cdot\bm{\delta}_{\text{NG}}=0$.
Therefore, the Nambu--Goldstone mode does not lead to a divergence in the Raman susceptibility in Eq.~(\ref{eq: raman2}).

\section{Symmetry operators in superconducting states and group-theoretical classification}\label{app: Symmetry operators in superconducting states and group theoretical classification}

We start with the Bogoliubov-de Gennes (BdG) Hamiltonian, which is given by
\begin{align}
H_{\text{BdG}} &= \frac{1}{2}\sum_{\bm{k}} \bm{c}^{\dagger}_{\bm{k}} H_{\text{BdG}}(\bm k) \bm{c}_{\bm{k}}, \notag \\
\nonumber
H_{\text{BdG}}(\bm{k}) &= \left[ 
\begin{array}{cc}
H_0(\bm{k}) & \Delta(\bm{k}) \\
\Delta^{\dagger}(\bm{k}) & -H_0^{\text{T}}(-\bm{k})
\end{array}
\right]_{2N\times2N}, \\
\Delta(\bm{k}) &= \sum_{\alpha=1,2,...,N_{\Delta}} \Delta^{\text{MF}}_{\alpha} \hat{\Delta}_\alpha(\bm{k}),
\end{align}
where $\bm{c}_{\bm{k}} = [c_{\bm{k}},c^{\dagger}_{-\bm{k}}]^{\text{T}}$, and the indices for the spin, orbital, and sublattice degrees of freedom are implicit. $H_0(\bm k)$ and $\hat{\Delta}_{\alpha}(\vv k)$ are the normal-state Hamiltonian and gap function, where the gap function satisfies $\hat{\Delta}^{\text{T}}_{\alpha}(\vv k) =-\hat{\Delta}_{\alpha}(-\vv k)$ due to the Fermi-Dirac statistics. The BdG Hamiltonian is invariant under PHS,
\begin{align}
 C H_{\text{BdG}}(\vv{k}) C^{-1} = -H_{\text{BdG}}(-\vv{k}), \ \ C=\tau_x K
\end{align}
where $\tau_i$ $(i=x,y,z)$ are the Pauli matrices in Nambu space, and $K$ is the complex conjugate operation.

We consider a normal state with point group symmetry $G$, with an element of $G$ acting on $H_0(\vv{k})$ as follows. For a given $g \in G$, the operations are defined by
\begin{align}
    &D(g) H_0(\bm k) D^{\dagger}(g) = H_0(g\bm k),  \label{eq:normalstateg}
\end{align}
where $g\vv{k}$ means the $O(3)$ transformation of $\vv{k}$ in terms of $g$, e.g., $2_z {\vv k} = (-k_x,-k_y,k_z)$, and $D(g)$ is a spinful unitary representation of $g$. Then, we assume that each gap function belongs to a one-dimensional (1D) IR of $G$, where   higher-dimensional IRs are regarded as 1D IRs of the subgroup of $G$:
\begin{align}
    &D(g) \hat{\Delta}_{\alpha}(\vv k) D^{\rm T}(g) = \eta^{\alpha}_g \hat{\Delta}_{\alpha}(g \vv k), \label{eq:gapfunctiong}
\end{align}
where the factor $\eta^{\alpha}_g \in U(1)$ is the character of 1D IRs that the gap function belongs to. For instance, for $G=C_2$ and $g=2_z \in C_2$, $\eta^{\alpha}_{2_z} = +1 (-1)$ for the $A$ ($B$) IR of $C_2$. If each order parameter belongs to the same IR of $G$, the BdG Hamiltonian also has the $G$ point group symmetry:
\begin{align}
    &\tilde{D}(g) H(\vv k) \tilde{D}^{\dagger}(g) = H(g\vv k), \notag \\
    &\tilde{D}(g) \equiv  \left[ \begin{array}{cc} D(g) & 0 \\ 0 & \eta_g D^{\ast}(g) \end{array} \right].
    \label{eq:nambug} 
\end{align}
Likewise, we assume that the Raman vertex $\gamma(\bm{k})$ transforms according to a 1D IR of $G$,
\begin{align}
D(g)\gamma(\bm{k})D^\dagger(g) = \eta^R_g\gamma(g\bm{k}),
\label{eq:RamannR} 
\end{align}
with a phase factor $\eta^R_g\in U(1)$ characteristic of the Raman symmetry channel (e.g., $A_{1g}$, $B_{1g}$, \dots).
For example, for $G = D_{2h}$, $\gamma(\bm{k})=\partial_{k_x}\partial_{k_y}H_0(\bm{k})$ belongs to $B_{1g}$. Also, for $G = D_{4h}$, $\gamma(\bm{k})=(\partial^2_{k_x}-\partial^2_{k_y})H_0(\bm{k})$ belongs to $B_{1g}$. 

From Fig.~\ref{fig: diagram} (c) and Eq.~(\ref{eq: diagram}), the nonzero coupling between the Raman source field and the collective modes requires
\begin{align}
[Q_{R,\Delta^{\dagger}}(i\Omega_n)]_{\alpha}=\sum_{\bm{k}}\sum_m\text{tr}\Big(U_{21}\hat{\Delta}^{\dagger}_{\alpha}(\bm{k})G_0(\bm{k},i\omega_m+i\Omega_n)\Gamma(\bm{k})G_0(\bm{k},i\omega_m)\Big) \neq 0,
\label{eq: condition-coupling}
\end{align}
at $\bm{q}=0$. The above equation can be rewritten as 
\begin{align}
\nonumber
[Q_{R,\Delta^{\dagger}}(i\Omega_n)]_{\alpha}=&\sum_{\bm{k}}\sum_m\text{tr}\Big(U_{21}\hat{\Delta}^{\dagger}_{\alpha}(\bm{k})G_0(\bm{k},i\omega_m+i\Omega_n)\Gamma(\bm{k})G_0(\bm{k},i\omega_m)\Big) \\ \nonumber
=&\sum_{\bm{k}}\sum_m\text{tr}\Big(\hat{\Delta}^{\dagger}_{\alpha}(\bm{k})G^{(11)}_0(\bm{k},i\omega_m+i\Omega_n)\gamma(\bm{k})G^{(12)}_0(\bm{k},i\omega_m)\Big) \\ \nonumber
&\hspace{20mm}+\sum_{\bm{k}}\sum_m\text{tr}\Big(\hat{\Delta}^{\dagger}_{\alpha}(\bm{k})U_{21}G_0(\bm{k},i\omega_m+i\Omega_n)(-\gamma^{\text{T}}(-\bm{k}))U_{22}G_0(\bm{k},i\omega_m)\Big) \\ \nonumber
=&\sum_{\bm{k}}\sum_m\text{tr}\Big(\hat{\Delta}^{\dagger}_{\alpha}(\bm{k})G^{(11)}_0(\bm{k},i\omega_m+i\Omega_n)\gamma(\bm{k})G^{(12)}_0(\bm{k},i\omega_m)\Big) \\ \nonumber
&\hspace{20mm}+\sum_{\bm{k}}\sum_m\text{tr}\Big(G^{\text{T}}_0(\bm{k},i\omega_m)U_{22}\gamma(-\bm{k})G^{\text{T}}_0(\bm{k},i\omega_m+i\Omega_n)U_{12}\hat{\Delta}^{\dagger}_{\alpha}(-\bm{k})\Big) \\ \nonumber
=&\sum_{\bm{k}}\sum_m\text{tr}\Big(\hat{\Delta}^{\dagger}_{\alpha}(\bm{k})G^{(11)}_0(\bm{k},i\omega_m+i\Omega_n)\gamma(\bm{k})G^{(12)}_0(\bm{k},i\omega_m)\Big) \\ \nonumber
&\hspace{20mm}+\sum_{\bm{k}}\sum_m\text{tr}\Big(\tau_xG_0(-\bm{k},-i\omega_m)\tau_xU_{22}\gamma(-\bm{k})\tau_xG_0(-\bm{k},-i\omega_m-i\Omega_n)\tau_xU_{12}\hat{\Delta}^{\dagger}_{\alpha}(-\bm{k})\Big) \\ \nonumber
=&\sum_{\bm{k}}\sum_m\text{tr}\Big(\hat{\Delta}^{\dagger}_{\alpha}(\bm{k})G^{(11)}_0(\bm{k},i\omega_m+i\Omega_n)\gamma(\bm{k})G^{(12)}_0(\bm{k},i\omega_m)\Big) \\ \nonumber
&\hspace{20mm}+\sum_{\bm{k}}\sum_m\text{tr}\Big(U_{21}\hat{\Delta}^{\dagger}_{\alpha}(\bm{k})G_0(\bm{k},i\omega_m+i\Omega_n)U_{11}\gamma(\bm{k})G_0(\bm{k},i\omega_m)\Big) \\
= 2&\sum_{\bm{k}}\sum_m\text{tr}\Big(\hat{\Delta}^{\dagger}_{\alpha}(\bm{k})G^{(11)}_0(\bm{k},i\omega_m+i\Omega_n)\gamma(\bm{k})G^{(12)}_0(\bm{k},i\omega_m)\Big),
\end{align}
where we use $\hat{\Delta}^{\text{T}}_{\alpha}(\vv k) =-\hat{\Delta}_{\alpha}(-\vv k)$ and the PHS of the Green's function: $\tau_x G_0(\bm{k},i\omega_m) \tau_x =-G^{\text{T}}_0(-\bm{k},-i\omega_m)$. As a result, to examine the condition of Eq.~(\ref{eq: condition-coupling}), it is enough to consider 
\begin{align}
\sum_{\bm{k}}\sum_m\text{tr}\Big(\hat{\Delta}^{\dagger}_{\alpha}(\bm{k})G^{(11)}_0(\bm{k},i\omega_m+i\Omega_n)\gamma(\bm{k})G^{(12)}_0(\bm{k},i\omega_m)\Big).
\label{eq: condition-coupling2}
\end{align}
Here, we consider the Green's function:
\begin{align}
G_{0}(\bm{k},i\omega_m)  = \frac{1}{i\omega_m1_{2N\times 2N}-H_{\text{BdG}}(\bm{k})} = \left[
\begin{array}{cc}
G^{(11)}_0(\bm{k},i\omega_m) & G^{(12)}_0(\bm{k},i\omega_m) \\
G^{(21)}_0(\bm{k},i\omega_m) & G^{(22)}_0(\bm{k},i\omega_m)
\end{array}
\right].
\end{align}
Then, we find 
\begin{align}
G^{(11)}_0(\bm{k},i\omega_m) &= S^{-1}(\bm{k},i\omega_m), \\ \nonumber
G^{(12)}_0(\bm{k},i\omega_m) &= S^{-1}(\bm{k},i\omega_m)\Delta(\bm{k})(i\omega_m1_{N\times N}+H_0^{\text{T}}(-\bm{k}))^{-1} \\
&=\sum_{\beta}[G^{(12)}_0(\bm{k},i\omega_m)]_{\beta}, \quad [G^{(12)}_0(\bm{k},i\omega_m)]_{\beta} \equiv S^{-1}(\bm{k},i\omega_m)\Delta^{\text{MF}}_{\beta}\hat{\Delta}_{\beta}(\bm{k})(i\omega_m1_{N\times N}+H_0^{\text{T}}(-\bm{k}))^{-1},
\end{align}
where
\begin{align}
S(\bm{k},i\omega_m) \equiv (i\omega_m1_{N\times N}-H_0(\bm{k}))-\Delta(\bm{k})(i\omega_m1_{N\times N}+H_0^{\text{T}}(-\bm{k}))^{-1}\Delta^{\dagger}(\bm{k}).
\end{align}
In what follows, we focus on the Ginzburg--Landau (GL) regime close to the superconducting transition,
where the condition $\beta \Delta^{\text{MF}}_{\alpha} \ll 1$ is satisfied. Then, we can simplify the above equation:
\begin{align}
S(\bm{k},i\omega_m) \approx (i\omega_m1_{N\times N}-H_0(\bm{k})).
\end{align}
In this case, we obtain
\begin{align}
&D(g)S(\bm{k},i\omega_m)D^{\dagger}(g) = S(g\bm{k},i\omega_m) \\ \leftrightarrow \quad  
&D(g)G^{(11)}_0(\bm{k},i\omega_m)D^{\dagger}(g) = G^{(11)}_0(g\bm{k},i\omega_m), \\ 
&D(g)[G^{(12)}_0(\bm{k},i\omega_m)]_{\beta}D^{\text{T}}(g) = \eta^{\beta}_g[G^{(12)}_0(g\bm{k},i\omega_m)]_{\beta}.
\end{align}
For later convenience, we define each $\beta$ contribution as
\begin{align}
I_{\alpha\beta}\equiv
\sum_{\bm{k}}\sum_m\text{tr}\Big(
\hat{\Delta}^{\dagger}_{\alpha}(\bm{k})\,
G^{(11)}_0(\bm{k},i\omega_m+i\Omega_n)\,
\gamma(\bm{k})\,
[G^{(12)}_0(\bm{k},i\omega_m)]_{\beta}
\Big),
\end{align}
so that Eq.~(\ref{eq: condition-coupling}) is given by $\sum_\beta I_{\alpha\beta} \neq 0$. Then, we obtain
\begin{align}
\nonumber
I_{\alpha \beta}
=&\frac{1}{|G|}\left[|G|\sum_{\bm{k}}\sum_m\text{tr}\Big(\hat{\Delta}^{\dagger}_{\alpha}(\bm{k})G^{(11)}_0(\bm{k},i\omega_m+i\Omega_n)\gamma(\bm{k})[G^{(12)}_0(\bm{k},i\omega_m)]_{\beta}\Big)\right]
\\ \nonumber
=&\frac{1}{|G|}\left[|G|\sum_{\bm{k}}\sum_m\text{tr}\Big(D^{\text{T}}(g)D^{*}(g)\hat{\Delta}^{\dagger}_{\alpha}(\bm{k})D^{\dagger}(g)D(g)G^{(11)}_0(\bm{k},i\omega_m+i\Omega_n)D^{\dagger}(g)D(g)\gamma(\bm{k})D^{\dagger}(g)D(g)[G^{(12)}_0(\bm{k},i\omega_m)]_{\beta}\Big)\right]
\\ \nonumber
=&\frac{1}{|G|}\left[\sum_{g\in G}(\eta^{\alpha}_g)^{*}\eta^R_g\eta^{\beta}_g\sum_{\bm{k}}\sum_m\text{tr}\Big(\hat{\Delta}^{\dagger}_{\alpha}(g\bm{k})G^{(11)}_0(g\bm{k},i\omega_m+i\Omega_n)\gamma(g\bm{k})[G^{(12)}_0(g\bm{k},i\omega_m)]_{\beta}\Big)\right]
\\
=&\frac{1}{|G|}\left[\sum_{g\in G}(\eta^{\alpha}_g)^{*}\eta^R_g\eta^{\beta}_g\sum_{\bm{k}}\sum_m\text{tr}\Big(\hat{\Delta}^{\dagger}_{\alpha}(\bm{k})G^{(11)}_0(\bm{k},i\omega_m+i\Omega_n)\gamma(\bm{k})[G^{(12)}_0(\bm{k},i\omega_m)]_{\beta}\Big)\right].
\label{eq: transformation-raman-collective}
\end{align}
Comparing the first and last terms in Eq.~(\ref{eq: transformation-raman-collective}), we find 
\begin{align}
I_{\alpha\beta}=\left[\frac{1}{|G|}\sum_{g\in G}(\eta^{\alpha}_g)^{*}\eta^R_g\eta^{\beta}_g\right]I_{\alpha\beta}.
\end{align}
Therefore, the condition Eq.~(\ref{eq: condition-coupling}) for a nonzero coupling between the Raman source field and the collective modes at $\bm q=0$ is equivalent to
\begin{align}
\frac{1}{|G|}\sum_{g\in G}(\eta^{\alpha}_g)^{*}\eta^R_g\eta^{\beta}_g =1.
\label{eq: eq: condition-coupling3}
\end{align}
In particular, for 1D IRs, Eq.~(\ref{eq: eq: condition-coupling3}) holds for $(\eta^{\alpha}_g)^{*}\eta^R_g\eta^{\beta}_g =1$ for all $g \in G$.
Equivalently, the product representation of $\hat{\Delta}^{\dagger}_{\alpha},\gamma$, and $\hat{\Delta}_{\beta}$, 
\begin{align}
\Gamma_{\hat{\Delta}^{\dagger}_\alpha} \otimes \Gamma_{\gamma} \otimes \Gamma_{\hat{\Delta}_\beta},
\end{align}
must be the totally symmetric representation of the point group $G$.

\section{Case of the $B_{3u}$ pairing}\label{app: B3u}

In this section, we consider collective modes and the Raman response of the UTe$_2$ superconductor for the $B_{3u}$ pairing, where the excitation becomes gapless. 
The $B_{3u}$ pairing is given by
\begin{align}
\Delta(\vv k) &= \sum_{\alpha=1,2,3}\Delta^{\text{MF}}_{\alpha}\hat{\Delta}_{\alpha}(\bm{k}),  
\end{align}
with basis functions:
\begin{align}
\hat\Delta_1(\bm k) = \bigl[s_y\sigma_y\bigr]i s_y, \quad
\hat\Delta_2(\bm k) = \bigl[\sin(k_y)s_{z}\bigr]i s_y, \quad
\hat\Delta_3(\bm k) = \bigl[\sin(k_z)s_{y}\bigr]i s_y.
\end{align}
where we set
\begin{align}
\Delta^{\text{MF}}_1=\Delta^{\text{MF}}_2 = 0.04,\ \Delta^{\text{MF}}_3 = 0.01,
\end{align}
and the normal part is the same as in the main text.

\begin{figure*}[t]
\centering
\includegraphics[scale=0.085]{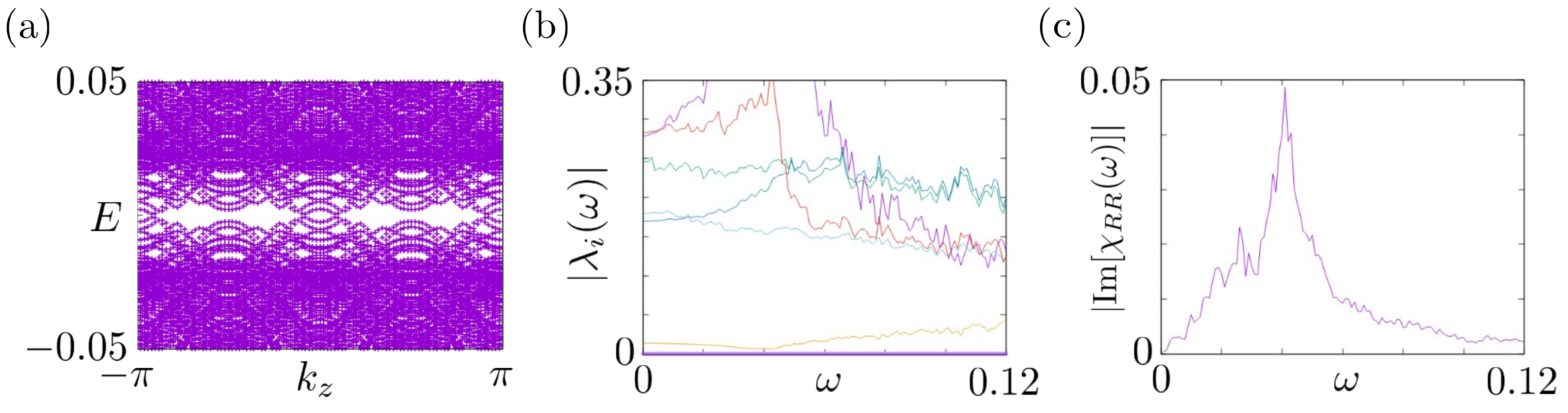}
\caption{(a) Energy spectrum of the BdG Hamiltonian in Eq.~(\ref{eq-main: BdG-Hamiltonian-multi}) plotted as a function of $k_z$, (b) fequency dependence of the eigenvalues of the inverse fluctuation matrix $U^{-1}_{\mathrm{eff},\phi}(\bm{q}=\bm{0},\omega)$ in Eq.~(\ref{eq: general-Anderson-Higgs-main}), (c) Raman susceptibility $\chi_{RR}(\bm{0},\omega)$ of Eq.~(\ref{eq: raman-numerical}) for $\bm{e}^{\text{in}}_x$ and $\bm{e}^{\text{out}}_x$ at $\bm{q}=0$.}
\label{fig: Results-B3u}
\end{figure*}

Fig.~\ref{fig: Results-B3u} (a) shows the energy spectrum in Eq.~(\ref{eq-main: BdG-Hamiltonian-multi}) for the $B_{3u}$ pairing. In contrast to the fully gapped case discussed in the main text, the present $B_{3u}$ state exhibits nodal structures, leading to low-energy quasiparticle excitations down to $\omega=0$. Fig.~\ref{fig: Results-B3u} (b) shows the eigenvalues of inverse fluctuation kernel calculated from Eq.~(\ref{eq: fluctuation-kernel-equation}). Since there is no finite gap that separates collective modes from quasiparticle excitations, collective modes are strongly damped by the quasiparticle continuum from $\omega = 0$.

As a result, the Raman response in Fig.~\ref{fig: Results-B3u} (c) is dominated by a broad continuum. Since the collective modes are embedded in the low-energy quasiparticle excitations, the Raman spectrum does not exhibit the sharp peak structures observed in the fully gapped case; instead it shows only a broadened response without a distinct collective-mode resonance. This comparison emphasizes that a quasiparticle continuum with a finite gap is a key ingredient for observing sharp in-gap Raman signatures arising from collective modes.

\twocolumngrid
\bibliography{main}

\begin{thebibliography}{84}%
\makeatletter
\providecommand \@ifxundefined [1]{%
 \@ifx{#1\undefined}
}%
\providecommand \@ifnum [1]{%
 \ifnum #1\expandafter \@firstoftwo
 \else \expandafter \@secondoftwo
 \fi
}%
\providecommand \@ifx [1]{%
 \ifx #1\expandafter \@firstoftwo
 \else \expandafter \@secondoftwo
 \fi
}%
\providecommand \natexlab [1]{#1}%
\providecommand \enquote  [1]{``#1''}%
\providecommand \bibnamefont  [1]{#1}%
\providecommand \bibfnamefont [1]{#1}%
\providecommand \citenamefont [1]{#1}%
\providecommand \href@noop [0]{\@secondoftwo}%
\providecommand \href [0]{\begingroup \@sanitize@url \@href}%
\providecommand \@href[1]{\@@startlink{#1}\@@href}%
\providecommand \@@href[1]{\endgroup#1\@@endlink}%
\providecommand \@sanitize@url [0]{\catcode `\\12\catcode `\$12\catcode `\&12\catcode `\#12\catcode `\^12\catcode `\_12\catcode `\%12\relax}%
\providecommand \@@startlink[1]{}%
\providecommand \@@endlink[0]{}%
\providecommand \url  [0]{\begingroup\@sanitize@url \@url }%
\providecommand \@url [1]{\endgroup\@href {#1}{\urlprefix }}%
\providecommand \urlprefix  [0]{URL }%
\providecommand \Eprint [0]{\href }%
\providecommand \doibase [0]{https://doi.org/}%
\providecommand \selectlanguage [0]{\@gobble}%
\providecommand \bibinfo  [0]{\@secondoftwo}%
\providecommand \bibfield  [0]{\@secondoftwo}%
\providecommand \translation [1]{[#1]}%
\providecommand \BibitemOpen [0]{}%
\providecommand \bibitemStop [0]{}%
\providecommand \bibitemNoStop [0]{.\EOS\space}%
\providecommand \EOS [0]{\spacefactor3000\relax}%
\providecommand \BibitemShut  [1]{\csname bibitem#1\endcsname}%
\let\auto@bib@innerbib\@empty
\bibitem [{\citenamefont {Anderson}(1958)}]{Anderson1900}%
  \BibitemOpen
  \bibfield  {author} {\bibinfo {author} {\bibfnamefont {P.~W.}\ \bibnamefont {Anderson}},\ }\bibfield  {title} {\bibinfo {title} {Random-{P}hase {A}pproximation in the {T}heory of {S}uperconductivity},\ }\href {https://doi.org/10.1103/PhysRev.112.1900} {\bibfield  {journal} {\bibinfo  {journal} {Phys. Rev.}\ }\textbf {\bibinfo {volume} {112}},\ \bibinfo {pages} {1900} (\bibinfo {year} {1958})}\BibitemShut {NoStop}%
\bibitem [{\citenamefont {Schmid}(1968)}]{Schmid129}%
  \BibitemOpen
  \bibfield  {author} {\bibinfo {author} {\bibfnamefont {A.}~\bibnamefont {Schmid}},\ }\bibfield  {title} {\bibinfo {title} {The approach to equilibrium in a pure superconductor the relaxation of the {C}ooper pair density},\ }\href {https://doi.org/10.1007/BF02422735} {\bibfield  {journal} {\bibinfo  {journal} {Phys. Kondens. Mater.}\ }\textbf {\bibinfo {volume} {8}},\ \bibinfo {pages} {129} (\bibinfo {year} {1968})}\BibitemShut {NoStop}%
\bibitem [{\citenamefont {Littlewood}\ and\ \citenamefont {Varma}(1981)}]{Littlewood811}%
  \BibitemOpen
  \bibfield  {author} {\bibinfo {author} {\bibfnamefont {P.~B.}\ \bibnamefont {Littlewood}}\ and\ \bibinfo {author} {\bibfnamefont {C.~M.}\ \bibnamefont {Varma}},\ }\bibfield  {title} {\bibinfo {title} {Gauge-{I}nvariant {T}heory of the {D}ynamical {I}nteraction of {C}harge {D}ensity {W}aves and {S}uperconductivity},\ }\href {https://doi.org/10.1103/PhysRevLett.47.811} {\bibfield  {journal} {\bibinfo  {journal} {Phys. Rev. Lett.}\ }\textbf {\bibinfo {volume} {47}},\ \bibinfo {pages} {811} (\bibinfo {year} {1981})}\BibitemShut {NoStop}%
\bibitem [{\citenamefont {Littlewood}\ and\ \citenamefont {Varma}(1982)}]{Littlewood4883}%
  \BibitemOpen
  \bibfield  {author} {\bibinfo {author} {\bibfnamefont {P.~B.}\ \bibnamefont {Littlewood}}\ and\ \bibinfo {author} {\bibfnamefont {C.~M.}\ \bibnamefont {Varma}},\ }\bibfield  {title} {\bibinfo {title} {Amplitude collective modes in superconductors and their coupling to charge-density waves},\ }\href {https://doi.org/10.1103/PhysRevB.26.4883} {\bibfield  {journal} {\bibinfo  {journal} {Phys. Rev. B}\ }\textbf {\bibinfo {volume} {26}},\ \bibinfo {pages} {4883} (\bibinfo {year} {1982})}\BibitemShut {NoStop}%
\bibitem [{\citenamefont {Pekker}\ and\ \citenamefont {Varma}(2015)}]{Pekker269}%
  \BibitemOpen
  \bibfield  {author} {\bibinfo {author} {\bibfnamefont {D.}~\bibnamefont {Pekker}}\ and\ \bibinfo {author} {\bibfnamefont {C.}~\bibnamefont {Varma}},\ }\bibfield  {title} {\bibinfo {title} {Amplitude/{H}iggs modes in condensed matter physics},\ }\href {https://doi.org/10.1146/annurev-conmatphys-031214-014350} {\bibfield  {journal} {\bibinfo  {journal} {Annu. Rev. Condens. Matter Phys.}\ }\textbf {\bibinfo {volume} {6}},\ \bibinfo {pages} {269} (\bibinfo {year} {2015})}\BibitemShut {NoStop}%
\bibitem [{\citenamefont {Shimano}\ and\ \citenamefont {Tsuji}(2020)}]{Shimano103}%
  \BibitemOpen
  \bibfield  {author} {\bibinfo {author} {\bibfnamefont {R.}~\bibnamefont {Shimano}}\ and\ \bibinfo {author} {\bibfnamefont {N.}~\bibnamefont {Tsuji}},\ }\bibfield  {title} {\bibinfo {title} {Higgs mode in superconductors},\ }\href {https://doi.org/10.1146/annurev-conmatphys-031119-050813} {\bibfield  {journal} {\bibinfo  {journal} {Annu. Rev. Condens. Matter Phys.}\ }\textbf {\bibinfo {volume} {11}},\ \bibinfo {pages} {103} (\bibinfo {year} {2020})}\BibitemShut {NoStop}%
\bibitem [{\citenamefont {N.~Tsuji}\ and\ \citenamefont {Tsuchiya}()}]{Tsuji}%
  \BibitemOpen
  \bibfield  {author} {\bibinfo {author} {\bibfnamefont {I.~D.}\ \bibnamefont {N.~Tsuji}}\ and\ \bibinfo {author} {\bibfnamefont {S.}~\bibnamefont {Tsuchiya}},\ }\bibfield  {title} {\bibinfo {title} {Higgs and {N}ambu--{G}oldstone modes in condensed matter physics},\ }\href@noop {} {\bibinfo  {journal} {in {\it Encyclopedia of Condensed Matter Physics}, 2nd ed. (Academic Press, Oxford, 2024)}\ }\BibitemShut {NoStop}%
\bibitem [{\citenamefont {Nambu}\ and\ \citenamefont {Jona-Lasinio}(1961)}]{Nambu345}%
  \BibitemOpen
\bibfield  {journal} {  }\bibfield  {author} {\bibinfo {author} {\bibfnamefont {Y.}~\bibnamefont {Nambu}}\ and\ \bibinfo {author} {\bibfnamefont {G.}~\bibnamefont {Jona-Lasinio}},\ }\bibfield  {title} {\bibinfo {title} {Dynamical {M}odel of {E}lementary {P}articles {B}ased on an {A}nalogy with {S}uperconductivity. i},\ }\href {https://doi.org/10.1103/PhysRev.122.345} {\bibfield  {journal} {\bibinfo  {journal} {Phys. Rev.}\ }\textbf {\bibinfo {volume} {122}},\ \bibinfo {pages} {345} (\bibinfo {year} {1961})}\BibitemShut {NoStop}%
\bibitem [{\citenamefont {Goldstone}(1961)}]{Goldstone154}%
  \BibitemOpen
  \bibfield  {author} {\bibinfo {author} {\bibfnamefont {J.}~\bibnamefont {Goldstone}},\ }\bibfield  {title} {\bibinfo {title} {Field {T}heories with “superconductor” solutions},\ }\href {https://doi.org/10.1007/BF02812722} {\bibfield  {journal} {\bibinfo  {journal} {Nuovo Cim.}\ }\textbf {\bibinfo {volume} {19}},\ \bibinfo {pages} {154} (\bibinfo {year} {1961})}\BibitemShut {NoStop}%
\bibitem [{\citenamefont {Anderson}(1963)}]{Anderson439}%
  \BibitemOpen
  \bibfield  {author} {\bibinfo {author} {\bibfnamefont {P.~W.}\ \bibnamefont {Anderson}},\ }\bibfield  {title} {\bibinfo {title} {Plasmons, {G}auge {I}nvariance, and {M}ass},\ }\href {https://doi.org/10.1103/PhysRev.130.439} {\bibfield  {journal} {\bibinfo  {journal} {Phys. Rev.}\ }\textbf {\bibinfo {volume} {130}},\ \bibinfo {pages} {439} (\bibinfo {year} {1963})}\BibitemShut {NoStop}%
\bibitem [{\citenamefont {Higgs}(1964)}]{Higgs508}%
  \BibitemOpen
  \bibfield  {author} {\bibinfo {author} {\bibfnamefont {P.~W.}\ \bibnamefont {Higgs}},\ }\bibfield  {title} {\bibinfo {title} {Broken {S}ymmetries and the {M}asses of {G}auge {B}osons},\ }\href {https://doi.org/10.1103/PhysRevLett.13.508} {\bibfield  {journal} {\bibinfo  {journal} {Phys. Rev. Lett.}\ }\textbf {\bibinfo {volume} {13}},\ \bibinfo {pages} {508} (\bibinfo {year} {1964})}\BibitemShut {NoStop}%
\bibitem [{\citenamefont {Leggett}(1966)}]{Leggett901}%
  \BibitemOpen
  \bibfield  {author} {\bibinfo {author} {\bibfnamefont {A.~J.}\ \bibnamefont {Leggett}},\ }\bibfield  {title} {\bibinfo {title} {Number-{P}hase {F}luctuations in {T}wo-{B}and {S}uperconductors},\ }\href {https://doi.org/10.1143/PTP.36.901} {\bibfield  {journal} {\bibinfo  {journal} {Prog. Theor. Phys.}\ }\textbf {\bibinfo {volume} {36}},\ \bibinfo {pages} {901} (\bibinfo {year} {1966})}\BibitemShut {NoStop}%
\bibitem [{\citenamefont {Bardasis}\ and\ \citenamefont {Schrieffer}(1961)}]{Bardasis1050}%
  \BibitemOpen
  \bibfield  {author} {\bibinfo {author} {\bibfnamefont {A.}~\bibnamefont {Bardasis}}\ and\ \bibinfo {author} {\bibfnamefont {J.~R.}\ \bibnamefont {Schrieffer}},\ }\bibfield  {title} {\bibinfo {title} {Excitons and {P}lasmons in {S}uperconductors},\ }\href {https://doi.org/10.1103/PhysRev.121.1050} {\bibfield  {journal} {\bibinfo  {journal} {Phys. Rev.}\ }\textbf {\bibinfo {volume} {121}},\ \bibinfo {pages} {1050} (\bibinfo {year} {1961})}\BibitemShut {NoStop}%
\bibitem [{\citenamefont {W\"olfle}(1976)}]{Wolfle1279}%
  \BibitemOpen
  \bibfield  {author} {\bibinfo {author} {\bibfnamefont {P.}~\bibnamefont {W\"olfle}},\ }\bibfield  {title} {\bibinfo {title} {Order-{P}arameter {C}ollective {M}odes in $^{3}${H}e-{A}},\ }\href {https://doi.org/10.1103/PhysRevLett.37.1279} {\bibfield  {journal} {\bibinfo  {journal} {Phys. Rev. Lett.}\ }\textbf {\bibinfo {volume} {37}},\ \bibinfo {pages} {1279} (\bibinfo {year} {1976})}\BibitemShut {NoStop}%
\bibitem [{\citenamefont {Tewordt}(1999)}]{Tewordt1007}%
  \BibitemOpen
  \bibfield  {author} {\bibinfo {author} {\bibfnamefont {L.}~\bibnamefont {Tewordt}},\ }\bibfield  {title} {\bibinfo {title} {Collective {O}rder {P}arameter {M}odes and {S}pin {F}luctuations for {S}pin-{T}riplet {S}uperconducting {S}tate in {S}r$_2${R}u{O}$_{4}$},\ }\href {https://doi.org/10.1103/PhysRevLett.83.1007} {\bibfield  {journal} {\bibinfo  {journal} {Phys. Rev. Lett.}\ }\textbf {\bibinfo {volume} {83}},\ \bibinfo {pages} {1007} (\bibinfo {year} {1999})}\BibitemShut {NoStop}%
\bibitem [{\citenamefont {Balatsky}\ \emph {et~al.}(2000)\citenamefont {Balatsky}, \citenamefont {Kumar},\ and\ \citenamefont {Schrieffer}}]{Balatsky4445}%
  \BibitemOpen
  \bibfield  {author} {\bibinfo {author} {\bibfnamefont {A.~V.}\ \bibnamefont {Balatsky}}, \bibinfo {author} {\bibfnamefont {P.}~\bibnamefont {Kumar}},\ and\ \bibinfo {author} {\bibfnamefont {J.~R.}\ \bibnamefont {Schrieffer}},\ }\bibfield  {title} {\bibinfo {title} {Collective {M}ode in a {S}uperconductor with {M}ixed-{S}ymmetry {O}rder {P}arameter {C}omponents},\ }\href {https://doi.org/10.1103/PhysRevLett.84.4445} {\bibfield  {journal} {\bibinfo  {journal} {Phys. Rev. Lett.}\ }\textbf {\bibinfo {volume} {84}},\ \bibinfo {pages} {4445} (\bibinfo {year} {2000})}\BibitemShut {NoStop}%
\bibitem [{\citenamefont {Poniatowski}\ \emph {et~al.}(2022)\citenamefont {Poniatowski}, \citenamefont {Curtis}, \citenamefont {Yacoby},\ and\ \citenamefont {Narang}}]{Nicholas44}%
  \BibitemOpen
  \bibfield  {author} {\bibinfo {author} {\bibfnamefont {N.~R.}\ \bibnamefont {Poniatowski}}, \bibinfo {author} {\bibfnamefont {J.~B.}\ \bibnamefont {Curtis}}, \bibinfo {author} {\bibfnamefont {A.}~\bibnamefont {Yacoby}},\ and\ \bibinfo {author} {\bibfnamefont {P.}~\bibnamefont {Narang}},\ }\bibfield  {title} {\bibinfo {title} {Spectroscopic signatures of time-reversal symmetry breaking superconductivity},\ }\href {https://doi.org/10.1038/s42005-022-00819-0} {\bibfield  {journal} {\bibinfo  {journal} {Commun. Phys.}\ }\textbf {\bibinfo {volume} {5}},\ \bibinfo {pages} {44} (\bibinfo {year} {2022})}\BibitemShut {NoStop}%
\bibitem [{\citenamefont {Blumberg}\ \emph {et~al.}(2007)\citenamefont {Blumberg}, \citenamefont {Mialitsin}, \citenamefont {Dennis}, \citenamefont {Klein}, \citenamefont {Zhigadlo},\ and\ \citenamefont {Karpinski}}]{Blumberg227002}%
  \BibitemOpen
  \bibfield  {author} {\bibinfo {author} {\bibfnamefont {G.}~\bibnamefont {Blumberg}}, \bibinfo {author} {\bibfnamefont {A.}~\bibnamefont {Mialitsin}}, \bibinfo {author} {\bibfnamefont {B.~S.}\ \bibnamefont {Dennis}}, \bibinfo {author} {\bibfnamefont {M.~V.}\ \bibnamefont {Klein}}, \bibinfo {author} {\bibfnamefont {N.~D.}\ \bibnamefont {Zhigadlo}},\ and\ \bibinfo {author} {\bibfnamefont {J.}~\bibnamefont {Karpinski}},\ }\bibfield  {title} {\bibinfo {title} {Observation of {L}eggett's {C}ollective {M}ode in a {M}ultiband {M}g{B}$_{2}$ {S}uperconductor},\ }\href {https://doi.org/10.1103/PhysRevLett.99.227002} {\bibfield  {journal} {\bibinfo  {journal} {Phys. Rev. Lett.}\ }\textbf {\bibinfo {volume} {99}},\ \bibinfo {pages} {227002} (\bibinfo {year} {2007})}\BibitemShut {NoStop}%
\bibitem [{\citenamefont {Kretzschmar}\ \emph {et~al.}(2013)\citenamefont {Kretzschmar}, \citenamefont {Muschler}, \citenamefont {B\"ohm}, \citenamefont {Baum}, \citenamefont {Hackl}, \citenamefont {Wen}, \citenamefont {Tsurkan}, \citenamefont {Deisenhofer},\ and\ \citenamefont {Loidl}}]{Kretzschmar187002}%
  \BibitemOpen
  \bibfield  {author} {\bibinfo {author} {\bibfnamefont {F.}~\bibnamefont {Kretzschmar}}, \bibinfo {author} {\bibfnamefont {B.}~\bibnamefont {Muschler}}, \bibinfo {author} {\bibfnamefont {T.}~\bibnamefont {B\"ohm}}, \bibinfo {author} {\bibfnamefont {A.}~\bibnamefont {Baum}}, \bibinfo {author} {\bibfnamefont {R.}~\bibnamefont {Hackl}}, \bibinfo {author} {\bibfnamefont {H.-H.}\ \bibnamefont {Wen}}, \bibinfo {author} {\bibfnamefont {V.}~\bibnamefont {Tsurkan}}, \bibinfo {author} {\bibfnamefont {J.}~\bibnamefont {Deisenhofer}},\ and\ \bibinfo {author} {\bibfnamefont {A.}~\bibnamefont {Loidl}},\ }\bibfield  {title} {\bibinfo {title} {Raman-{S}cattering {D}etection of {N}early {D}egenerate $s$-{W}ave and $d$-{W}ave {P}airing {C}hannels in {I}ron-{B}ased {B}a$_{0.6}${K}$_{0.4}${F}e$_{2}${A}s$_{2}$ and {R}b$_{0.8}${F}e$_{1.6}${S}e$_{2}$ {S}uperconductors},\ }\href {https://doi.org/10.1103/PhysRevLett.110.187002} {\bibfield  {journal} {\bibinfo  {journal} {Phys. Rev. Lett.}\ }\textbf {\bibinfo {volume} {110}},\
  \bibinfo {pages} {187002} (\bibinfo {year} {2013})}\BibitemShut {NoStop}%
\bibitem [{\citenamefont {B\"ohm}\ \emph {et~al.}(2014)\citenamefont {B\"ohm}, \citenamefont {Kemper}, \citenamefont {Moritz}, \citenamefont {Kretzschmar}, \citenamefont {Muschler}, \citenamefont {Eiter}, \citenamefont {Hackl}, \citenamefont {Devereaux}, \citenamefont {Scalapino},\ and\ \citenamefont {Wen}}]{Bohm041046}%
  \BibitemOpen
  \bibfield  {author} {\bibinfo {author} {\bibfnamefont {T.}~\bibnamefont {B\"ohm}}, \bibinfo {author} {\bibfnamefont {A.~F.}\ \bibnamefont {Kemper}}, \bibinfo {author} {\bibfnamefont {B.}~\bibnamefont {Moritz}}, \bibinfo {author} {\bibfnamefont {F.}~\bibnamefont {Kretzschmar}}, \bibinfo {author} {\bibfnamefont {B.}~\bibnamefont {Muschler}}, \bibinfo {author} {\bibfnamefont {H.-M.}\ \bibnamefont {Eiter}}, \bibinfo {author} {\bibfnamefont {R.}~\bibnamefont {Hackl}}, \bibinfo {author} {\bibfnamefont {T.~P.}\ \bibnamefont {Devereaux}}, \bibinfo {author} {\bibfnamefont {D.~J.}\ \bibnamefont {Scalapino}},\ and\ \bibinfo {author} {\bibfnamefont {H.-H.}\ \bibnamefont {Wen}},\ }\bibfield  {title} {\bibinfo {title} {Balancing {A}ct: {E}vidence for a {S}trong {S}ubdominant $d$-{W}ave {P}airing {C}hannel in {B}a$_{0.6}${K}$_{0.4}${F}e$_{2}${A}s$_{2}$},\ }\href {https://doi.org/10.1103/PhysRevX.4.041046} {\bibfield  {journal} {\bibinfo  {journal} {Phys. Rev. X}\ }\textbf {\bibinfo {volume} {4}},\ \bibinfo {pages}
  {041046} (\bibinfo {year} {2014})}\BibitemShut {NoStop}%
\bibitem [{\citenamefont {Jost}\ \emph {et~al.}(2018)\citenamefont {Jost}, \citenamefont {Scholz}, \citenamefont {Zweck}, \citenamefont {Meier}, \citenamefont {B\"ohmer}, \citenamefont {Canfield}, \citenamefont {Lazarevi\ifmmode~\acute{c}\else \'{c}\fi{}},\ and\ \citenamefont {Hackl}}]{Jost020504}%
  \BibitemOpen
  \bibfield  {author} {\bibinfo {author} {\bibfnamefont {D.}~\bibnamefont {Jost}}, \bibinfo {author} {\bibfnamefont {J.-R.}\ \bibnamefont {Scholz}}, \bibinfo {author} {\bibfnamefont {U.}~\bibnamefont {Zweck}}, \bibinfo {author} {\bibfnamefont {W.~R.}\ \bibnamefont {Meier}}, \bibinfo {author} {\bibfnamefont {A.~E.}\ \bibnamefont {B\"ohmer}}, \bibinfo {author} {\bibfnamefont {P.~C.}\ \bibnamefont {Canfield}}, \bibinfo {author} {\bibfnamefont {N.}~\bibnamefont {Lazarevi\ifmmode~\acute{c}\else \'{c}\fi{}}},\ and\ \bibinfo {author} {\bibfnamefont {R.}~\bibnamefont {Hackl}},\ }\bibfield  {title} {\bibinfo {title} {Indication of subdominant $d$-wave interaction in superconducting {C}a{K}{F}e$_{4}${A}s$_{4}$},\ }\href {https://doi.org/10.1103/PhysRevB.98.020504} {\bibfield  {journal} {\bibinfo  {journal} {Phys. Rev. B}\ }\textbf {\bibinfo {volume} {98}},\ \bibinfo {pages} {020504} (\bibinfo {year} {2018})}\BibitemShut {NoStop}%
\bibitem [{\citenamefont {He}\ \emph {et~al.}(2020)\citenamefont {He}, \citenamefont {Li}, \citenamefont {Jost}, \citenamefont {Baum}, \citenamefont {Shen}, \citenamefont {Dong}, \citenamefont {Zhao},\ and\ \citenamefont {Hackl}}]{He217002}%
  \BibitemOpen
  \bibfield  {author} {\bibinfo {author} {\bibfnamefont {G.}~\bibnamefont {He}}, \bibinfo {author} {\bibfnamefont {D.}~\bibnamefont {Li}}, \bibinfo {author} {\bibfnamefont {D.}~\bibnamefont {Jost}}, \bibinfo {author} {\bibfnamefont {A.}~\bibnamefont {Baum}}, \bibinfo {author} {\bibfnamefont {P.~P.}\ \bibnamefont {Shen}}, \bibinfo {author} {\bibfnamefont {X.~L.}\ \bibnamefont {Dong}}, \bibinfo {author} {\bibfnamefont {Z.~X.}\ \bibnamefont {Zhao}},\ and\ \bibinfo {author} {\bibfnamefont {R.}~\bibnamefont {Hackl}},\ }\bibfield  {title} {\bibinfo {title} {Raman {S}tudy of {C}ooper {P}airing {I}nstabilities in {L}i$_{1-x}${F}e$_{x}${O}{H}{F}e{S}e},\ }\href {https://doi.org/10.1103/PhysRevLett.125.217002} {\bibfield  {journal} {\bibinfo  {journal} {Phys. Rev. Lett.}\ }\textbf {\bibinfo {volume} {125}},\ \bibinfo {pages} {217002} (\bibinfo {year} {2020})}\BibitemShut {NoStop}%
\bibitem [{\citenamefont {Matsumoto}\ \emph {et~al.}()\citenamefont {Matsumoto}, \citenamefont {Neri}, \citenamefont {Kobayashi}, \citenamefont {Maeda}, \citenamefont {Manske},\ and\ \citenamefont {Shimano}}]{Matsumoto2025}%
  \BibitemOpen
  \bibfield  {author} {\bibinfo {author} {\bibfnamefont {H.}~\bibnamefont {Matsumoto}}, \bibinfo {author} {\bibfnamefont {S.}~\bibnamefont {Neri}}, \bibinfo {author} {\bibfnamefont {T.}~\bibnamefont {Kobayashi}}, \bibinfo {author} {\bibfnamefont {A.}~\bibnamefont {Maeda}}, \bibinfo {author} {\bibfnamefont {D.}~\bibnamefont {Manske}},\ and\ \bibinfo {author} {\bibfnamefont {R.}~\bibnamefont {Shimano}},\ }\bibfield  {title} {\bibinfo {title} {{A new collective mode in an iron-based superconductor with electronic nematicity}},\ }\href {https://arxiv.org/abs/2507.14466} {\bibinfo  {journal} {arXiv:2507.14466}\ }\BibitemShut {NoStop}%
\bibitem [{\citenamefont {Varma}(2002)}]{Varma901}%
  \BibitemOpen
\bibfield  {journal} {  }\bibfield  {author} {\bibinfo {author} {\bibfnamefont {C.~M.}\ \bibnamefont {Varma}},\ }\bibfield  {title} {\bibinfo {title} {Higgs {B}oson in {S}uperconductors},\ }\href {https://doi.org/10.1023/A:1013890507658} {\bibfield  {journal} {\bibinfo  {journal} {J. Low Temp. Phys.}\ }\textbf {\bibinfo {volume} {126}},\ \bibinfo {pages} {901} (\bibinfo {year} {2002})}\BibitemShut {NoStop}%
\bibitem [{\citenamefont {Tsuji}\ and\ \citenamefont {Aoki}(2015)}]{Tsuji064508}%
  \BibitemOpen
  \bibfield  {author} {\bibinfo {author} {\bibfnamefont {N.}~\bibnamefont {Tsuji}}\ and\ \bibinfo {author} {\bibfnamefont {H.}~\bibnamefont {Aoki}},\ }\bibfield  {title} {\bibinfo {title} {Theory of {A}nderson pseudospin resonance with {H}iggs mode in superconductors},\ }\href {https://doi.org/10.1103/PhysRevB.92.064508} {\bibfield  {journal} {\bibinfo  {journal} {Phys. Rev. B}\ }\textbf {\bibinfo {volume} {92}},\ \bibinfo {pages} {064508} (\bibinfo {year} {2015})}\BibitemShut {NoStop}%
\bibitem [{\citenamefont {Kemper}\ \emph {et~al.}(2015)\citenamefont {Kemper}, \citenamefont {Sentef}, \citenamefont {Moritz}, \citenamefont {Freericks},\ and\ \citenamefont {Devereaux}}]{Kemper224517}%
  \BibitemOpen
  \bibfield  {author} {\bibinfo {author} {\bibfnamefont {A.~F.}\ \bibnamefont {Kemper}}, \bibinfo {author} {\bibfnamefont {M.~A.}\ \bibnamefont {Sentef}}, \bibinfo {author} {\bibfnamefont {B.}~\bibnamefont {Moritz}}, \bibinfo {author} {\bibfnamefont {J.~K.}\ \bibnamefont {Freericks}},\ and\ \bibinfo {author} {\bibfnamefont {T.~P.}\ \bibnamefont {Devereaux}},\ }\bibfield  {title} {\bibinfo {title} {Direct observation of {H}iggs mode oscillations in the pump-probe photoemission spectra of electron-phonon mediated superconductors},\ }\href {https://doi.org/10.1103/PhysRevB.92.224517} {\bibfield  {journal} {\bibinfo  {journal} {Phys. Rev. B}\ }\textbf {\bibinfo {volume} {92}},\ \bibinfo {pages} {224517} (\bibinfo {year} {2015})}\BibitemShut {NoStop}%
\bibitem [{\citenamefont {Cea}\ \emph {et~al.}(2016)\citenamefont {Cea}, \citenamefont {Castellani},\ and\ \citenamefont {Benfatto}}]{Cea180507}%
  \BibitemOpen
  \bibfield  {author} {\bibinfo {author} {\bibfnamefont {T.}~\bibnamefont {Cea}}, \bibinfo {author} {\bibfnamefont {C.}~\bibnamefont {Castellani}},\ and\ \bibinfo {author} {\bibfnamefont {L.}~\bibnamefont {Benfatto}},\ }\bibfield  {title} {\bibinfo {title} {Nonlinear optical effects and third-harmonic generation in superconductors: {C}ooper pairs versus {H}iggs mode contribution},\ }\href {https://doi.org/10.1103/PhysRevB.93.180507} {\bibfield  {journal} {\bibinfo  {journal} {Phys. Rev. B}\ }\textbf {\bibinfo {volume} {93}},\ \bibinfo {pages} {180507} (\bibinfo {year} {2016})}\BibitemShut {NoStop}%
\bibitem [{\citenamefont {Tsuji}\ \emph {et~al.}(2016)\citenamefont {Tsuji}, \citenamefont {Murakami},\ and\ \citenamefont {Aoki}}]{Tsuji224519}%
  \BibitemOpen
  \bibfield  {author} {\bibinfo {author} {\bibfnamefont {N.}~\bibnamefont {Tsuji}}, \bibinfo {author} {\bibfnamefont {Y.}~\bibnamefont {Murakami}},\ and\ \bibinfo {author} {\bibfnamefont {H.}~\bibnamefont {Aoki}},\ }\bibfield  {title} {\bibinfo {title} {Nonlinear light--{H}iggs coupling in superconductors beyond {B}{C}{S}: {E}ffects of the retarded phonon-mediated interaction},\ }\href {https://doi.org/10.1103/PhysRevB.94.224519} {\bibfield  {journal} {\bibinfo  {journal} {Phys. Rev. B}\ }\textbf {\bibinfo {volume} {94}},\ \bibinfo {pages} {224519} (\bibinfo {year} {2016})}\BibitemShut {NoStop}%
\bibitem [{\citenamefont {Jujo}(2018)}]{Jujo024704}%
  \BibitemOpen
  \bibfield  {author} {\bibinfo {author} {\bibfnamefont {T.}~\bibnamefont {Jujo}},\ }\bibfield  {title} {\bibinfo {title} {Quasiclassical {T}heory on {T}hird-{H}armonic {G}eneration in {C}onventional {S}uperconductors with {P}aramagnetic {I}mpurities},\ }\href {https://doi.org/10.7566/JPSJ.87.024704} {\bibfield  {journal} {\bibinfo  {journal} {J. Phys. Soc. Jpn.}\ }\textbf {\bibinfo {volume} {87}},\ \bibinfo {pages} {024704} (\bibinfo {year} {2018})}\BibitemShut {NoStop}%
\bibitem [{\citenamefont {Silaev}(2019)}]{Silaevl224511}%
  \BibitemOpen
  \bibfield  {author} {\bibinfo {author} {\bibfnamefont {M.}~\bibnamefont {Silaev}},\ }\bibfield  {title} {\bibinfo {title} {Nonlinear electromagnetic response and {H}iggs-mode excitation in {B}{C}{S} superconductors with impurities},\ }\href {https://doi.org/10.1103/PhysRevB.99.224511} {\bibfield  {journal} {\bibinfo  {journal} {Phys. Rev. B}\ }\textbf {\bibinfo {volume} {99}},\ \bibinfo {pages} {224511} (\bibinfo {year} {2019})}\BibitemShut {NoStop}%
\bibitem [{\citenamefont {Schwarz}\ \emph {et~al.}(2020)\citenamefont {Schwarz}, \citenamefont {Fauseweh}, \citenamefont {Tsuji}, \citenamefont {Cheng}, \citenamefont {Bittner}, \citenamefont {Krull}, \citenamefont {Berciu}, \citenamefont {Uhrig}, \citenamefont {Schnyder}, \citenamefont {Kaiser},\ and\ \citenamefont {Manske}}]{Schwarz287}%
  \BibitemOpen
  \bibfield  {author} {\bibinfo {author} {\bibfnamefont {L.}~\bibnamefont {Schwarz}}, \bibinfo {author} {\bibfnamefont {B.}~\bibnamefont {Fauseweh}}, \bibinfo {author} {\bibfnamefont {N.}~\bibnamefont {Tsuji}}, \bibinfo {author} {\bibfnamefont {N.}~\bibnamefont {Cheng}}, \bibinfo {author} {\bibfnamefont {N.}~\bibnamefont {Bittner}}, \bibinfo {author} {\bibfnamefont {H.}~\bibnamefont {Krull}}, \bibinfo {author} {\bibfnamefont {M.}~\bibnamefont {Berciu}}, \bibinfo {author} {\bibfnamefont {G.~S.}\ \bibnamefont {Uhrig}}, \bibinfo {author} {\bibfnamefont {A.~P.}\ \bibnamefont {Schnyder}}, \bibinfo {author} {\bibfnamefont {S.}~\bibnamefont {Kaiser}},\ and\ \bibinfo {author} {\bibfnamefont {D.}~\bibnamefont {Manske}},\ }\bibfield  {title} {\bibinfo {title} {Classification and characterization of nonequilibrium {H}iggs modes in unconventional superconductors},\ }\href {https://doi.org/10.1038/s41467-019-13763-5} {\bibfield  {journal} {\bibinfo  {journal} {Nat. Commun.}\ }\textbf {\bibinfo {volume} {11}},\ \bibinfo
  {pages} {287} (\bibinfo {year} {2020})}\BibitemShut {NoStop}%
\bibitem [{\citenamefont {Tsuji}\ and\ \citenamefont {Nomura}(2020)}]{Tsuji043029}%
  \BibitemOpen
  \bibfield  {author} {\bibinfo {author} {\bibfnamefont {N.}~\bibnamefont {Tsuji}}\ and\ \bibinfo {author} {\bibfnamefont {Y.}~\bibnamefont {Nomura}},\ }\bibfield  {title} {\bibinfo {title} {Higgs-mode resonance in third harmonic generation in {N}b{N} superconductors: {M}ultiband electron-phonon coupling, impurity scattering, and polarization-angle dependence},\ }\href {https://doi.org/10.1103/PhysRevResearch.2.043029} {\bibfield  {journal} {\bibinfo  {journal} {Phys. Rev. Res.}\ }\textbf {\bibinfo {volume} {2}},\ \bibinfo {pages} {043029} (\bibinfo {year} {2020})}\BibitemShut {NoStop}%
\bibitem [{\citenamefont {Seibold}\ \emph {et~al.}(2021)\citenamefont {Seibold}, \citenamefont {Udina}, \citenamefont {Castellani},\ and\ \citenamefont {Benfatto}}]{Seibold014512}%
  \BibitemOpen
  \bibfield  {author} {\bibinfo {author} {\bibfnamefont {G.}~\bibnamefont {Seibold}}, \bibinfo {author} {\bibfnamefont {M.}~\bibnamefont {Udina}}, \bibinfo {author} {\bibfnamefont {C.}~\bibnamefont {Castellani}},\ and\ \bibinfo {author} {\bibfnamefont {L.}~\bibnamefont {Benfatto}},\ }\bibfield  {title} {\bibinfo {title} {Third harmonic generation from collective modes in disordered superconductors},\ }\href {https://doi.org/10.1103/PhysRevB.103.014512} {\bibfield  {journal} {\bibinfo  {journal} {Phys. Rev. B}\ }\textbf {\bibinfo {volume} {103}},\ \bibinfo {pages} {014512} (\bibinfo {year} {2021})}\BibitemShut {NoStop}%
\bibitem [{\citenamefont {Haenel}\ \emph {et~al.}(2021)\citenamefont {Haenel}, \citenamefont {Froese}, \citenamefont {Manske},\ and\ \citenamefont {Schwarz}}]{Haenel134504}%
  \BibitemOpen
  \bibfield  {author} {\bibinfo {author} {\bibfnamefont {R.}~\bibnamefont {Haenel}}, \bibinfo {author} {\bibfnamefont {P.}~\bibnamefont {Froese}}, \bibinfo {author} {\bibfnamefont {D.}~\bibnamefont {Manske}},\ and\ \bibinfo {author} {\bibfnamefont {L.}~\bibnamefont {Schwarz}},\ }\bibfield  {title} {\bibinfo {title} {Time-resolved optical conductivity and {H}iggs oscillations in two-band dirty superconductors},\ }\href {https://doi.org/10.1103/PhysRevB.104.134504} {\bibfield  {journal} {\bibinfo  {journal} {Phys. Rev. B}\ }\textbf {\bibinfo {volume} {104}},\ \bibinfo {pages} {134504} (\bibinfo {year} {2021})}\BibitemShut {NoStop}%
\bibitem [{\citenamefont {Udina}\ \emph {et~al.}(2022)\citenamefont {Udina}, \citenamefont {Fiore}, \citenamefont {Cea}, \citenamefont {Castellani}, \citenamefont {Seibold},\ and\ \citenamefont {Benfatto}}]{Udina168}%
  \BibitemOpen
  \bibfield  {author} {\bibinfo {author} {\bibfnamefont {M.}~\bibnamefont {Udina}}, \bibinfo {author} {\bibfnamefont {J.}~\bibnamefont {Fiore}}, \bibinfo {author} {\bibfnamefont {T.}~\bibnamefont {Cea}}, \bibinfo {author} {\bibfnamefont {C.}~\bibnamefont {Castellani}}, \bibinfo {author} {\bibfnamefont {G.}~\bibnamefont {Seibold}},\ and\ \bibinfo {author} {\bibfnamefont {L.}~\bibnamefont {Benfatto}},\ }\bibfield  {title} {\bibinfo {title} {T{H}z non-linear optical response in cuprates: {P}redominance of the {B}{C}{S} response over the {H}iggs mode},\ }\href {https://doi.org/10.1039/D2FD00016D} {\bibfield  {journal} {\bibinfo  {journal} {Faraday Discuss.}\ }\textbf {\bibinfo {volume} {237}},\ \bibinfo {pages} {168} (\bibinfo {year} {2022})}\BibitemShut {NoStop}%
\bibitem [{\citenamefont {Oh}\ \emph {et~al.}()\citenamefont {Oh}, \citenamefont {Watanabe},\ and\ \citenamefont {Tsuji}}]{Oh2025}%
  \BibitemOpen
  \bibfield  {author} {\bibinfo {author} {\bibfnamefont {C.-g.}\ \bibnamefont {Oh}}, \bibinfo {author} {\bibfnamefont {H.}~\bibnamefont {Watanabe}},\ and\ \bibinfo {author} {\bibfnamefont {N.}~\bibnamefont {Tsuji}},\ }\bibfield  {title} {\bibinfo {title} {{Role of {Q}uantum {G}eometry in the {C}ompetition between {H}iggs {M}ode and {Q}uasiparticles in {T}hird-{H}armonic {G}eneration of {S}uperconductors}},\ }\href {https://arxiv.org/abs/2512.01200} {\bibinfo  {journal} {arXiv:2512.01200}\ }\BibitemShut {NoStop}%
\bibitem [{\citenamefont {Matsunaga}\ \emph {et~al.}(2013)\citenamefont {Matsunaga}, \citenamefont {Hamada}, \citenamefont {Makise}, \citenamefont {Uzawa}, \citenamefont {Terai}, \citenamefont {Wang},\ and\ \citenamefont {Shimano}}]{Matsunaga057002}%
  \BibitemOpen
\bibfield  {journal} {  }\bibfield  {author} {\bibinfo {author} {\bibfnamefont {R.}~\bibnamefont {Matsunaga}}, \bibinfo {author} {\bibfnamefont {Y.~I.}\ \bibnamefont {Hamada}}, \bibinfo {author} {\bibfnamefont {K.}~\bibnamefont {Makise}}, \bibinfo {author} {\bibfnamefont {Y.}~\bibnamefont {Uzawa}}, \bibinfo {author} {\bibfnamefont {H.}~\bibnamefont {Terai}}, \bibinfo {author} {\bibfnamefont {Z.}~\bibnamefont {Wang}},\ and\ \bibinfo {author} {\bibfnamefont {R.}~\bibnamefont {Shimano}},\ }\bibfield  {title} {\bibinfo {title} {Higgs {A}mplitude {M}ode in the {B}{C}{S} {S}uperconductors {N}b$_{1-x}${T}i$_{x}${N} {I}nduced by {T}erahertz {P}ulse {E}xcitation},\ }\href {https://doi.org/10.1103/PhysRevLett.111.057002} {\bibfield  {journal} {\bibinfo  {journal} {Phys. Rev. Lett.}\ }\textbf {\bibinfo {volume} {111}},\ \bibinfo {pages} {057002} (\bibinfo {year} {2013})}\BibitemShut {NoStop}%
\bibitem [{\citenamefont {Matsunaga}\ \emph {et~al.}(2014)\citenamefont {Matsunaga}, \citenamefont {Tsuji}, \citenamefont {Fujita}, \citenamefont {Sugioka}, \citenamefont {Makise}, \citenamefont {Uzawa}, \citenamefont {Terai}, \citenamefont {Wang}, \citenamefont {Aoki},\ and\ \citenamefont {Shimano}}]{Matsunaga1254697}%
  \BibitemOpen
  \bibfield  {author} {\bibinfo {author} {\bibfnamefont {R.}~\bibnamefont {Matsunaga}}, \bibinfo {author} {\bibfnamefont {N.}~\bibnamefont {Tsuji}}, \bibinfo {author} {\bibfnamefont {H.}~\bibnamefont {Fujita}}, \bibinfo {author} {\bibfnamefont {A.}~\bibnamefont {Sugioka}}, \bibinfo {author} {\bibfnamefont {K.}~\bibnamefont {Makise}}, \bibinfo {author} {\bibfnamefont {Y.}~\bibnamefont {Uzawa}}, \bibinfo {author} {\bibfnamefont {H.}~\bibnamefont {Terai}}, \bibinfo {author} {\bibfnamefont {Z.}~\bibnamefont {Wang}}, \bibinfo {author} {\bibfnamefont {H.}~\bibnamefont {Aoki}},\ and\ \bibinfo {author} {\bibfnamefont {R.}~\bibnamefont {Shimano}},\ }\bibfield  {title} {\bibinfo {title} {Light-induced collective pseudospin precession resonating with {H}iggs mode in a superconductor},\ }\href {https://doi.org/10.1126/science.1254697} {\bibfield  {journal} {\bibinfo  {journal} {Science}\ }\textbf {\bibinfo {volume} {345}},\ \bibinfo {pages} {1145} (\bibinfo {year} {2014})}\BibitemShut {NoStop}%
\bibitem [{\citenamefont {Matsunaga}\ \emph {et~al.}(2017)\citenamefont {Matsunaga}, \citenamefont {Tsuji}, \citenamefont {Makise}, \citenamefont {Terai}, \citenamefont {Aoki},\ and\ \citenamefont {Shimano}}]{Matsunaga020505}%
  \BibitemOpen
  \bibfield  {author} {\bibinfo {author} {\bibfnamefont {R.}~\bibnamefont {Matsunaga}}, \bibinfo {author} {\bibfnamefont {N.}~\bibnamefont {Tsuji}}, \bibinfo {author} {\bibfnamefont {K.}~\bibnamefont {Makise}}, \bibinfo {author} {\bibfnamefont {H.}~\bibnamefont {Terai}}, \bibinfo {author} {\bibfnamefont {H.}~\bibnamefont {Aoki}},\ and\ \bibinfo {author} {\bibfnamefont {R.}~\bibnamefont {Shimano}},\ }\bibfield  {title} {\bibinfo {title} {Polarization-resolved terahertz third-harmonic generation in a single-crystal superconductor {N}b{N}: {D}ominance of the {H}iggs mode beyond the {B}{C}{S} approximation},\ }\href {https://doi.org/10.1103/PhysRevB.96.020505} {\bibfield  {journal} {\bibinfo  {journal} {Phys. Rev. B}\ }\textbf {\bibinfo {volume} {96}},\ \bibinfo {pages} {020505} (\bibinfo {year} {2017})}\BibitemShut {NoStop}%
\bibitem [{\citenamefont {Katsumi}\ \emph {et~al.}(2018)\citenamefont {Katsumi}, \citenamefont {Tsuji}, \citenamefont {Hamada}, \citenamefont {Matsunaga}, \citenamefont {Schneeloch}, \citenamefont {Zhong}, \citenamefont {Gu}, \citenamefont {Aoki}, \citenamefont {Gallais},\ and\ \citenamefont {Shimano}}]{Katsumi117001}%
  \BibitemOpen
  \bibfield  {author} {\bibinfo {author} {\bibfnamefont {K.}~\bibnamefont {Katsumi}}, \bibinfo {author} {\bibfnamefont {N.}~\bibnamefont {Tsuji}}, \bibinfo {author} {\bibfnamefont {Y.~I.}\ \bibnamefont {Hamada}}, \bibinfo {author} {\bibfnamefont {R.}~\bibnamefont {Matsunaga}}, \bibinfo {author} {\bibfnamefont {J.}~\bibnamefont {Schneeloch}}, \bibinfo {author} {\bibfnamefont {R.~D.}\ \bibnamefont {Zhong}}, \bibinfo {author} {\bibfnamefont {G.~D.}\ \bibnamefont {Gu}}, \bibinfo {author} {\bibfnamefont {H.}~\bibnamefont {Aoki}}, \bibinfo {author} {\bibfnamefont {Y.}~\bibnamefont {Gallais}},\ and\ \bibinfo {author} {\bibfnamefont {R.}~\bibnamefont {Shimano}},\ }\bibfield  {title} {\bibinfo {title} {Higgs {M}ode in the $d$-wave {S}uperconductor {B}i$_2${S}r$_2${C}a{C}u$_2${O}$_{8+x}$ {D}riven by an {I}ntense {T}erahertz {P}ulse},\ }\href {https://doi.org/10.1103/PhysRevLett.120.117001} {\bibfield  {journal} {\bibinfo  {journal} {Phys. Rev. Lett.}\ }\textbf {\bibinfo {volume} {120}},\ \bibinfo {pages} {117001}
  (\bibinfo {year} {2018})}\BibitemShut {NoStop}%
\bibitem [{\citenamefont {Chu}\ \emph {et~al.}(2020)\citenamefont {Chu}, \citenamefont {Kim}, \citenamefont {Katsumi} \emph {et~al.}}]{Chu1793}%
  \BibitemOpen
  \bibfield  {author} {\bibinfo {author} {\bibfnamefont {H.}~\bibnamefont {Chu}}, \bibinfo {author} {\bibfnamefont {M.}~\bibnamefont {Kim}}, \bibinfo {author} {\bibfnamefont {K.}~\bibnamefont {Katsumi}}, \emph {et~al.},\ }\bibfield  {title} {\bibinfo {title} {Phase-resolved {H}iggs response in superconducting cuprates},\ }\href {https://doi.org/10.1038/s41467-020-15613-1} {\bibfield  {journal} {\bibinfo  {journal} {Nat. Commun.}\ }\textbf {\bibinfo {volume} {11}},\ \bibinfo {pages} {1793} (\bibinfo {year} {2020})}\BibitemShut {NoStop}%
\bibitem [{\citenamefont {Kamatani}\ \emph {et~al.}(2022)\citenamefont {Kamatani}, \citenamefont {Kitamura}, \citenamefont {Tsuji}, \citenamefont {Shimano},\ and\ \citenamefont {Morimoto}}]{Kamatani094520}%
  \BibitemOpen
  \bibfield  {author} {\bibinfo {author} {\bibfnamefont {T.}~\bibnamefont {Kamatani}}, \bibinfo {author} {\bibfnamefont {S.}~\bibnamefont {Kitamura}}, \bibinfo {author} {\bibfnamefont {N.}~\bibnamefont {Tsuji}}, \bibinfo {author} {\bibfnamefont {R.}~\bibnamefont {Shimano}},\ and\ \bibinfo {author} {\bibfnamefont {T.}~\bibnamefont {Morimoto}},\ }\bibfield  {title} {\bibinfo {title} {Optical response of the {L}eggett mode in multiband superconductors in the linear response regime},\ }\href {https://doi.org/10.1103/PhysRevB.105.094520} {\bibfield  {journal} {\bibinfo  {journal} {Phys. Rev. B}\ }\textbf {\bibinfo {volume} {105}},\ \bibinfo {pages} {094520} (\bibinfo {year} {2022})}\BibitemShut {NoStop}%
\bibitem [{\citenamefont {Nagashima}\ \emph {et~al.}(2024)\citenamefont {Nagashima}, \citenamefont {Tian}, \citenamefont {Haenel}, \citenamefont {Tsuji},\ and\ \citenamefont {Manske}}]{Nagashima013120}%
  \BibitemOpen
  \bibfield  {author} {\bibinfo {author} {\bibfnamefont {R.}~\bibnamefont {Nagashima}}, \bibinfo {author} {\bibfnamefont {S.}~\bibnamefont {Tian}}, \bibinfo {author} {\bibfnamefont {R.}~\bibnamefont {Haenel}}, \bibinfo {author} {\bibfnamefont {N.}~\bibnamefont {Tsuji}},\ and\ \bibinfo {author} {\bibfnamefont {D.}~\bibnamefont {Manske}},\ }\bibfield  {title} {\bibinfo {title} {Classification of {L}ifshitz invariant in multiband superconductors: {A}n application to {L}eggett modes in the linear response regime in {K}agome lattice models},\ }\href {https://doi.org/10.1103/PhysRevResearch.6.013120} {\bibfield  {journal} {\bibinfo  {journal} {Phys. Rev. Res.}\ }\textbf {\bibinfo {volume} {6}},\ \bibinfo {pages} {013120} (\bibinfo {year} {2024})}\BibitemShut {NoStop}%
\bibitem [{\citenamefont {Lee}\ and\ \citenamefont {Chung}(2023)}]{Lee307}%
  \BibitemOpen
  \bibfield  {author} {\bibinfo {author} {\bibfnamefont {C.}~\bibnamefont {Lee}}\ and\ \bibinfo {author} {\bibfnamefont {S.~B.}\ \bibnamefont {Chung}},\ }\bibfield  {title} {\bibinfo {title} {Linear optical response from the odd-parity {B}ardasis-{S}chrieffer mode in locally non-centrosymmetric superconductors},\ }\href {https://doi.org/10.1038/s42005-023-01421-8} {\bibfield  {journal} {\bibinfo  {journal} {Commun. Phys.}\ }\textbf {\bibinfo {volume} {6}},\ \bibinfo {pages} {307} (\bibinfo {year} {2023})}\BibitemShut {NoStop}%
\bibitem [{\citenamefont {Matsushita}\ \emph {et~al.}()\citenamefont {Matsushita}, \citenamefont {Ieda}, \citenamefont {Araki}, \citenamefont {Morimoto}, \citenamefont {Vekhter},\ and\ \citenamefont {Yanase}}]{Matsushita2026}%
  \BibitemOpen
  \bibfield  {author} {\bibinfo {author} {\bibfnamefont {T.}~\bibnamefont {Matsushita}}, \bibinfo {author} {\bibfnamefont {J.}~\bibnamefont {Ieda}}, \bibinfo {author} {\bibfnamefont {Y.}~\bibnamefont {Araki}}, \bibinfo {author} {\bibfnamefont {T.}~\bibnamefont {Morimoto}}, \bibinfo {author} {\bibfnamefont {I.}~\bibnamefont {Vekhter}},\ and\ \bibinfo {author} {\bibfnamefont {Y.}~\bibnamefont {Yanase}},\ }\bibfield  {title} {\bibinfo {title} {{Microwave {K}err/{F}araday {R}esonance in {T}wo-dimensional {C}hiral {S}uperconductors}},\ }\href {https://arxiv.org/abs/2601.10151} {\bibinfo  {journal} {arXiv:2601.10151}\ }\BibitemShut {NoStop}%
\bibitem [{\citenamefont {Landau}\ and\ \citenamefont {Lifshitz}()}]{Lifshitz}%
  \BibitemOpen
\bibfield  {journal} {  }\bibfield  {author} {\bibinfo {author} {\bibfnamefont {L.~D.}\ \bibnamefont {Landau}}\ and\ \bibinfo {author} {\bibfnamefont {E.~M.}\ \bibnamefont {Lifshitz}},\ }\href@noop {} {\bibinfo  {journal} {Statistical Physics (Pergamon Press, Oxford, 1969)}\ }\BibitemShut {NoStop}%
\bibitem [{\citenamefont {Klein}\ and\ \citenamefont {Dierker}(1984)}]{Klein4976}%
  \BibitemOpen
\bibfield  {journal} {  }\bibfield  {author} {\bibinfo {author} {\bibfnamefont {M.~V.}\ \bibnamefont {Klein}}\ and\ \bibinfo {author} {\bibfnamefont {S.~B.}\ \bibnamefont {Dierker}},\ }\bibfield  {title} {\bibinfo {title} {Theory of {R}aman scattering in superconductors},\ }\href {https://doi.org/10.1103/PhysRevB.29.4976} {\bibfield  {journal} {\bibinfo  {journal} {Phys. Rev. B}\ }\textbf {\bibinfo {volume} {29}},\ \bibinfo {pages} {4976} (\bibinfo {year} {1984})}\BibitemShut {NoStop}%
\bibitem [{\citenamefont {Devereaux}\ and\ \citenamefont {Hackl}(2007)}]{Devereaux175}%
  \BibitemOpen
  \bibfield  {author} {\bibinfo {author} {\bibfnamefont {T.~P.}\ \bibnamefont {Devereaux}}\ and\ \bibinfo {author} {\bibfnamefont {R.}~\bibnamefont {Hackl}},\ }\bibfield  {title} {\bibinfo {title} {Inelastic light scattering from correlated electrons},\ }\href {https://doi.org/10.1103/RevModPhys.79.175} {\bibfield  {journal} {\bibinfo  {journal} {Rev. Mod. Phys.}\ }\textbf {\bibinfo {volume} {79}},\ \bibinfo {pages} {175} (\bibinfo {year} {2007})}\BibitemShut {NoStop}%
\bibitem [{\citenamefont {Sooryakumar}\ and\ \citenamefont {Klein}(1980)}]{Sooryakumar660}%
  \BibitemOpen
  \bibfield  {author} {\bibinfo {author} {\bibfnamefont {R.}~\bibnamefont {Sooryakumar}}\ and\ \bibinfo {author} {\bibfnamefont {M.~V.}\ \bibnamefont {Klein}},\ }\bibfield  {title} {\bibinfo {title} {Raman {S}cattering by {S}uperconducting-{G}ap {E}xcitations and {T}heir {C}oupling to {C}harge-{D}ensity {W}aves},\ }\href {https://doi.org/10.1103/PhysRevLett.45.660} {\bibfield  {journal} {\bibinfo  {journal} {Phys. Rev. Lett.}\ }\textbf {\bibinfo {volume} {45}},\ \bibinfo {pages} {660} (\bibinfo {year} {1980})}\BibitemShut {NoStop}%
\bibitem [{\citenamefont {M\'easson}\ \emph {et~al.}(2014)\citenamefont {M\'easson}, \citenamefont {Gallais}, \citenamefont {Cazayous}, \citenamefont {Clair}, \citenamefont {Rodi\`ere}, \citenamefont {Cario},\ and\ \citenamefont {Sacuto}}]{Measson060503}%
  \BibitemOpen
  \bibfield  {author} {\bibinfo {author} {\bibfnamefont {M.-A.}\ \bibnamefont {M\'easson}}, \bibinfo {author} {\bibfnamefont {Y.}~\bibnamefont {Gallais}}, \bibinfo {author} {\bibfnamefont {M.}~\bibnamefont {Cazayous}}, \bibinfo {author} {\bibfnamefont {B.}~\bibnamefont {Clair}}, \bibinfo {author} {\bibfnamefont {P.}~\bibnamefont {Rodi\`ere}}, \bibinfo {author} {\bibfnamefont {L.}~\bibnamefont {Cario}},\ and\ \bibinfo {author} {\bibfnamefont {A.}~\bibnamefont {Sacuto}},\ }\bibfield  {title} {\bibinfo {title} {Amplitude {H}iggs mode in the $2${H}-{N}b{S}e$_{2}$ superconductor},\ }\href {https://doi.org/10.1103/PhysRevB.89.060503} {\bibfield  {journal} {\bibinfo  {journal} {Phys. Rev. B}\ }\textbf {\bibinfo {volume} {89}},\ \bibinfo {pages} {060503} (\bibinfo {year} {2014})}\BibitemShut {NoStop}%
\bibitem [{\citenamefont {Grasset}\ \emph {et~al.}(2018)\citenamefont {Grasset}, \citenamefont {Cea}, \citenamefont {Gallais}, \citenamefont {Cazayous}, \citenamefont {Sacuto}, \citenamefont {Cario}, \citenamefont {Benfatto},\ and\ \citenamefont {M\'easson}}]{Grasset094502}%
  \BibitemOpen
  \bibfield  {author} {\bibinfo {author} {\bibfnamefont {R.}~\bibnamefont {Grasset}}, \bibinfo {author} {\bibfnamefont {T.}~\bibnamefont {Cea}}, \bibinfo {author} {\bibfnamefont {Y.}~\bibnamefont {Gallais}}, \bibinfo {author} {\bibfnamefont {M.}~\bibnamefont {Cazayous}}, \bibinfo {author} {\bibfnamefont {A.}~\bibnamefont {Sacuto}}, \bibinfo {author} {\bibfnamefont {L.}~\bibnamefont {Cario}}, \bibinfo {author} {\bibfnamefont {L.}~\bibnamefont {Benfatto}},\ and\ \bibinfo {author} {\bibfnamefont {M.-A.}\ \bibnamefont {M\'easson}},\ }\bibfield  {title} {\bibinfo {title} {Higgs-mode radiance and charge-density-wave order in $2${H}-{N}b{S}e$_{2}$},\ }\href {https://doi.org/10.1103/PhysRevB.97.094502} {\bibfield  {journal} {\bibinfo  {journal} {Phys. Rev. B}\ }\textbf {\bibinfo {volume} {97}},\ \bibinfo {pages} {094502} (\bibinfo {year} {2018})}\BibitemShut {NoStop}%
\bibitem [{\citenamefont {Grasset}\ \emph {et~al.}(2019)\citenamefont {Grasset}, \citenamefont {Gallais}, \citenamefont {Sacuto}, \citenamefont {Cazayous}, \citenamefont {Ma\~nas Valero}, \citenamefont {Coronado},\ and\ \citenamefont {M\'easson}}]{Grasset127001}%
  \BibitemOpen
  \bibfield  {author} {\bibinfo {author} {\bibfnamefont {R.}~\bibnamefont {Grasset}}, \bibinfo {author} {\bibfnamefont {Y.}~\bibnamefont {Gallais}}, \bibinfo {author} {\bibfnamefont {A.}~\bibnamefont {Sacuto}}, \bibinfo {author} {\bibfnamefont {M.}~\bibnamefont {Cazayous}}, \bibinfo {author} {\bibfnamefont {S.}~\bibnamefont {Ma\~nas Valero}}, \bibinfo {author} {\bibfnamefont {E.}~\bibnamefont {Coronado}},\ and\ \bibinfo {author} {\bibfnamefont {M.-A.}\ \bibnamefont {M\'easson}},\ }\bibfield  {title} {\bibinfo {title} {Pressure-{I}nduced {C}ollapse of the {C}harge {D}ensity {W}ave and {H}iggs {M}ode {V}isibility in $2${H}-{T}a{S}$_{2}$},\ }\href {https://doi.org/10.1103/PhysRevLett.122.127001} {\bibfield  {journal} {\bibinfo  {journal} {Phys. Rev. Lett.}\ }\textbf {\bibinfo {volume} {122}},\ \bibinfo {pages} {127001} (\bibinfo {year} {2019})}\BibitemShut {NoStop}%
\bibitem [{\citenamefont {Majumdar}\ \emph {et~al.}(2020)\citenamefont {Majumdar}, \citenamefont {VanGennep}, \citenamefont {Brisbois}, \citenamefont {Chareev}, \citenamefont {Sadakov}, \citenamefont {Usoltsev}, \citenamefont {Mito}, \citenamefont {Silhanek}, \citenamefont {Sarkar}, \citenamefont {Hassan}, \citenamefont {Karis}, \citenamefont {Ahuja},\ and\ \citenamefont {Abdel-Hafiez}}]{Majumdar084005}%
  \BibitemOpen
  \bibfield  {author} {\bibinfo {author} {\bibfnamefont {A.}~\bibnamefont {Majumdar}}, \bibinfo {author} {\bibfnamefont {D.}~\bibnamefont {VanGennep}}, \bibinfo {author} {\bibfnamefont {J.}~\bibnamefont {Brisbois}}, \bibinfo {author} {\bibfnamefont {D.}~\bibnamefont {Chareev}}, \bibinfo {author} {\bibfnamefont {A.~V.}\ \bibnamefont {Sadakov}}, \bibinfo {author} {\bibfnamefont {A.~S.}\ \bibnamefont {Usoltsev}}, \bibinfo {author} {\bibfnamefont {M.}~\bibnamefont {Mito}}, \bibinfo {author} {\bibfnamefont {A.~V.}\ \bibnamefont {Silhanek}}, \bibinfo {author} {\bibfnamefont {T.}~\bibnamefont {Sarkar}}, \bibinfo {author} {\bibfnamefont {A.}~\bibnamefont {Hassan}}, \bibinfo {author} {\bibfnamefont {O.}~\bibnamefont {Karis}}, \bibinfo {author} {\bibfnamefont {R.}~\bibnamefont {Ahuja}},\ and\ \bibinfo {author} {\bibfnamefont {M.}~\bibnamefont {Abdel-Hafiez}},\ }\bibfield  {title} {\bibinfo {title} {Interplay of charge density wave and multiband superconductivity in layered quasi-two-dimensional materials: The case of
  $2${H}-{N}b{S}$_{2}$ and $2${H}-{N}b{S}e$_{2}$},\ }\href {https://doi.org/10.1103/PhysRevMaterials.4.084005} {\bibfield  {journal} {\bibinfo  {journal} {Phys. Rev. Mater.}\ }\textbf {\bibinfo {volume} {4}},\ \bibinfo {pages} {084005} (\bibinfo {year} {2020})}\BibitemShut {NoStop}%
\bibitem [{\citenamefont {Scalapino}\ and\ \citenamefont {Devereaux}(2009)}]{Scalapino140512}%
  \BibitemOpen
  \bibfield  {author} {\bibinfo {author} {\bibfnamefont {D.~J.}\ \bibnamefont {Scalapino}}\ and\ \bibinfo {author} {\bibfnamefont {T.~P.}\ \bibnamefont {Devereaux}},\ }\bibfield  {title} {\bibinfo {title} {Collective $d$-wave exciton modes in the calculated {R}aman spectrum of {F}e-based superconductors},\ }\href {https://doi.org/10.1103/PhysRevB.80.140512} {\bibfield  {journal} {\bibinfo  {journal} {Phys. Rev. B}\ }\textbf {\bibinfo {volume} {80}},\ \bibinfo {pages} {140512} (\bibinfo {year} {2009})}\BibitemShut {NoStop}%
\bibitem [{\citenamefont {Lee}\ and\ \citenamefont {Choi}(2009)}]{Lee445701}%
  \BibitemOpen
  \bibfield  {author} {\bibinfo {author} {\bibfnamefont {H.~C.}\ \bibnamefont {Lee}}\ and\ \bibinfo {author} {\bibfnamefont {H.~Y.}\ \bibnamefont {Choi}},\ }\bibfield  {title} {\bibinfo {title} {Electronic {R}aman scattering of two-band superconductors: a time-dependent {L}andau–{G}inzburg theory approach},\ }\href {https://doi.org/10.1088/0953-8984/21/44/445701} {\bibfield  {journal} {\bibinfo  {journal} {J. Phys.: Condens. Matter}\ }\textbf {\bibinfo {volume} {21}},\ \bibinfo {pages} {445701} (\bibinfo {year} {2009})}\BibitemShut {NoStop}%
\bibitem [{\citenamefont {Klein}(2010)}]{Klein014507}%
  \BibitemOpen
  \bibfield  {author} {\bibinfo {author} {\bibfnamefont {M.~V.}\ \bibnamefont {Klein}},\ }\bibfield  {title} {\bibinfo {title} {Theory of {R}aman scattering from {L}eggett's collective mode in a multiband superconductor: {A}pplication to {M}g{B}$_{2}$},\ }\href {https://doi.org/10.1103/PhysRevB.82.014507} {\bibfield  {journal} {\bibinfo  {journal} {Phys. Rev. B}\ }\textbf {\bibinfo {volume} {82}},\ \bibinfo {pages} {014507} (\bibinfo {year} {2010})}\BibitemShut {NoStop}%
\bibitem [{\citenamefont {Khodas}\ \emph {et~al.}(2014)\citenamefont {Khodas}, \citenamefont {Chubukov},\ and\ \citenamefont {Blumberg}}]{Khodas245134}%
  \BibitemOpen
  \bibfield  {author} {\bibinfo {author} {\bibfnamefont {M.}~\bibnamefont {Khodas}}, \bibinfo {author} {\bibfnamefont {A.~V.}\ \bibnamefont {Chubukov}},\ and\ \bibinfo {author} {\bibfnamefont {G.}~\bibnamefont {Blumberg}},\ }\bibfield  {title} {\bibinfo {title} {Collective modes in multiband superconductors: {R}aman scattering in iron selenides},\ }\href {https://doi.org/10.1103/PhysRevB.89.245134} {\bibfield  {journal} {\bibinfo  {journal} {Phys. Rev. B}\ }\textbf {\bibinfo {volume} {89}},\ \bibinfo {pages} {245134} (\bibinfo {year} {2014})}\BibitemShut {NoStop}%
\bibitem [{\citenamefont {Maiti}\ \emph {et~al.}(2016)\citenamefont {Maiti}, \citenamefont {Maier}, \citenamefont {B\"ohm}, \citenamefont {Hackl},\ and\ \citenamefont {Hirschfeld}}]{Maiti257001}%
  \BibitemOpen
  \bibfield  {author} {\bibinfo {author} {\bibfnamefont {S.}~\bibnamefont {Maiti}}, \bibinfo {author} {\bibfnamefont {T.~A.}\ \bibnamefont {Maier}}, \bibinfo {author} {\bibfnamefont {T.}~\bibnamefont {B\"ohm}}, \bibinfo {author} {\bibfnamefont {R.}~\bibnamefont {Hackl}},\ and\ \bibinfo {author} {\bibfnamefont {P.~J.}\ \bibnamefont {Hirschfeld}},\ }\bibfield  {title} {\bibinfo {title} {Probing the {P}airing {I}nteraction and {M}ultiple {B}ardasis-{S}chrieffer {M}odes {U}sing {R}aman {S}pectroscopy},\ }\href {https://doi.org/10.1103/PhysRevLett.117.257001} {\bibfield  {journal} {\bibinfo  {journal} {Phys. Rev. Lett.}\ }\textbf {\bibinfo {volume} {117}},\ \bibinfo {pages} {257001} (\bibinfo {year} {2016})}\BibitemShut {NoStop}%
\bibitem [{\citenamefont {Cea}\ and\ \citenamefont {Benfatto}(2016)}]{Cea064512}%
  \BibitemOpen
  \bibfield  {author} {\bibinfo {author} {\bibfnamefont {T.}~\bibnamefont {Cea}}\ and\ \bibinfo {author} {\bibfnamefont {L.}~\bibnamefont {Benfatto}},\ }\bibfield  {title} {\bibinfo {title} {Signature of the {L}eggett mode in the ${A}_{1g}$ {R}aman response: {F}rom {M}g{B}$_{2}$ to iron-based superconductors},\ }\href {https://doi.org/10.1103/PhysRevB.94.064512} {\bibfield  {journal} {\bibinfo  {journal} {Phys. Rev. B}\ }\textbf {\bibinfo {volume} {94}},\ \bibinfo {pages} {064512} (\bibinfo {year} {2016})}\BibitemShut {NoStop}%
\bibitem [{\citenamefont {Maiti}\ \emph {et~al.}(2017)\citenamefont {Maiti}, \citenamefont {Chubukov},\ and\ \citenamefont {Hirschfeld}}]{Maiti014503}%
  \BibitemOpen
  \bibfield  {author} {\bibinfo {author} {\bibfnamefont {S.}~\bibnamefont {Maiti}}, \bibinfo {author} {\bibfnamefont {A.~V.}\ \bibnamefont {Chubukov}},\ and\ \bibinfo {author} {\bibfnamefont {P.~J.}\ \bibnamefont {Hirschfeld}},\ }\bibfield  {title} {\bibinfo {title} {Conservation laws, vertex corrections, and screening in {R}aman spectroscopy},\ }\href {https://doi.org/10.1103/PhysRevB.96.014503} {\bibfield  {journal} {\bibinfo  {journal} {Phys. Rev. B}\ }\textbf {\bibinfo {volume} {96}},\ \bibinfo {pages} {014503} (\bibinfo {year} {2017})}\BibitemShut {NoStop}%
\bibitem [{\citenamefont {Benek-Lins}\ and\ \citenamefont {Maiti}(2024)}]{Igor104505}%
  \BibitemOpen
  \bibfield  {author} {\bibinfo {author} {\bibfnamefont {I.}~\bibnamefont {Benek-Lins}}\ and\ \bibinfo {author} {\bibfnamefont {S.}~\bibnamefont {Maiti}},\ }\bibfield  {title} {\bibinfo {title} {Many-body physics-induced selection rules: {A}pplication to {R}aman spectroscopy},\ }\href {https://doi.org/10.1103/PhysRevB.109.104505} {\bibfield  {journal} {\bibinfo  {journal} {Phys. Rev. B}\ }\textbf {\bibinfo {volume} {109}},\ \bibinfo {pages} {104505} (\bibinfo {year} {2024})}\BibitemShut {NoStop}%
\bibitem [{\citenamefont {Sarkar}\ and\ \citenamefont {Maiti}(2024)}]{Sarkar094515}%
  \BibitemOpen
  \bibfield  {author} {\bibinfo {author} {\bibfnamefont {S.}~\bibnamefont {Sarkar}}\ and\ \bibinfo {author} {\bibfnamefont {S.}~\bibnamefont {Maiti}},\ }\bibfield  {title} {\bibinfo {title} {Electronic {R}aman response of a superconductor across a time reversal symmetry breaking phase transition},\ }\href {https://doi.org/10.1103/PhysRevB.109.094515} {\bibfield  {journal} {\bibinfo  {journal} {Phys. Rev. B}\ }\textbf {\bibinfo {volume} {109}},\ \bibinfo {pages} {094515} (\bibinfo {year} {2024})}\BibitemShut {NoStop}%
\bibitem [{\citenamefont {Takasan}\ and\ \citenamefont {Tsuji}()}]{Takasan2025}%
  \BibitemOpen
  \bibfield  {author} {\bibinfo {author} {\bibfnamefont {K.}~\bibnamefont {Takasan}}\ and\ \bibinfo {author} {\bibfnamefont {N.}~\bibnamefont {Tsuji}},\ }\bibfield  {title} {\bibinfo {title} {{Superconducting nonlinear {H}all effect induced by geometric phases}},\ }\href {https://arxiv.org/abs/2503.14589} {\bibinfo  {journal} {arXiv:2503.14589}\ }\BibitemShut {NoStop}%
\bibitem [{\citenamefont {Ran}\ \emph {et~al.}(2019)\citenamefont {Ran}, \citenamefont {Eckberg}, \citenamefont {Ding}, \citenamefont {Furukawa}, \citenamefont {Metz}, \citenamefont {Saha}, \citenamefont {Liu}, \citenamefont {Zic}, \citenamefont {Kim}, \citenamefont {Paglione},\ and\ \citenamefont {Butch}}]{Ran684}%
  \BibitemOpen
\bibfield  {journal} {  }\bibfield  {author} {\bibinfo {author} {\bibfnamefont {S.}~\bibnamefont {Ran}}, \bibinfo {author} {\bibfnamefont {C.}~\bibnamefont {Eckberg}}, \bibinfo {author} {\bibfnamefont {Q.-P.}\ \bibnamefont {Ding}}, \bibinfo {author} {\bibfnamefont {Y.}~\bibnamefont {Furukawa}}, \bibinfo {author} {\bibfnamefont {T.}~\bibnamefont {Metz}}, \bibinfo {author} {\bibfnamefont {S.~R.}\ \bibnamefont {Saha}}, \bibinfo {author} {\bibfnamefont {I.-L.}\ \bibnamefont {Liu}}, \bibinfo {author} {\bibfnamefont {M.}~\bibnamefont {Zic}}, \bibinfo {author} {\bibfnamefont {H.}~\bibnamefont {Kim}}, \bibinfo {author} {\bibfnamefont {J.}~\bibnamefont {Paglione}},\ and\ \bibinfo {author} {\bibfnamefont {N.~P.}\ \bibnamefont {Butch}},\ }\bibfield  {title} {\bibinfo {title} {{Nearly ferromagnetic spin-triplet superconductivity}},\ }\href {https://doi.org/10.1126/science.aav8645} {\bibfield  {journal} {\bibinfo  {journal} {Science}\ }\textbf {\bibinfo {volume} {365}},\ \bibinfo {pages} {684} (\bibinfo {year}
  {2019})}\BibitemShut {NoStop}%
\bibitem [{\citenamefont {Aoki}\ \emph {et~al.}(2022{\natexlab{a}})\citenamefont {Aoki}, \citenamefont {Brison}, \citenamefont {Flouquet}, \citenamefont {Ishida}, \citenamefont {Knebel}, \citenamefont {Tokunaga},\ and\ \citenamefont {Yanase}}]{aoki2022}%
  \BibitemOpen
  \bibfield  {author} {\bibinfo {author} {\bibfnamefont {D.}~\bibnamefont {Aoki}}, \bibinfo {author} {\bibfnamefont {J.-P.}\ \bibnamefont {Brison}}, \bibinfo {author} {\bibfnamefont {J.}~\bibnamefont {Flouquet}}, \bibinfo {author} {\bibfnamefont {K.}~\bibnamefont {Ishida}}, \bibinfo {author} {\bibfnamefont {G.}~\bibnamefont {Knebel}}, \bibinfo {author} {\bibfnamefont {Y.}~\bibnamefont {Tokunaga}},\ and\ \bibinfo {author} {\bibfnamefont {Y.}~\bibnamefont {Yanase}},\ }\bibfield  {title} {\bibinfo {title} {Unconventional superconductivity in {{U}{T}e$_2$}},\ }\href {https://doi.org/10.1088/1361-648x/ac5863} {\bibfield  {journal} {\bibinfo  {journal} {Journal of Physics: Condensed Matter}\ }\textbf {\bibinfo {volume} {34}},\ \bibinfo {pages} {243002} (\bibinfo {year} {2022}{\natexlab{a}})}\BibitemShut {NoStop}%
\bibitem [{\citenamefont {Matsumura}\ \emph {et~al.}(2023)\citenamefont {Matsumura}, \citenamefont {Fujibayashi}, \citenamefont {Kinjo}, \citenamefont {Kitagawa}, \citenamefont {Ishida}, \citenamefont {Tokunaga}, \citenamefont {Sakai}, \citenamefont {Kambe}, \citenamefont {Nakamura}, \citenamefont {Shimizu}, \citenamefont {Homma}, \citenamefont {Li}, \citenamefont {Honda},\ and\ \citenamefont {Aoki}}]{Matsumura063701}%
  \BibitemOpen
  \bibfield  {author} {\bibinfo {author} {\bibfnamefont {H.}~\bibnamefont {Matsumura}}, \bibinfo {author} {\bibfnamefont {H.}~\bibnamefont {Fujibayashi}}, \bibinfo {author} {\bibfnamefont {K.}~\bibnamefont {Kinjo}}, \bibinfo {author} {\bibfnamefont {S.}~\bibnamefont {Kitagawa}}, \bibinfo {author} {\bibfnamefont {K.}~\bibnamefont {Ishida}}, \bibinfo {author} {\bibfnamefont {Y.}~\bibnamefont {Tokunaga}}, \bibinfo {author} {\bibfnamefont {H.}~\bibnamefont {Sakai}}, \bibinfo {author} {\bibfnamefont {S.}~\bibnamefont {Kambe}}, \bibinfo {author} {\bibfnamefont {A.}~\bibnamefont {Nakamura}}, \bibinfo {author} {\bibfnamefont {Y.}~\bibnamefont {Shimizu}}, \bibinfo {author} {\bibfnamefont {Y.}~\bibnamefont {Homma}}, \bibinfo {author} {\bibfnamefont {D.}~\bibnamefont {Li}}, \bibinfo {author} {\bibfnamefont {F.}~\bibnamefont {Honda}},\ and\ \bibinfo {author} {\bibfnamefont {D.}~\bibnamefont {Aoki}},\ }\bibfield  {title} {\bibinfo {title} {Large {R}eduction in the a-axis {K}night {S}hift on {U}{T}e$_2$ with {T}$_c$ = 2.1
  {K}},\ }\href {https://doi.org/10.7566/JPSJ.92.063701} {\bibfield  {journal} {\bibinfo  {journal} {J. Phys. Soc. Jpn.}\ }\textbf {\bibinfo {volume} {92}},\ \bibinfo {pages} {063701} (\bibinfo {year} {2023})}\BibitemShut {NoStop}%
\bibitem [{\citenamefont {Theuss}\ \emph {et~al.}(2024)\citenamefont {Theuss}, \citenamefont {Shragai}, \citenamefont {Grissonnanche}, \citenamefont {Hayes}, \citenamefont {Saha}, \citenamefont {Eo}, \citenamefont {Suarez}, \citenamefont {Shishidou}, \citenamefont {Butch}, \citenamefont {Paglione},\ and\ \citenamefont {Ramshaw}}]{theuss2024single}%
  \BibitemOpen
  \bibfield  {author} {\bibinfo {author} {\bibfnamefont {F.}~\bibnamefont {Theuss}}, \bibinfo {author} {\bibfnamefont {A.}~\bibnamefont {Shragai}}, \bibinfo {author} {\bibfnamefont {G.}~\bibnamefont {Grissonnanche}}, \bibinfo {author} {\bibfnamefont {I.~M.}\ \bibnamefont {Hayes}}, \bibinfo {author} {\bibfnamefont {S.~R.}\ \bibnamefont {Saha}}, \bibinfo {author} {\bibfnamefont {Y.~S.}\ \bibnamefont {Eo}}, \bibinfo {author} {\bibfnamefont {A.}~\bibnamefont {Suarez}}, \bibinfo {author} {\bibfnamefont {T.}~\bibnamefont {Shishidou}}, \bibinfo {author} {\bibfnamefont {N.~P.}\ \bibnamefont {Butch}}, \bibinfo {author} {\bibfnamefont {J.}~\bibnamefont {Paglione}},\ and\ \bibinfo {author} {\bibfnamefont {B.~J.}\ \bibnamefont {Ramshaw}},\ }\bibfield  {title} {\bibinfo {title} {Single-component superconductivity in {UTe$_2$} at ambient pressure},\ }\href {https://doi.org/10.1038/s41567-024-02493-1} {\bibfield  {journal} {\bibinfo  {journal} {Nature Physics}\ }\textbf {\bibinfo {volume} {20}},\ \bibinfo {pages} {1124}
  (\bibinfo {year} {2024})}\BibitemShut {NoStop}%
\bibitem [{\citenamefont {Suetsugu}\ \emph {et~al.}(2024)\citenamefont {Suetsugu}, \citenamefont {Shimomura}, \citenamefont {Kamimura}, \citenamefont {Asaba}, \citenamefont {Asaeda}, \citenamefont {Kosuge}, \citenamefont {Sekino}, \citenamefont {Ikemori}, \citenamefont {Kasahara}, \citenamefont {Kohsaka}, \citenamefont {Lee}, \citenamefont {Yanase}, \citenamefont {Sakai}, \citenamefont {Opletal}, \citenamefont {Tokiwa}, \citenamefont {Haga},\ and\ \citenamefont {Matsuda}}]{Suetsugu3772}%
  \BibitemOpen
  \bibfield  {author} {\bibinfo {author} {\bibfnamefont {S.}~\bibnamefont {Suetsugu}}, \bibinfo {author} {\bibfnamefont {M.}~\bibnamefont {Shimomura}}, \bibinfo {author} {\bibfnamefont {M.}~\bibnamefont {Kamimura}}, \bibinfo {author} {\bibfnamefont {T.}~\bibnamefont {Asaba}}, \bibinfo {author} {\bibfnamefont {H.}~\bibnamefont {Asaeda}}, \bibinfo {author} {\bibfnamefont {Y.}~\bibnamefont {Kosuge}}, \bibinfo {author} {\bibfnamefont {Y.}~\bibnamefont {Sekino}}, \bibinfo {author} {\bibfnamefont {S.}~\bibnamefont {Ikemori}}, \bibinfo {author} {\bibfnamefont {Y.}~\bibnamefont {Kasahara}}, \bibinfo {author} {\bibfnamefont {Y.}~\bibnamefont {Kohsaka}}, \bibinfo {author} {\bibfnamefont {M.}~\bibnamefont {Lee}}, \bibinfo {author} {\bibfnamefont {Y.}~\bibnamefont {Yanase}}, \bibinfo {author} {\bibfnamefont {H.}~\bibnamefont {Sakai}}, \bibinfo {author} {\bibfnamefont {P.}~\bibnamefont {Opletal}}, \bibinfo {author} {\bibfnamefont {Y.}~\bibnamefont {Tokiwa}}, \bibinfo {author} {\bibfnamefont {Y.}~\bibnamefont {Haga}},\
  and\ \bibinfo {author} {\bibfnamefont {Y.}~\bibnamefont {Matsuda}},\ }\bibfield  {title} {\bibinfo {title} {Fully gapped pairing state in spin-triplet superconductor {U}{T}e$_2$},\ }\href {https://doi.org/10.1126/sciadv.adk3772} {\bibfield  {journal} {\bibinfo  {journal} {Sci. Adv.}\ }\textbf {\bibinfo {volume} {10}},\ \bibinfo {pages} {eadk3772} (\bibinfo {year} {2024})}\BibitemShut {NoStop}%
\bibitem [{\citenamefont {Li}\ \emph {et~al.}(2025)\citenamefont {Li}, \citenamefont {Moir}, \citenamefont {McKee}, \citenamefont {Lee-Wong}, \citenamefont {Baumbach}, \citenamefont {Maple},\ and\ \citenamefont {Liu}}]{Li2419734122}%
  \BibitemOpen
  \bibfield  {author} {\bibinfo {author} {\bibfnamefont {Z.}~\bibnamefont {Li}}, \bibinfo {author} {\bibfnamefont {C.~M.}\ \bibnamefont {Moir}}, \bibinfo {author} {\bibfnamefont {N.~J.}\ \bibnamefont {McKee}}, \bibinfo {author} {\bibfnamefont {E.}~\bibnamefont {Lee-Wong}}, \bibinfo {author} {\bibfnamefont {R.~E.}\ \bibnamefont {Baumbach}}, \bibinfo {author} {\bibfnamefont {M.~B.}\ \bibnamefont {Maple}},\ and\ \bibinfo {author} {\bibfnamefont {Y.}~\bibnamefont {Liu}},\ }\bibfield  {title} {\bibinfo {title} {Observation of odd-parity superconductivity in {U}{T}e$_2$},\ }\href {https://doi.org/10.1073/pnas.2419734122} {\bibfield  {journal} {\bibinfo  {journal} {Proceedings of the National Academy of Sciences}\ }\textbf {\bibinfo {volume} {122}},\ \bibinfo {pages} {e2419734122} (\bibinfo {year} {2025})}\BibitemShut {NoStop}%
\bibitem [{\citenamefont {Hayes}\ \emph {et~al.}(2025)\citenamefont {Hayes}, \citenamefont {Metz}, \citenamefont {Frank}, \citenamefont {Saha}, \citenamefont {Butch}, \citenamefont {Mishra}, \citenamefont {Hirschfeld},\ and\ \citenamefont {Paglione}}]{Hayes021029}%
  \BibitemOpen
  \bibfield  {author} {\bibinfo {author} {\bibfnamefont {I.~M.}\ \bibnamefont {Hayes}}, \bibinfo {author} {\bibfnamefont {T.~E.}\ \bibnamefont {Metz}}, \bibinfo {author} {\bibfnamefont {C.~E.}\ \bibnamefont {Frank}}, \bibinfo {author} {\bibfnamefont {S.~R.}\ \bibnamefont {Saha}}, \bibinfo {author} {\bibfnamefont {N.~P.}\ \bibnamefont {Butch}}, \bibinfo {author} {\bibfnamefont {V.}~\bibnamefont {Mishra}}, \bibinfo {author} {\bibfnamefont {P.~J.}\ \bibnamefont {Hirschfeld}},\ and\ \bibinfo {author} {\bibfnamefont {J.}~\bibnamefont {Paglione}},\ }\bibfield  {title} {\bibinfo {title} {Robust {N}odal {B}ehavior in the {T}hermal {C}onductivity of {S}uperconducting {U}{T}e$_2$},\ }\href {https://doi.org/10.1103/PhysRevX.15.021029} {\bibfield  {journal} {\bibinfo  {journal} {Phys. Rev. X}\ }\textbf {\bibinfo {volume} {15}},\ \bibinfo {pages} {021029} (\bibinfo {year} {2025})}\BibitemShut {NoStop}%
\bibitem [{\citenamefont {Gu}\ \emph {et~al.}(2025)\citenamefont {Gu}, \citenamefont {Wang}, \citenamefont {Carroll}, \citenamefont {Zhussupbekov}, \citenamefont {Broyles}, \citenamefont {Ran}, \citenamefont {Butch}, \citenamefont {Horn}, \citenamefont {Saha}, \citenamefont {Paglione}, \citenamefont {Liu}, \citenamefont {Davis},\ and\ \citenamefont {Lee}}]{QGu2025}%
  \BibitemOpen
  \bibfield  {author} {\bibinfo {author} {\bibfnamefont {Q.}~\bibnamefont {Gu}}, \bibinfo {author} {\bibfnamefont {S.}~\bibnamefont {Wang}}, \bibinfo {author} {\bibfnamefont {J.~P.}\ \bibnamefont {Carroll}}, \bibinfo {author} {\bibfnamefont {K.}~\bibnamefont {Zhussupbekov}}, \bibinfo {author} {\bibfnamefont {C.}~\bibnamefont {Broyles}}, \bibinfo {author} {\bibfnamefont {S.}~\bibnamefont {Ran}}, \bibinfo {author} {\bibfnamefont {N.~P.}\ \bibnamefont {Butch}}, \bibinfo {author} {\bibfnamefont {J.~A.}\ \bibnamefont {Horn}}, \bibinfo {author} {\bibfnamefont {S.}~\bibnamefont {Saha}}, \bibinfo {author} {\bibfnamefont {J.}~\bibnamefont {Paglione}}, \bibinfo {author} {\bibfnamefont {X.}~\bibnamefont {Liu}}, \bibinfo {author} {\bibfnamefont {J.~C.~S.}\ \bibnamefont {Davis}},\ and\ \bibinfo {author} {\bibfnamefont {D.-H.}\ \bibnamefont {Lee}},\ }\bibfield  {title} {\bibinfo {title} {Pair wave function symmetry in {UTe$_2$} from zero-energy surface state visualization},\ }\href {https://doi.org/10.1126/science.adk7219}
  {\bibfield  {journal} {\bibinfo  {journal} {Science}\ }\textbf {\bibinfo {volume} {388}},\ \bibinfo {pages} {938} (\bibinfo {year} {2025})}\BibitemShut {NoStop}%
\bibitem [{\citenamefont {Wang}\ \emph {et~al.}(2025)\citenamefont {Wang}, \citenamefont {Zhussupbekov},\ and\ \citenamefont {Carroll}}]{Wang1555}%
  \BibitemOpen
  \bibfield  {author} {\bibinfo {author} {\bibfnamefont {S.}~\bibnamefont {Wang}}, \bibinfo {author} {\bibfnamefont {K.}~\bibnamefont {Zhussupbekov}},\ and\ \bibinfo {author} {\bibfnamefont {J.~e.~a.}\ \bibnamefont {Carroll}},\ }\bibfield  {title} {\bibinfo {title} {Odd-parity quasiparticle interference in the superconductive surface state of {U}{T}e$_2$},\ }\href {https://doi.org/10.1038/s41567-025-03000-w} {\bibfield  {journal} {\bibinfo  {journal} {Nat. Phys.}\ }\textbf {\bibinfo {volume} {21}},\ \bibinfo {pages} {1555} (\bibinfo {year} {2025})}\BibitemShut {NoStop}%
\bibitem [{\citenamefont {Ishizuka}\ \emph {et~al.}(2019)\citenamefont {Ishizuka}, \citenamefont {Sumita}, \citenamefont {Daido},\ and\ \citenamefont {Yanase}}]{Ishizuka217001}%
  \BibitemOpen
  \bibfield  {author} {\bibinfo {author} {\bibfnamefont {J.}~\bibnamefont {Ishizuka}}, \bibinfo {author} {\bibfnamefont {S.}~\bibnamefont {Sumita}}, \bibinfo {author} {\bibfnamefont {A.}~\bibnamefont {Daido}},\ and\ \bibinfo {author} {\bibfnamefont {Y.}~\bibnamefont {Yanase}},\ }\bibfield  {title} {\bibinfo {title} {{I}nsulator-{M}etal {T}ransition and {T}opological {S}uperconductivity in {U}{T}e$_2$ from a {F}irst-{P}rinciples {C}alculation},\ }\href {https://doi.org/10.1103/PhysRevLett.123.217001} {\bibfield  {journal} {\bibinfo  {journal} {Phys. Rev. Lett.}\ }\textbf {\bibinfo {volume} {123}},\ \bibinfo {pages} {217001} (\bibinfo {year} {2019})}\BibitemShut {NoStop}%
\bibitem [{\citenamefont {Tei}\ \emph {et~al.}(2023)\citenamefont {Tei}, \citenamefont {Mizushima},\ and\ \citenamefont {Fujimoto}}]{Tei144517}%
  \BibitemOpen
  \bibfield  {author} {\bibinfo {author} {\bibfnamefont {J.}~\bibnamefont {Tei}}, \bibinfo {author} {\bibfnamefont {T.}~\bibnamefont {Mizushima}},\ and\ \bibinfo {author} {\bibfnamefont {S.}~\bibnamefont {Fujimoto}},\ }\bibfield  {title} {\bibinfo {title} {Possible realization of topological crystalline superconductivity with time-reversal symmetry in {U}{T}e$_2$},\ }\href {https://doi.org/10.1103/PhysRevB.107.144517} {\bibfield  {journal} {\bibinfo  {journal} {Phys. Rev. B}\ }\textbf {\bibinfo {volume} {107}},\ \bibinfo {pages} {144517} (\bibinfo {year} {2023})}\BibitemShut {NoStop}%
\bibitem [{\citenamefont {Yamazaki}\ and\ \citenamefont {Kobayashi}(2025)}]{Yamazaki174507}%
  \BibitemOpen
  \bibfield  {author} {\bibinfo {author} {\bibfnamefont {Y.}~\bibnamefont {Yamazaki}}\ and\ \bibinfo {author} {\bibfnamefont {S.}~\bibnamefont {Kobayashi}},\ }\bibfield  {title} {\bibinfo {title} {Higher-order topological phases for time-reversal-symmetry breaking superconductivity in {U}{T}e$_2$},\ }\href {https://doi.org/10.1103/7wcw-svtz} {\bibfield  {journal} {\bibinfo  {journal} {Phys. Rev. B}\ }\textbf {\bibinfo {volume} {112}},\ \bibinfo {pages} {174507} (\bibinfo {year} {2025})}\BibitemShut {NoStop}%
\bibitem [{\citenamefont {Devereaux}\ \emph {et~al.}(1996)\citenamefont {Devereaux}, \citenamefont {Virosztek},\ and\ \citenamefont {Zawadowski}}]{Devereaux12523}%
  \BibitemOpen
  \bibfield  {author} {\bibinfo {author} {\bibfnamefont {T.~P.}\ \bibnamefont {Devereaux}}, \bibinfo {author} {\bibfnamefont {A.}~\bibnamefont {Virosztek}},\ and\ \bibinfo {author} {\bibfnamefont {A.}~\bibnamefont {Zawadowski}},\ }\bibfield  {title} {\bibinfo {title} {Multiband electronic {R}aman scattering in bilayer superconductors},\ }\href {https://doi.org/10.1103/PhysRevB.54.12523} {\bibfield  {journal} {\bibinfo  {journal} {Phys. Rev. B}\ }\textbf {\bibinfo {volume} {54}},\ \bibinfo {pages} {12523} (\bibinfo {year} {1996})}\BibitemShut {NoStop}%
\bibitem [{\citenamefont {Boyd}\ \emph {et~al.}(2009)\citenamefont {Boyd}, \citenamefont {Devereaux}, \citenamefont {Hirschfeld}, \citenamefont {Mishra},\ and\ \citenamefont {Scalapino}}]{Boyd174521}%
  \BibitemOpen
  \bibfield  {author} {\bibinfo {author} {\bibfnamefont {G.~R.}\ \bibnamefont {Boyd}}, \bibinfo {author} {\bibfnamefont {T.~P.}\ \bibnamefont {Devereaux}}, \bibinfo {author} {\bibfnamefont {P.~J.}\ \bibnamefont {Hirschfeld}}, \bibinfo {author} {\bibfnamefont {V.}~\bibnamefont {Mishra}},\ and\ \bibinfo {author} {\bibfnamefont {D.~J.}\ \bibnamefont {Scalapino}},\ }\bibfield  {title} {\bibinfo {title} {Probing the pairing symmetry of the iron pnictides with electronic {R}aman scattering},\ }\href {https://doi.org/10.1103/PhysRevB.79.174521} {\bibfield  {journal} {\bibinfo  {journal} {Phys. Rev. B}\ }\textbf {\bibinfo {volume} {79}},\ \bibinfo {pages} {174521} (\bibinfo {year} {2009})}\BibitemShut {NoStop}%
\bibitem [{\citenamefont {Sauer}\ and\ \citenamefont {Blumberg}(2010)}]{Sauer014525}%
  \BibitemOpen
  \bibfield  {author} {\bibinfo {author} {\bibfnamefont {C.}~\bibnamefont {Sauer}}\ and\ \bibinfo {author} {\bibfnamefont {G.}~\bibnamefont {Blumberg}},\ }\bibfield  {title} {\bibinfo {title} {Screening of the {R}aman response in multiband superconductors: {A}pplication to iron pnictides},\ }\href {https://doi.org/10.1103/PhysRevB.82.014525} {\bibfield  {journal} {\bibinfo  {journal} {Phys. Rev. B}\ }\textbf {\bibinfo {volume} {82}},\ \bibinfo {pages} {014525} (\bibinfo {year} {2010})}\BibitemShut {NoStop}%
\bibitem [{Note1()}]{Note1}%
  \BibitemOpen
  \bibinfo {note} {Higher-dimensional IRs are regarded as 1D IRs of the subgroup of $G$}\BibitemShut {NoStop}%
\bibitem [{Note2()}]{Note2}%
  \BibitemOpen
  \bibinfo {note} {While our explicit calculations focus on fluctuations of the condensed components included in the mean-field BdG Hamiltonian, the selection rule and the classification tables are more general. In particular, by taking one of $(\Delta _\alpha ,\Delta _\beta )$ to be condensed and the other to represent a fluctuation in a subdominant (noncondensed) pairing channel, the same classification also contains Raman-active noncondensed pairing modes (with the corresponding subdominant coupling kept as an additional input in the kernel).}\BibitemShut {Stop}%
\bibitem [{\citenamefont {Shishidou}\ \emph {et~al.}(2021)\citenamefont {Shishidou}, \citenamefont {Suh}, \citenamefont {Brydon}, \citenamefont {Weinert},\ and\ \citenamefont {Agterberg}}]{Shishidou104504}%
  \BibitemOpen
  \bibfield  {author} {\bibinfo {author} {\bibfnamefont {T.}~\bibnamefont {Shishidou}}, \bibinfo {author} {\bibfnamefont {H.~G.}\ \bibnamefont {Suh}}, \bibinfo {author} {\bibfnamefont {P.~M.~R.}\ \bibnamefont {Brydon}}, \bibinfo {author} {\bibfnamefont {M.}~\bibnamefont {Weinert}},\ and\ \bibinfo {author} {\bibfnamefont {D.~F.}\ \bibnamefont {Agterberg}},\ }\bibfield  {title} {\bibinfo {title} {Topological band and superconductivity in {U}{T}e$_2$},\ }\href {https://doi.org/10.1103/PhysRevB.103.104504} {\bibfield  {journal} {\bibinfo  {journal} {Phys. Rev. B}\ }\textbf {\bibinfo {volume} {103}},\ \bibinfo {pages} {104504} (\bibinfo {year} {2021})}\BibitemShut {NoStop}%
\bibitem [{\citenamefont {Aoki}\ \emph {et~al.}(2022{\natexlab{b}})\citenamefont {Aoki}, \citenamefont {Sakai}, \citenamefont {Opletal}, \citenamefont {Tokiwa}, \citenamefont {Ishizuka}, \citenamefont {Yanase}, \citenamefont {Harima}, \citenamefont {Nakamura}, \citenamefont {Li}, \citenamefont {Homma}, \citenamefont {Shimizu}, \citenamefont {Knebe}, \citenamefont {Flouquet},\ and\ \citenamefont {Haga}}]{Aoki083704}%
  \BibitemOpen
  \bibfield  {author} {\bibinfo {author} {\bibfnamefont {D.}~\bibnamefont {Aoki}}, \bibinfo {author} {\bibfnamefont {H.}~\bibnamefont {Sakai}}, \bibinfo {author} {\bibfnamefont {P.}~\bibnamefont {Opletal}}, \bibinfo {author} {\bibfnamefont {Y.}~\bibnamefont {Tokiwa}}, \bibinfo {author} {\bibfnamefont {J.}~\bibnamefont {Ishizuka}}, \bibinfo {author} {\bibfnamefont {Y.}~\bibnamefont {Yanase}}, \bibinfo {author} {\bibfnamefont {H.}~\bibnamefont {Harima}}, \bibinfo {author} {\bibfnamefont {A.}~\bibnamefont {Nakamura}}, \bibinfo {author} {\bibfnamefont {D.}~\bibnamefont {Li}}, \bibinfo {author} {\bibfnamefont {Y.}~\bibnamefont {Homma}}, \bibinfo {author} {\bibfnamefont {Y.}~\bibnamefont {Shimizu}}, \bibinfo {author} {\bibfnamefont {G.}~\bibnamefont {Knebe}}, \bibinfo {author} {\bibfnamefont {J.}~\bibnamefont {Flouquet}},\ and\ \bibinfo {author} {\bibfnamefont {Y.}~\bibnamefont {Haga}},\ }\bibfield  {title} {\bibinfo {title} {First {O}bservation of the de {H}aas–van {A}lphen {E}ffect and {F}ermi {S}urfaces in the
  {U}nconventional {S}uperconductor {U}{T}e$_2$},\ }\href {https://doi.org/10.7566/JPSJ.91.083704} {\bibfield  {journal} {\bibinfo  {journal} {J. Phys. Soc. Jpn.}\ }\textbf {\bibinfo {volume} {91}},\ \bibinfo {pages} {083704} (\bibinfo {year} {2022}{\natexlab{b}})}\BibitemShut {NoStop}%
\bibitem [{\citenamefont {Weinberger}\ \emph {et~al.}(2024)\citenamefont {Weinberger}, \citenamefont {Wu}, \citenamefont {Graf}, \citenamefont {Skourski}, \citenamefont {Cabala}, \citenamefont {Posp\'{\i}\ifmmode~\check{s}\else \v{s}\fi{}il}, \citenamefont {Prokle\ifmmode~\check{s}\else \v{s}\fi{}ka}, \citenamefont {Haidamak}, \citenamefont {Bastien}, \citenamefont {Sechovsk\'y}, \citenamefont {Lonzarich}, \citenamefont {Vali\ifmmode~\check{s}\else \v{s}\fi{}ka}, \citenamefont {Grosche},\ and\ \citenamefont {Eaton}}]{Weinberger2024}%
  \BibitemOpen
  \bibfield  {author} {\bibinfo {author} {\bibfnamefont {T.~I.}\ \bibnamefont {Weinberger}}, \bibinfo {author} {\bibfnamefont {Z.}~\bibnamefont {Wu}}, \bibinfo {author} {\bibfnamefont {D.~E.}\ \bibnamefont {Graf}}, \bibinfo {author} {\bibfnamefont {Y.}~\bibnamefont {Skourski}}, \bibinfo {author} {\bibfnamefont {A.}~\bibnamefont {Cabala}}, \bibinfo {author} {\bibfnamefont {J.}~\bibnamefont {Posp\'{\i}\ifmmode~\check{s}\else \v{s}\fi{}il}}, \bibinfo {author} {\bibfnamefont {J.}~\bibnamefont {Prokle\ifmmode~\check{s}\else \v{s}\fi{}ka}}, \bibinfo {author} {\bibfnamefont {T.}~\bibnamefont {Haidamak}}, \bibinfo {author} {\bibfnamefont {G.}~\bibnamefont {Bastien}}, \bibinfo {author} {\bibfnamefont {V.}~\bibnamefont {Sechovsk\'y}}, \bibinfo {author} {\bibfnamefont {G.~G.}\ \bibnamefont {Lonzarich}}, \bibinfo {author} {\bibfnamefont {M.}~\bibnamefont {Vali\ifmmode~\check{s}\else \v{s}\fi{}ka}}, \bibinfo {author} {\bibfnamefont {F.~M.}\ \bibnamefont {Grosche}},\ and\ \bibinfo {author} {\bibfnamefont {A.~G.}\
  \bibnamefont {Eaton}},\ }\bibfield  {title} {\bibinfo {title} {Quantum {I}nterference between {Q}uasi-2{D} {F}ermi {S}urface {S}heets in {U}{T}e$_2$},\ }\href {https://doi.org/10.1103/PhysRevLett.132.266503} {\bibfield  {journal} {\bibinfo  {journal} {Phys. Rev. Lett.}\ }\textbf {\bibinfo {volume} {132}},\ \bibinfo {pages} {266503} (\bibinfo {year} {2024})}\BibitemShut {NoStop}%
\bibitem [{\citenamefont {Aoki}\ \emph {et~al.}(2024)\citenamefont {Aoki}, \citenamefont {Sheikin}, \citenamefont {Marquardt}, \citenamefont {Lapertot}, \citenamefont {Flouquet},\ and\ \citenamefont {Knebel}}]{Aoki123702}%
  \BibitemOpen
  \bibfield  {author} {\bibinfo {author} {\bibfnamefont {D.}~\bibnamefont {Aoki}}, \bibinfo {author} {\bibfnamefont {I.}~\bibnamefont {Sheikin}}, \bibinfo {author} {\bibfnamefont {N.}~\bibnamefont {Marquardt}}, \bibinfo {author} {\bibfnamefont {G.}~\bibnamefont {Lapertot}}, \bibinfo {author} {\bibfnamefont {J.}~\bibnamefont {Flouquet}},\ and\ \bibinfo {author} {\bibfnamefont {G.}~\bibnamefont {Knebel}},\ }\bibfield  {title} {\bibinfo {title} {High {F}ield {S}uperconducting {P}hases of {U}ltra {C}lean {S}ingle {C}rystal {U}{T}e$_2$},\ }\href {https://doi.org/10.7566/JPSJ.93.123702} {\bibfield  {journal} {\bibinfo  {journal} {J. Phys. Soc. Jpn.}\ }\textbf {\bibinfo {volume} {93}},\ \bibinfo {pages} {123702} (\bibinfo {year} {2024})}\BibitemShut {NoStop}%
\end{thebibliography}%

\end{document}